%% file: conference_101719.tex
\documentclass[conference,onecolumn,11pt]{IEEEtran}

\usepackage{cite}
\usepackage{amsmath,amssymb,amsfonts}
\usepackage{algorithmic}
\usepackage{graphicx}
\usepackage{caption}
\usepackage{subcaption}
\usepackage{textcomp}
\usepackage{xcolor}
\usepackage{braket}[qcircuit]
\usepackage[hidelinks]{hyperref}
\usepackage{booktabs}
\usepackage{siunitx}
\usepackage{booktabs}
\usepackage{multirow}

\usepackage[utf8]{inputenc}
\usepackage{pgfplots}
\DeclareUnicodeCharacter{2212}{−}
\usepgfplotslibrary{groupplots,dateplot}
\usetikzlibrary{patterns,shapes.arrows}
\pgfplotsset{compat=newest}

\DeclareCaptionLabelSeparator{periodspace}{.\quad}
\DeclareSIUnit{\hartree}{\text{\ensuremath{E_{\mathrm{h}}}}}
\DeclareSIUnit{\atomicunit}{a.u.}
\def\BibTeX{{\rm B\kern-.05em{\sc i\kern-.025em b}\kern-.08em
    T\kern-.1667em\lower.7ex\hbox{E}\kern-.125emX}}

\definecolor{crimson2143940}{RGB}{214,39,40}
\definecolor{darkgray176}{RGB}{176,176,176}
\definecolor{darkorange25512714}{RGB}{255,127,14}
\definecolor{forestgreen4416044}{RGB}{44,160,44}
\definecolor{lightgray204}{RGB}{204,204,204}
\definecolor{mediumpurple148103189}{RGB}{148,103,189}
\definecolor{steelblue31119180}{RGB}{31,119,180}
\definecolor{gray127}{RGB}{127,127,127}
\definecolor{orchid227119194}{RGB}{227,119,194}
\definecolor{sienna1408675}{RGB}{140,86,75}

\begin{document}
\bstctlcite{IEEEexample:BSTcontrol}
\title{Towards large-scale ab initio Molecular Dynamics Simulation with Efficient Sampling-based Quantum Computing Algorithms}

\author{
\IEEEauthorblockN{
Ankit Khandelwal\textsuperscript{1}\IEEEauthorrefmark{2}\IEEEauthorrefmark{1},
Nirmal M R\textsuperscript{1}\IEEEauthorrefmark{1},
Aaryav Mishra\textsuperscript{2},\\
Debasmita Bhoumik\textsuperscript{2},
Dhinakaran Vinayagamurthy\textsuperscript{2} and
M Girish Chandra\textsuperscript{1},
}

\IEEEauthorblockA{\textsuperscript{1}TCS Research, Tata Consultancy Services Limited, India\\
\textsuperscript{2}IBM Research, India\\
\IEEEauthorrefmark{2}khandelwal.ankit3@tcs.com
}
}

\maketitle

\begingroup
    \renewcommand\thefootnote{*}
    \footnotetext{These authors contributed equally to this work.}
\endgroup

\begin{abstract}
Ab initio molecular dynamics (AIMD) requires the calculation of highly accurate electronic-structure energies and nuclear forces at every time step, which can quickly become computationally expensive. Sample-based quantum diagonalization offers a route to such calculations on quantum hardware, but its use in AIMD simulations has so far been limited to active spaces of about 20 qubits.
In these spaces, exact diagonalization is tractable, and the quantum circuits are typically built from standard coupled-cluster amplitudes. 
Here, we introduce SqDRIFT-AIMD, which combines randomized sample-based Krylov quantum diagonalization (SqDRIFT) with analytical nuclear gradients. With SqDRIFT-AIMD, we obtain more accurate ground-state energies than the conventional coupled-cluster singles and doubles (CCSD) driven AIMD simulation for the 40-qubit active space.  
We apply SqDRIFT-AIMD to study the dynamics of proton transfer in the Zundel cation, using circuits executed on IBM Heron r3 quantum processors.
SqDRIFT energies fall below CCSD across the $40$-qubit double-well PES and stay within a few $\si{\milli\hartree}$ of CCSD at 52 qubits.
The $40$-qubit SqDRIFT-AIMD trajectory tracks the SHCI-driven proton transfer to within $\SI{1.33}{\milli\angstrom}$ and exhibits excellent total energy conservation, with a standard deviation of only $\SI{0.245}{\milli\hartree}$.
These results extend quantum-centric AIMD to active spaces beyond the reach of exact diagonalization.
\end{abstract}


\section{Introduction}

Ab initio molecular dynamics (AIMD) has been widely adopted to study various chemical and biological phenomena, including electrochemical reactions in fuel cells, solvation and liquid-phase dynamics, enzyme catalysis, and drug-protein interactions~\cite{marx2009ab}. In general, the computational subroutines required to implement any AIMD simulation workflow are $(i)$ an eigensolver to solve the electronic time-independent Schrödinger equation (TISE) within the Born-Oppenheimer (BO) approximation to compute the electronic energies, $(ii)$ a method to compute nuclear forces as first-order derivatives of the total energy of the molecule, and  $(iii)$ an integration algorithm to solve the classical Hamilton's equation of motion. The electronic energies are computed on-the-fly at every time step, and the nuclei are propagated on the electronic ground-state potential energy surface (PES). Therefore, the quality of AIMD trajectories and of the molecular properties estimated at the end of the simulation depends strongly on the accuracy of the ground-state eigensolver used to calculate the electronic energies.

Over the past many decades, AIMD simulations on classical computers relied on various post-Hartree-Fock (HF) electronic structure methods, including, but not limited to, configuration interaction with singles and doubles (CISD), selected configuration interaction (SCI), and coupled cluster singles and doubles (CCSD). These methods have different computational complexities and accuracies, with the standard coupled cluster singles and doubles with perturbative triples (CCSD(T)), often regarded as the ``gold standard" method of quantum chemistry, exhibiting a computational scaling of $O(M_O^7)$, where $ M_O$ is the number of molecular orbitals. For practical, real-world applications, AIMD simulations are performed for very long durations, typically 10 to 100 picoseconds, with femtosecond time steps. Therefore, the unfavorable scaling with system size of high-accuracy methods, specifically, CCSD and CCSD(T), has limited their applicability in AIMD simulation workflows to molecules with only a few tens of atoms.

With the advent of quantum computers, numerous quantum algorithms have been developed to compute ground-state properties of many-electron systems. The most prominent classes of these algorithms include the quantum phase estimation (QPE) and its variants~\cite{qpe1,qpe2}, the hybrid variational quantum eigensolver (VQE)~\cite{vqe1,vqe2,vqe3}, and the more recent quantum selected configuration interaction (QSCI) methods~\cite{qsci1, kqd1}. Although QPE promises exponential advantage in solving the electronic structure problem, its implementation requires deep quantum circuits, making it impractical on near-term quantum processors. Compared to QPE, VQE allows the use of shallower quantum circuits (also known as ansatzes); hence, early efforts to perform AIMD on quantum computers utilized VQE to compute ground-state energies~\cite{vqe-aimd1,vqe_aimd2,vqe-aimd3}.
However, VQE requires classical optimization of circuit parameters at every geometry and many measurements to estimate the energy, both of which are difficult to sustain over the thousands of time-steps in an AIMD trajectory on present-day noisy hardware~\cite{gonthier2022measurements}. 

Sample-based quantum diagonalization (SQD)~\cite{sqd1} avoids these bottlenecks by using the quantum processor only to sample electronic configurations, while classical high-performance computers obtain the energy and wavefunction by projecting and diagonalizing the Hamiltonian in the sampled subspace. SQD has recently been used to drive AIMD simulations~\cite{clevelandclinic}, including proton transfer in the Zundel cation~\cite{iitmpaper} $\text{H}_\text{5}\text{O}_\text{2}^+$ in a $(10\text{e},10\text{o})$ active space~ and dynamics of $\text{CH}_\text{4}$ in a $(10\text{e},9\text{o})$, $21$-qubit active space~\cite{clevelandclinic}. These studies, however, are limited to active spaces of about $20$ qubits, where exact diagonalization of the full active-space Hamiltonian is still tractable on classical computers, and they rely on quantum circuits whose parameters are taken from classical coupled-cluster amplitudes. Moreover, these works used the minimal STO-nG basis sets to describe the chemical systems, which may not be well-suited to describe the PES of strongly correlated systems even qualitatively.

A further difficulty is that the accuracy of SQD depends on whether the quantum circuit actually samples the dominant determinants of the true ground state. Ansatzes that approximate the ground-state wavefunction, such as unitary coupled cluster (UCC)~\cite{romero2019strategies} or local unitary cluster Jastrow (LUCJ)~\cite{lucj}, are not designed for this purpose and may miss important configurations~\cite{sqdrift, skqd}. Sample-based Krylov quantum diagonalization (SKQD)~\cite{skqd} addresses this by sampling from time-evolved states, and comes with convergence guarantees under sparsity assumptions on the ground state. Its randomized variant, SqDRIFT~\cite{sqdrift}, compiles the time evolution using the qDRIFT protocol~\cite{qdrift}, which reduces the circuit depth compared to Trotterization and therefore makes large active spaces amenable on current quantum hardware. Neither approach has, to the best of our knowledge, been used for AIMD.

In this work, we introduce SqDRIFT-AIMD, an AIMD workflow that combines SqDRIFT ground-state energies and wavefunctions with analytical nuclear forces computed from the subspace-reduced density matrices~\cite{analytic_sci_grad}, and propagates the nuclei with the velocity-Verlet integrator~\cite{velocity_verlet1,velocity_verlet2}. The sampling circuits require neither variational optimization nor amplitudes from classical correlated methods. We apply SqDRIFT-AIMD to simulate proton transfer dynamics in the Zundel cation, using circuits executed on the IBM Heron r3 processor \texttt{ibm\_boston}. 
Our main contributions are:
\begin{itemize}
    \item To the best of our knowledge, our work is the first AIMD simulation driven by randomized sample-based Krylov quantum diagonalization (SqDRIFT), 
    with a system size of $10^{10}$, which is over $10^5$ times the earlier quantum sampling with AIMD work. 
    \item Potential energy surfaces of $\text{H}_\text{5}\text{O}_\text{2}^+$ along the proton-transfer coordinate in a $40$-qubit $(16\text{e},20\text{o})$ and a $52$-qubit $(16\text{e},26\text{o})$ active space, benchmarked against CISD, CCSD, CCSD(T), and SHCI + PT2 references. In the $40$-qubit space, SqDRIFT energies fall below CCSD at every geometry, using a subspace containing only $0.039\%$ of the full determinant space.
    \item A $40$-qubit SqDRIFT-AIMD trajectory, about twice the size of the active spaces in previous sample-based AIMD studies, that reproduces a full forward-and-back proton-transfer cycle. It agrees with a SHCI-driven reference trajectory to within $\SI{1.33}{\milli\angstrom}$ in $\delta$, and conserves the total energy with a standard deviation of $\SI{0.245}{\milli\hartree}$.
    \item We demonstrate that highly accurate ground-state properties can be estimated with the SqDRIFT-AIMD workflow for active spaces that are beyond the scale of exact diagonalization.
\end{itemize}

The remainder of the paper is organized as follows. Section~\ref{sec:related} gives the related works. Section~\ref{sec:backg} reviews sample-based quantum diagonalization and sample-based Krylov quantum diagonalization. Section~\ref{sec:methods} describes the SqDRIFT ground-state solver, the analytical nuclear-force calculation, the velocity-Verlet integrator, and the full SqDRIFT-AIMD workflow. Section~\ref{sec:computational_details} gives the computational details, including the molecular system, active spaces, hardware execution, and classical references. Section~\ref{sec:results} presents the potential energy surfaces in the $40$- and $52$-qubit active spaces and the proton-transfer dynamics. Section~\ref{sec:conclusion} concludes the paper and discusses future directions.

\section{Related Work}\label{sec:related}

The quantum mechanical treatment of the shared proton in the Zundel cation was established early on. Ab initio path-integral simulations by Tuckerman et al.~\cite{tuckermanreview}\ showed that the proton in $\text{H}_\text{5}\text{O}_\text{2}^+$ behaves essentially classically, whereas in the low-barrier $\text{H}_\text{3}\text{O}_\text{2}^-$ bond zero-point motion dominates even at room temperature, a difference that is attributed to a small change in the O--O separation~\cite{tuckermanreview}. Correlated forces computed on-the-fly followed much later. Mouhat et al.\ used variational quantum Monte Carlo forces to drive path-integral Langevin dynamics of $\text{H}_\text{5}\text{O}_\text{2}^+$, absorbing the statistical noise of the forces into a thermostat matched to their covariance, and found that nuclear quantum effects outweigh thermal ones well beyond room temperature~\cite{mouhat2017fully}. The same approach was later extended to the protonated water hexamer, where delocalized, short Zundel-like configurations persist up to \SI{300}{\kelvin} and proton transfer is most efficient between 250 and \SI{300}{\kelvin}~\cite{mouhat2023thermal}.

Most current work takes a different route, fitting the potential energy surface a priori. For $\text{H}_\text{5}\text{O}_\text{2}^+$, the reference surface is a refit of $48{,}199$ CCSD(T)/aug-cc-pVTZ energies with fast analytic gradients~\cite{kotaru2026new}. On a neural-network surface of comparable quality, full-dimensional quantum nuclear dynamics in $51$ coupled degrees of freedom now reproduces the infrared spectrum of the extended Zundel complex $\text{H}^+(\text{H}_\text{2}\text{O})_\text{6}$~\cite{mendivetapiaNN}.
For hydrated Nafion, neural-network potentials trained on density-functional data now simulate $10000-20000$-atom models for \SI{31}{\nano\second} and reproduce the measured hydrogen and hydronium self-diffusion, but their accuracy stays tied to the training method: the density is underestimated, which the authors partly attribute to the revPBE functional~\cite{yoshimotoXXL}.

Classical simulations can therefore reach the membrane scale, but only by using potentials fitted to a fixed reference method. For this system, the only available source of on-the-fly correlated forces is stochastic sampling, placing it in the same regime as a quantum sampler.

Quantum-selected Configuration Interaction (QSCI) introduced the idea of using a quantum processor to sample electronic configurations, and subsequently diagonalizing the Hamiltonian classically within the subspace spanned by those configurations ~\cite{ qsci1}. SQD added configuration recovery and supercomputer-scale diagonalization, reaching circuits of up to $77$ qubits~\cite{sqd1}, and SKQD and SqDRIFT replaced the heuristic sampling circuit with time evolution, the latter compiled shallowly through qDRIFT~\cite{skqd,sqdrift,qdrift}. 
 These methods now operate beyond routine exact diagonalization, though not beyond exact configuration interaction, which has reached $10^{12}$ and $10^{15}$ determinants~\cite{gaoistributedfci,shayitfci}. All of them, however, have been applied at single molecular geometries. 
Section \ref{sec:backg} describes SQD and SKQD in further detail.

Reaction dynamics is widely seen as the next frontier for quantum chemistry, beyond static ground-state energies~\cite{bidartMD}. Simulating electrons and nuclei fully quantum mechanically remains a fault-tolerant problem, estimated to need more than a thousand logical qubits even for a small reaction~\cite{pocrnicFT}, so near-term work follows the Born--Oppenheimer route.

Sample-based methods have entered this landscape in the past year. Das et al.\ drove every step of gas-phase and solvated trajectories with SQD on IBM hardware, obtaining analytic gradients from the subspace reduced density matrices in STO-3G active spaces of at most $21$ qubits~\cite{clevelandclinic}. Kumar et al.\ applied SQD-AIMD to proton transfer between two water molecules through a transient Zundel cation, in a $(10\text{e},10\text{o})$ STO-3G active space, with LUCJ circuits, but using an IBM noise model~\cite{iitmpaper}. Meanwhile, geometry-dependent SQD reused quantum samples across neighboring geometries, aligning the molecular orbitals and resampling only when they drift, and computed forces on circuits of up to $124$ qubits, but only for geometry optimizations and reaction paths~\cite{fortindeschenes}.

SQD-driven dynamics has therefore been demonstrated only in active spaces small enough for exact diagonalization, while larger SQD force calculations have been used only for geometry optimizations and fixed reaction paths, without propagating the nuclei in time. SqDRIFT-AIMD builds on these threads: it extends SqDRIFT from static, single-point energies to molecular dynamics, obtains forces from subspace reduced density matrices as in Ref.~\cite{clevelandclinic}, and reuses samples between neighboring time steps in the spirit of Ref.~\cite{fortindeschenes}. The nuclei are propagated classically, which Ref.~\cite{tuckermanreview} supports for $\text{H}_\text{5}\text{O}_\text{2}^+$.

\section{Background} \label{sec:backg}
\begin{figure}
    \centering
    \includegraphics[width=0.9\linewidth]{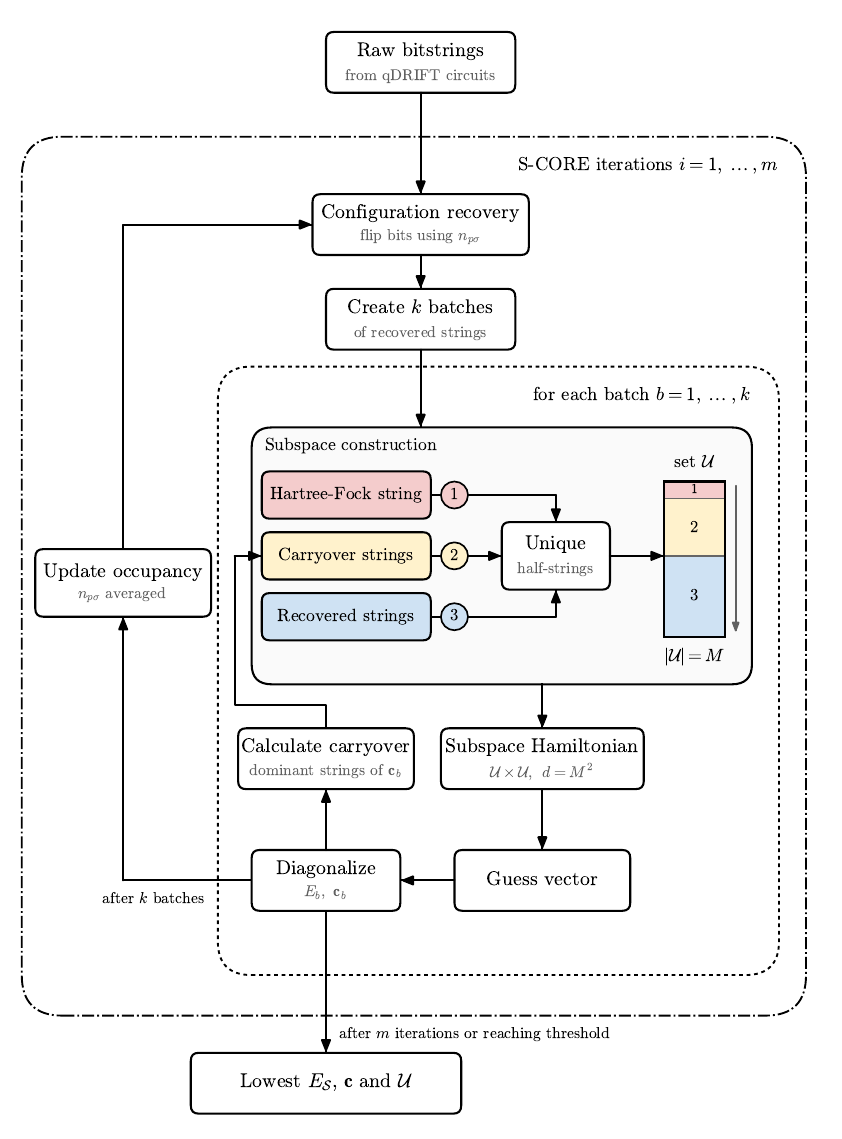}
    \caption{S-CORE workflow used to calculate the ground-state energy.}
    \label{fig:sqdrift_workflow}
\end{figure}
\subsection{Sample-based Quantum Diagonalization}\label{SQD}
In this section, we briefly describe the generic workflow for calculating the ground-state energy of a many-electron Hamiltonian using the framework of sample-based quantum diagonalization (SQD)~\cite{sqd1, sqd2}. The non-relativistic many-electron Hamiltonian under the Born-Oppenheimer approximation can be written in second-quantized formalism as
\begin{equation}
    \hat{H}_{ele} = \sum_{pq,\sigma} h_{pq}\hat{a}_{p\sigma}^{\dagger}\hat{a}_{q\sigma} + \frac{1}{2}\sum_{pqrs, \sigma \tau} \braket{pr|qs} \hat{a}_{p\sigma}^{\dagger}\hat{a}_{q\tau}^{\dagger}\hat{a}_{s\tau}\hat{a}_{r\sigma}
    \label{eqn: electronic Hamiltonian}
\end{equation}
where $p$, $q$, $r$, and $s$ label the spatial orbitals, and $\sigma$, $\tau$ $\in \{\alpha, \beta\}$ denote the associated spin components. The real coefficients $h_{pq}$ and $\braket{pr|qs}$ represent the one- and two-electron integrals, respectively, and $\hat{a}_{p\sigma}$ ($\hat{a}_{p\sigma}^{\dagger}$) denotes the fermionic annihilation (creation) operator associated with the spin-orbital $p\sigma$.

To simulate the above Hamiltonian on a quantum processor, we map the fermionic operators onto Pauli operators acting on a set of $N$ qubits using the Jordan-Wigner transformation\cite{jordan1928pauli}. Here, $N$ is twice the number of spatial orbitals $M_O$ used to describe the Hamiltonian in Eq.~\eqref{eqn: electronic Hamiltonian}, and the single-qubit basis states $\ket{0}$ and $\ket{1}$ represent the occupation of the corresponding molecular spin-orbital. Hence, the many-electron ground-state wavefunction can be described as a linear combination of $N$-qubit computational basis states, each representing a Slater determinant in the second-quantization formalism.

The SQD workflow begins by preparing an approximate ground-state wavefunction of the molecule of interest on a quantum computer using an $N$-qubit quantum circuit. This quantum circuit (often referred to as an ansatz) is repeatedly executed and measured in the computational basis for a finite number of times (shots) to obtain a set of bitstrings $\{x\}$ that follows some probability distribution $P_{\psi}(x)$. The sampled bitstrings represent the many-electron configurations (or Slater determinants) in the occupation-number basis. 

Due to the effect of various forms of quantum noise and qubit decoherence in present-day quantum hardware, not all the sampled bitstrings satisfy the particle-number and spin symmetries of the studied chemical system. The sampled bitstrings with their corresponding probabilities are fed to a classical computer, where a self-consistent configuration recovery (S-CORE) procedure is performed iteratively to correct the unphysical bitstrings. 
In the first iteration of the S-CORE procedure, $k$ batches of $d$ configurations are sampled from the set of valid bitstrings, where $d$ is the dimension of each batch (subspace). The original electronic Hamiltonian (in Eq.~\eqref{eqn: electronic Hamiltonian}) is then projected onto each subspace to yield a lower-dimension Hamiltonian $\hat{H}_{k}$. By leveraging the capabilities of classical high-performance computing (HPC) systems, the projected Hamiltonian $\hat{H}_{k}$ is diagonalized exactly to obtain an approximate ground-state energy $\tilde{E}^{k}_{0}$ and wavefunction $\tilde{\Psi}^{k}_{0}$ for each batch. An initial estimate of the average orbital occupation for each spin-orbital is then computed using all subspace ground-state eigenvectors as
\begin{equation}
    n_{p\sigma} = \frac{1}{K}\sum_k\bra{\Psi^{k}}\hat{n}_{p\sigma}\ket{\Psi^{k}}
    \label{eqn: average orbital occupation}
\end{equation}
This is used in subsequent iterations of S-CORE to probabilistically flip a certain number of bits in each unphysical bitstring. Note that correcting the unphysical bitstrings based on average orbital occupation doesn't guarantee the recovery of important bitstrings, and the resulting valid bitstring could be any one of the possible bitstrings that span an exponentially large Hilbert space. It is, therefore, crucial that the bitstrings obtained through sampling the quantum circuit are the actual dominant set of Slater determinants in the ground-state wavefunction to yield very accurate ground-state energies.

These iterations, which involve sampling batches of recovered configurations, projecting, and diagonalizing, are repeated until the ground-state energy reaches a predefined convergence criterion. The accuracy of ground-state energies derived from an SQD calculation largely depends on the quality of the subspaces used for projection and diagonalization of the active-space electronic Hamiltonian. Therefore, SQD holds the potential to yield near-exact ground-state energies when the number of important determinants in the exact ground-state wavefunction is much smaller than the full Fock-space dimension $D$.

\begin{table}[]
\centering
\caption{Active spaces, determinant-space dimensions and sampled subspace sizes used in this work, compared against previous SQD-based AIMD studies. Values not reported in the original works are marked `-'.}
\label{tab:subspace_comparison}
\resizebox{.95\textwidth}{!}{
\begin{tabular}{@{}ccccccc@{}}
\toprule
System                                        & Active Space                                       & Qubits & Full Determinant Space ($D$) & Samples per Batch ($M$) & $d_{\max}$          & $d_{\max}/D$ \\ \midrule
\multirow{3}{*}{$\text{H}_\text{5}\text{O}_\text{2}^+$} & $(16\text{e},20\text{o})$                                             & $40$     & $1.59\cdot10^{10}$             & $1000, 2000, 2500  $             & $6.25\cdot10^6$ & $0.039\%$      \\
                                             &  $(16\text{e},26\text{o})$                                             & $52$     & $2.44\cdot10^{12}$             & $5000$                           & $2.50\cdot10^7$ & $0.001\%$      \\
                                             &  $(10\text{e},10\text{o})$$~$\cite{iitmpaper}       & -      & $6.35\cdot10^4$              & $700-900$                              & -               & -            \\ \midrule
$\text{CH}_\text{4}$                                        &  $(10\text{e},9\text{o})$$~$\cite{clevelandclinic} & $21$     & $1.59\cdot10^4$              & $200,400,800$                              & -               & $5\%$          \\ \bottomrule
\end{tabular}
}
\end{table}

\subsection{Sample-based Krylov quantum diagonalization}\label{SKQD}
We provide a brief overview of the SKQD algorithm proposed in reference~\cite{skqd}. In principle, one can use any parametrized quantum circuit to sample the bitstrings for subspace projection and diagonalization of the target Hamiltonian $\hat{H}_{ele}$. However, the accuracy of the SQD method largely depends on the ability of the quantum circuit to generate all the important determinants of the true ground-state wavefunction for a given finite number of shots. In other words, the bitstrings sampled from the chosen quantum circuit should ideally follow a uniform probability distribution over all the important bitstrings to guarantee that none of the important bitstrings are missed. In this regard, conventional quantum chemistry ansatzes, such as UCC~\cite{uccsd}, parallelized Givens singles and doubles (PGSD)~\cite{pgsd}, LUCJ~\cite{lucj}, that approximate the ground-state wavefunction may not be well-suited for sample-based protocols, as discussed in detail in references~\cite{sqdrift, skqd}. 

In contrast, the SKQD approach, inspired by Krylov quantum diagonalization (KQD) methods~\cite{kqd1,kqd2}, obtains the relevant bitstrings for subspace projection and diagonalization via sampling time-evolved quantum states. These states are prepared by repeated application of the unitary time-evolution operator $e^{-{\iota}Ht}$ for a time $t$ on a reference wavefunction $\ket{\Psi_{ref}}$ as given by
\begin{equation}
    \ket{\Psi_k} = (\prod_{j=1}^{k}e^{-iHt})\ket{\Psi_{ref}} = e^{-iHkt}\ket{\Psi_{ref}}
\end{equation}

where $H$ is the $N$-qubit Hamiltonian obtained through the JW transformation applied to $\hat{H}_{ele}$. It is important to note that the reference state $\ket{\Psi_{ref}}$ should have a non-zero overlap with the exact ground state of $H$, and it can be prepared with shallow circuits on present-day quantum hardware. The relevant bitstrings spanning the target subspaces are obtained by measuring the time-evolution circuits corresponding to each $k$ in the computational basis. Under sparsity assumptions on the exact ground state, it has been shown in~\cite{skqd} that SKQD provides convergence guarantees for efficiently estimating highly accurate ground-state energies. 

\section{Methods}
\label{sec:methods}

\begin{figure}
    \centering
    \includegraphics[width=0.8\linewidth]{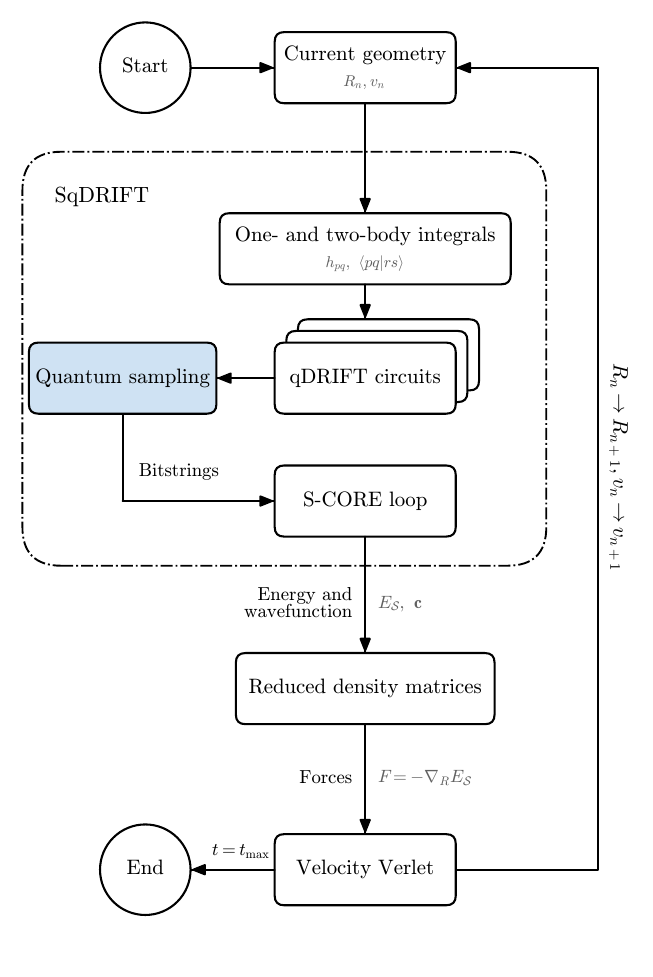}
    \caption{SqDRIFT-AIMD workflow. Only the sampling step is performed on a QPU.}
    \label{fig:sqdrift_aimd_workflow}
\end{figure}

\subsection{Ground-state energy calculation with SqDRIFT}\label{SqDRIFT}
In this section, we describe the SqDRIFT algorithm\cite{sqdrift} used to compute the ground-state energies and wavefunctions in our AIMD simulations. As discussed in Section~\ref{SKQD}, applying the SKQD method requires the implementation of the time-evolved unitaries $e^{-iHkt}$ on quantum processors. Canonical approaches for compiling these time-evolved unitaries, such as Trotterization~\cite{quantumalgosfordynamics} and linear combination of unitaries (LCUs)~\cite{lcu}, often yield deep quantum circuits, particularly for large active spaces, that cannot be executed reliably on near-term quantum computers. To minimize the overhead on circuit depth, the SqDRIFT approach compiles the unitaries $e^{-iHkt}$ through the randomized quantum stochastic drift (qDRIFT) technique proposed in reference~\cite{qdrift}.

In Fig.~\ref{fig:sqdrift_workflow} we show the workflow followed in this work to calculate the ground-state energy using SqDRIFT. Raw bitstrings sampled after execution of qDRIFT circuits on quantum processors are corrected by configuration recovery using the average orbital occupations $n_{p\sigma}$ and split into $k$ batches. For each batch, the set $\mathcal{U}$ of $M$ unique half-strings is filled in priority order: ($1$) the Hartree-Fock string, ($2$) carryover strings from the previous diagonalization, and ($3$) recovered strings. The Hamiltonian is then projected onto $\mathcal{U}\times\mathcal{U}$ ($d=M^2$) and diagonalized. The diagonalization gives the batch energy $E_b$ and the eigenvector coefficients $\mathbf{c}_b$. The dominant strings from $\mathbf{c}_b$ ($>10^{-4}$) are carried over to the next batch. After $k(=5)$ batches, the occupations are updated, and the cycle repeats for $m(=11)$ iterations. An energy improvement threshold of $10^{-5}\,\si{\hartree}$ is used for early stopping. 
The final subspace energy $E_S$, coefficients $\mathbf{c}$ and the final set $\mathcal{U}$ are then returned. This loop is followed for all batches except the first batch of a new geometry. For the first batch, the average occupancy vector does not exist, so only valid strings (i.e., those with the correct number of $\alpha$ and $\beta$ electrons) are passed on. The first iteration is then stopped after the first batch and configuration recovery is performed. Thus, a maximum of $51(1+10\cdot5)$ diagonalizations are used. For the first batch of all geometries, except the first one in a PES, we also follow a warm start process as described in Section~\ref{sec:40pes}. 
SqDRIFT inherits a convergence guarantee from SKQD~\cite{sqdrift}: if the ground state is concentrated on a polynomial number of determinants and has non-negligible overlap with the reference state, these determinants are sampled with high probability, and the subspace energy then converges to the ground-state energy. The proof, however, assumes noiseless circuits, so hardware noise and configuration recovery lie outside it. Nevertheless, the resulting energies remain variational upper bounds on the energy.

\subsection{Nuclear force calculation}
In AIMD simulation workflows, the motion of the nuclei is governed by the classical Hamilton's equations of motion given by 

 \begin{gather}
    \frac{dR_I}{dt} = \frac{P_I}{M_I}  \notag \\
    \frac{dP_I}{dt} = F_I = -\nabla_I(E_0(R)+V_{NN}(R))    
    \label{eqn: Hamilton's eqn}
\end{gather}   

where $R_I$, $P_I$, $M_I$ denote the positions, momenta and masses along the nuclear coordinate $I$, respectively, and $R$ is the set of all nuclear coordinates representing the geometry of the molecule. The force $F_I$ is expressed as the negative of the first-order gradient of the total energy of the molecule (i.e., the sum of the electronic ground-state energy $E_0(R)$ and nuclear repulsion energy $V_{NN}(R)$) with respect to $R_I$. The accuracy of AIMD trajectories depends heavily on the accuracy of the nuclear forces, and in our simulations, we compute the forces as analytical gradients of the total energy.

We follow the theory discussed in reference~\cite{analytic_sci_grad} to compute analytic energy gradients. As given in Eq.~(7) of this article, the analytic derivative of the electronic energy $E_0(R)$ along the nuclear coordinate $R_I$ is given by
\begin{equation}
    \frac{\partial E_0(R)}{\partial R_I} = \sum_{pq}\frac{\partial h_{pq}}{\partial R_I}\gamma_{pq} + \frac{1}{2}\sum_{pqrs}\frac{\partial \braket{pr|qs}}{\partial R_I}\Gamma_{prsq} + 2\sum_{tp}F_{tp}U_{tp}^{R_I}
    \label{eqn: analytic energy derivative}
\end{equation}

where $\gamma_{pq}$ and $\Gamma_{prsq}$ denote the one- and two-particle reduced density matrices (RDMs), respectively. Here, $U_{tp}^{R_I}$ represents the change in the molecular orbital (MO) coefficients with respect to the change in the nuclear coordinate $R_I$, and is obtained by solving the coupled-perturbed Hartree-Fock (CPHF) equations~\cite{cphf}. The CI Lagrangian $F_{tp}$ is written as 
\begin{equation}
    F_{tp} = \sum_q\gamma_{pq}h_{tq} + \sum_{qrs}\Gamma_{prsq}\braket{tr|qs}
\end{equation}

We compute the electronic ground-state energy $E_0(R)$ and the associated wavefunction $\ket{\Psi_0(R)}$ using the SqDRIFT method discussed in Sec.~\ref{SqDRIFT}. The ground-state wavefunction as a function of molecular geometry $R$ can be expressed as
\begin{equation}
   \ket{\Psi_0(R)} = \sum_i c_i(R)\ket{\phi_i}
   \label{eqn: SCI wavefunction}
\end{equation}

where $\ket{\phi_i}$ represents a Slater determinant in the many-electron basis and $c_i(R)$ is its corresponding coefficient. The reduced density matrices (RDMs) in Eq.~\eqref{eqn: analytic energy derivative} depend only on the coefficients $c_i(R)$ and the occupation of the corresponding determinants. We used PySCF's \texttt{fci.selected\_ci}~\cite{pyscf1,pyscf2} module to calculate these RDMs from the eigenvector $\{c_i(R)\}$ obtained from the SqDRIFT method. The computed RDMs were then passed to the analytical gradient calculation routine implemented in PySCF's \texttt{mcscf.CASCI} module to obtain the energy gradients. 

In contrast to calculating forces as analytical energy gradients, a common practical approach in AIMD workflows is to compute them numerically using finite-difference methods. In this context, the Hellmann-Feynman theorem~\cite{hellmann-feynman_theorem} is applied to evaluate forces as
\begin{equation}
    F_I = -\frac{\partial}{\partial R_I}\bra{\Psi} \hat{H}_{ele}\ket{\Psi} = -\bra{\Psi} \frac{\partial \hat{H}_{ele}}{\partial R_I}\ket{\Psi}
    \label{eqn: HFT}
\end{equation}

where $\ket{\Psi}$ is the exact electronic wavefunction. The derivative $\frac{\partial\hat{H}_{ele}}{\partial R_I}$ is usually computed using the central finite-difference method as $\frac{\hat{H}_{ele}(R+\Delta e_I)-\hat{H}_{ele}(R-\Delta e_I)}{2\Delta d}$, where $\Delta d$ is a small displacement and $e_I$ is the unit vector along $R_I$. The major drawbacks of this approach are $(i)$ as the computed wavefunction is only an approximation to the exact wavefunction, the Pulay contributions ($\bra{\frac{\partial \Psi}{\partial R_I}}\hat{H}_{ele}\ket{\Psi}$, $\bra{\frac{\partial \Psi}{\partial R_I}}\hat{H}_{ele}\ket{\Psi}$) due to the dependence of the wavefunction on the nuclear coordinates could have a non-negligible contribution, $(ii)$ the error in the calculation of the derivative $\frac{\partial\hat{H}_{ele}}{\partial R_I}$ is $O(\Delta d^2)$ with the central finite-difference method. As analytical energy gradients are not subject to these approximations, they are expected to yield more accurate nuclear forces and therefore lead to more stable AIMD trajectories.

\begin{figure}
    \centering
    \begin{subfigure}[t]{0.4\textwidth}
            \centering
            \parbox[t][4cm][t]{\linewidth}{
            \input{ccsd_energy}
            }
            \par
            \caption{}
            \label{fig:ccsd_dynamics_energy}
    \end{subfigure}
    \hspace{19mm}
    \begin{subfigure}[t]{0.4\textwidth}
            \centering  
            \parbox[t][4cm][t]{\linewidth}{
            \input{ccsd_delta}
            }
            \par
            \caption{}
            \label{fig:ccsd_dynamics_delta}
    \end{subfigure}
    \caption{CCSD-driven AIMD of H$_\text{5}$O$_\text{2}^+$ in the $40$-qubit $(16e,20o)$ active space used as reference. (\subref{fig:ccsd_dynamics_energy}) Potential, kinetic, and total energies. (\subref{fig:ccsd_dynamics_energy}) Proton-transfer coordinate $\delta$.}
    \label{fig:ccsd_baseline}
\end{figure}
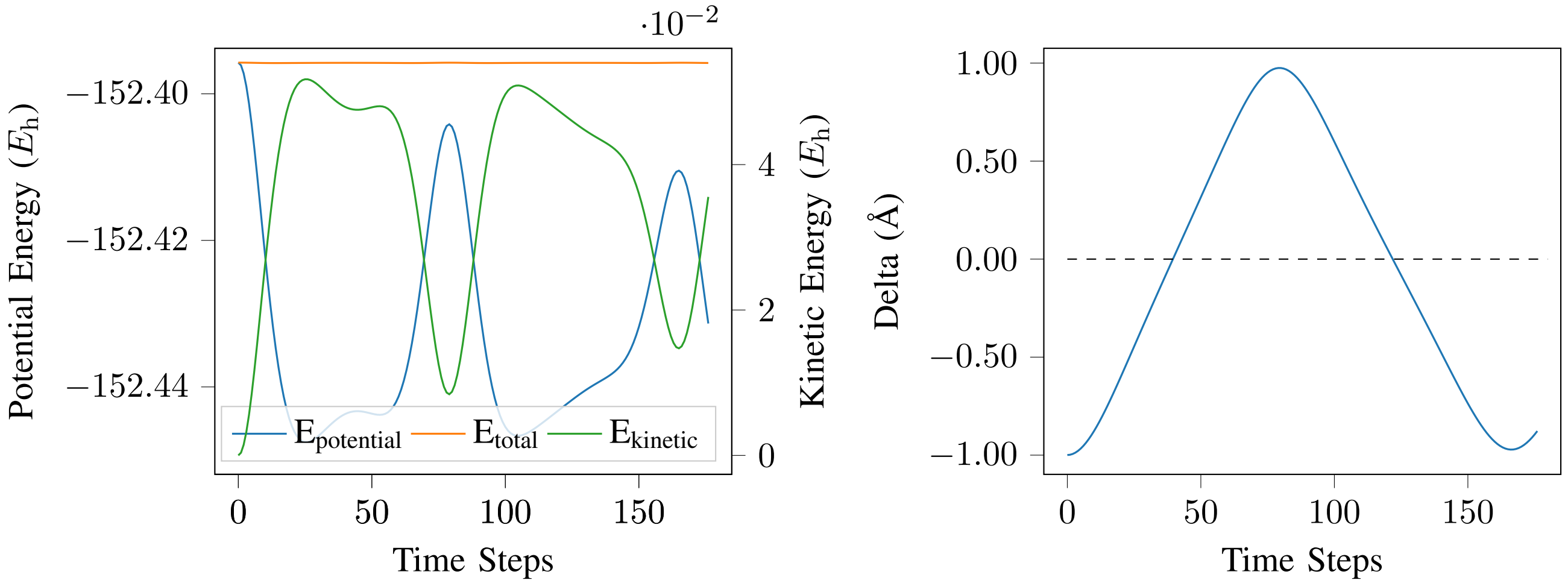

\subsection{Velocity-Verlet integrator}\label{sec:velocity}
The AIMD trajectories are obtained by solving Eq.~\eqref{eqn: Hamilton's eqn} to determine the positions and velocities of all nuclei at every time step. In this work, we use the velocity-Verlet integration algorithm~\cite{velocity_verlet1, velocity_verlet2} to update the position $R_I$ and velocity $v_I(t)$ of nuclear coordinate $I$ as
\begin{equation}
    R_I(t+\Delta t) = R_I(t) + v_I(t)\Delta t + \frac{1}{2M_I}F_I(t)\Delta t^2 
    \label{eqn: position update}
\end{equation}
\begin{equation}
    v_I(t+\Delta t) = v_I(t) + \frac{1}{2M_I}(F_I(t) + F_I(t+\Delta t))\Delta t 
    \label{eqn: velocity update}
\end{equation}
Note that only one evaluation of forces per time step is required by this method in the simulation of AIMD trajectories. Compared to other numerical integration techniques, such as the Runge-Kutta method~\cite{runge-kutta}, the Verlet method offers better numerical stability and conservation of total energy.

\subsection{The SqDRIFT-AIMD workflow}

Figure~\ref{fig:sqdrift_aimd_workflow} describes the SqDRIFT-AIMD workflow. At each geometry $R_n$, the active-space integrals define the qDRIFT circuits that are sampled on quantum hardware. The sampled bitstrings go through the S-CORE loop, which gives the subspace energy and eigenvector. Analytical forces are calculated from the reduced density matrices and propagated using the velocity-Verlet method.

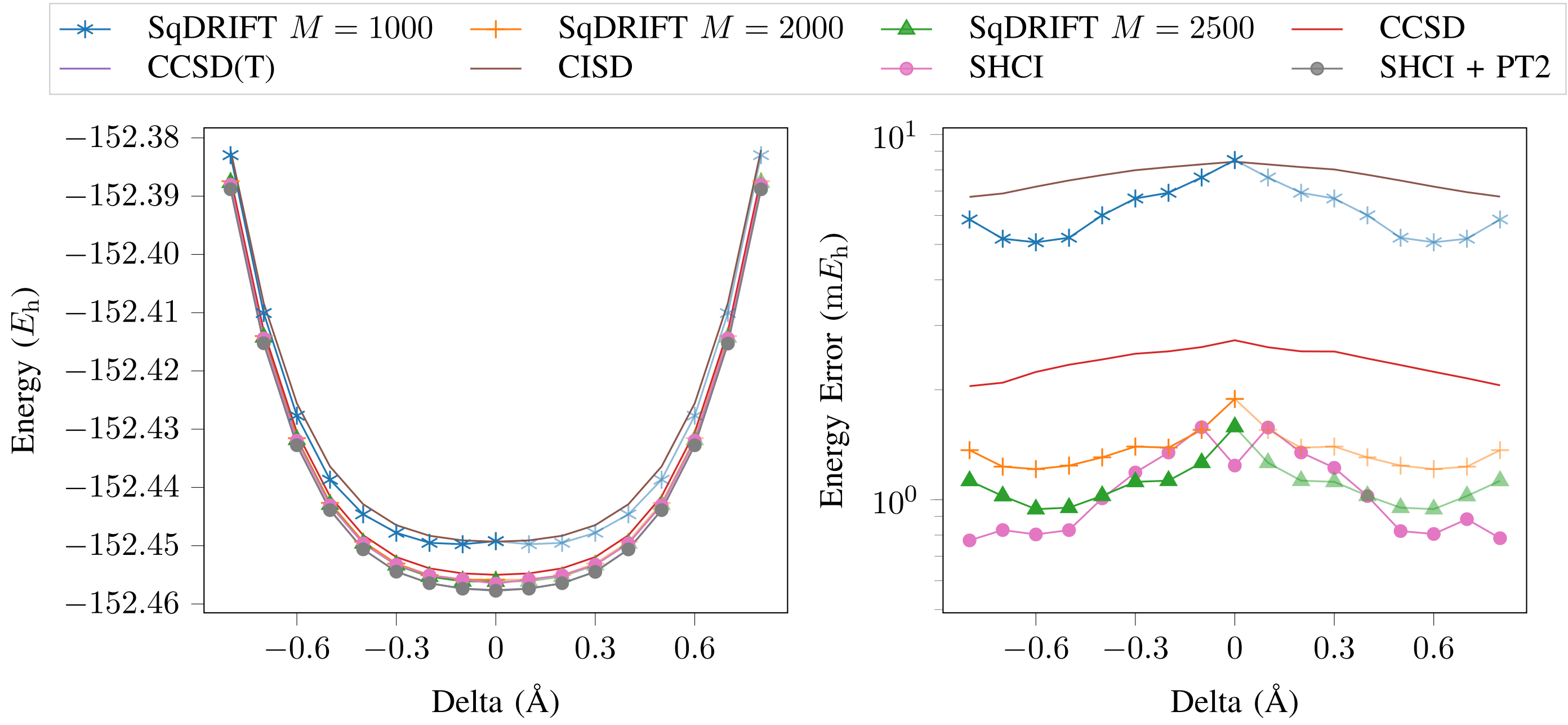
\begin{figure}
    \centering
    \makebox[\textwidth][c]{
    \hspace{20pt}
    \pgfplotslegendfromname{sharedlegend_40q_2.4017}
    }
    \par\vspace{5pt}
    \begin{subfigure}[t]{0.48\textwidth}
        \centering
        \parbox[t][6.5cm][t]{\linewidth}{
        \input{40q_2.4017_energies}
        }
        \par
        \caption{}
        \label{fig:40q_2.4017_energies}
    \end{subfigure}
    \hfill
    \begin{subfigure}[t]{0.48\textwidth}
        \centering
        \input{40q_2.4017_error}
       \caption{}
        \label{fig:40q_2.4017_error}
    \end{subfigure}
    \caption{PES of H$_\text{5}$O$_\text{2}^+$ in the $40$-qubit active space at $r(\text{O}_a\text{O}_b)=\SI{2.4017}{\angstrom}$. (\subref{fig:40q_2.4017_energies}) Energies along $\delta$ from SqDRIFT and the classical benchmarks. At this separation, the PES has a single well centered at $\delta=0$. (\subref{fig:40q_2.4017_error}) Errors relative to SHCI + PT2.}
    \label{fig:40q_2.4017}
\end{figure}
\section{Computational Details}
\label{sec:computational_details}
This section gives the full computational protocol behind the results in Section~\ref{sec:results}. We first describe the molecular system under consideration. We then describe how the active spaces were built, the parameters for SqDRIFT sampling and subspace diagonalization, the classical reference calculations, and the molecular dynamics protocol. All one- and two-electron integrals, RHF orbitals, reference energies, and analytical gradients were computed with PySCF~\cite{pyscf1,pyscf2}. At each geometry, every method uses the same basis set, frozen core, and active orbitals, and therefore the same active-space Hamiltonian.

The Zundel cation $\text{H}_\text{5}\text{O}_\text{2}^+$ contains $20$ electrons and has a closed-shell singlet ground state. It was described with the 6-31G basis set. We label the two Oxygen atoms O$_a$ and O$_b$ and the shared proton H$^*$, and define the proton-transfer coordinate
\begin{equation}
    \delta=r(\text{O}_a\text{H}^*)-r(\text{O}_b\text{H}^*)
\end{equation}
which is negative when the proton is bound to O$_a$, positive when it is bound to O$_b$, and zero in the symmetric, proton-shared configuration at $\delta=0$. At the equilibrium $r($O$_a$O$_b)$ separation, the shared proton of the cation sits in a single energy well. A double well appears along $\delta$ once $r($O$_a$O$_b)$ is elongated~\cite{marx2006proton}. We therefore perform PES scans at fixed $r($O$_a$O$_b)\in\{2.4017,2.536,2.6418,2.786\}\,\si{\angstrom}$, moving H$^*$ along the O-O axis with all other atoms frozen, for $\delta\in[-0.8,0.8]\,\si{\angstrom}$ in steps of $\SI{0.1}{\angstrom}$.

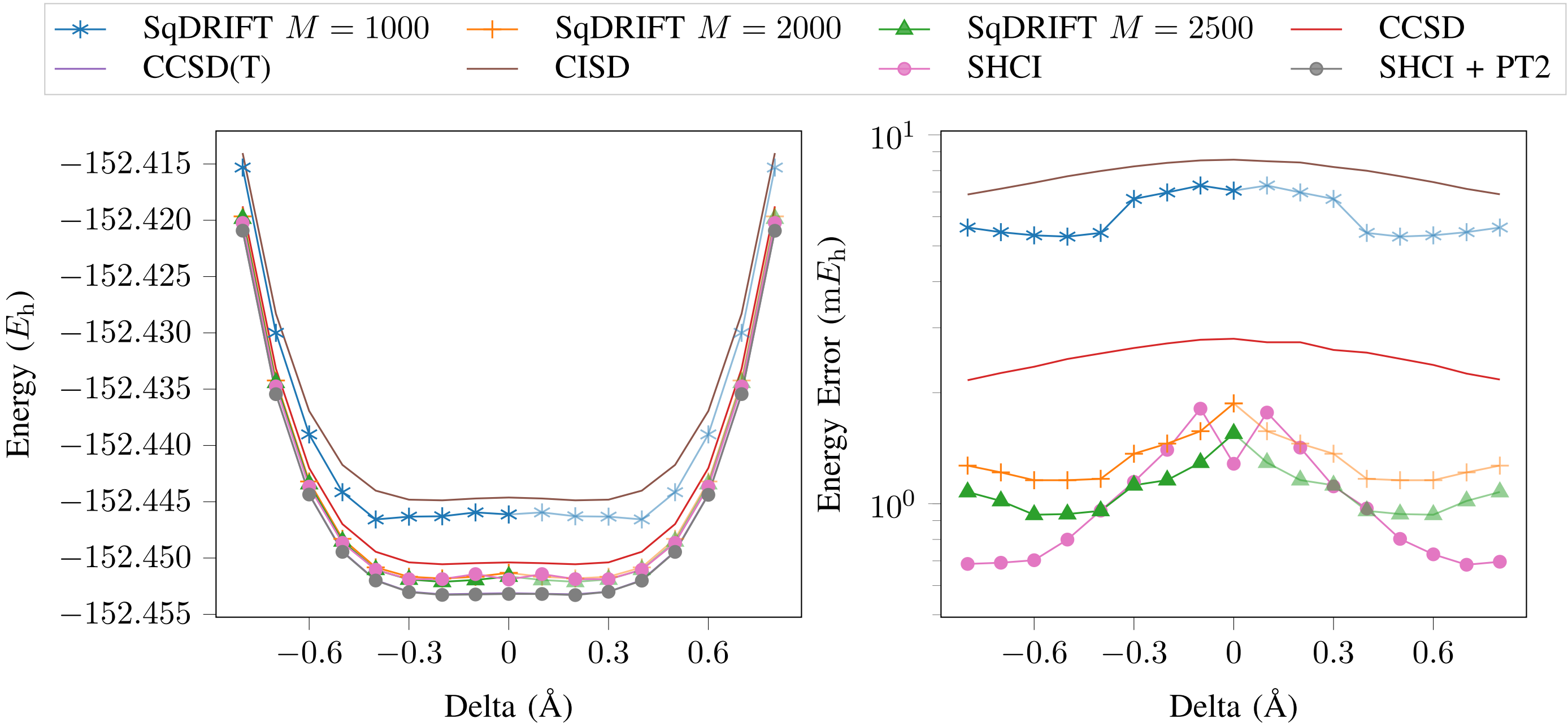
\begin{figure}
    \centering
    \makebox[\textwidth][c]{
    \hspace{20pt}
    \pgfplotslegendfromname{sharedlegend_40q_2.536}
    }
    \par\vspace{5pt}
    \begin{subfigure}[t]{0.48\textwidth}
        \centering
        \parbox[t][6.5cm][t]{\linewidth}{
        \input{40q_2.536_energies}
        }
        \par
        \caption{}
        \label{fig:40q_2.536_energies}
    \end{subfigure}
    \hfill
    \begin{subfigure}[t]{0.48\textwidth}
        \centering
        \input{40q_2.536_error}
       \caption{}
        \label{fig:40q_2.536_error}
    \end{subfigure}
    \caption{PES of H$_\text{5}$O$_\text{2}^+$ in the $40$-qubit active space at $r(\text{O}_a\text{O}_b)=\SI{2.536}{\angstrom}$. (\subref{fig:40q_2.536_energies}) Energies along $\delta$ from SqDRIFT and the classical benchmarks. A very shallow double well appears with minima at $\delta=\SI{-0.2}{\angstrom}$ with $\Delta E^\ddagger=\SI{0.103}{\milli\hartree}$ for SHCI + PT2. (\subref{fig:40q_2.536_error}) Errors relative to SHCI + PT2.}
    \label{fig:40q_2.536}
\end{figure}

We consider two active spaces built from canonical RHF orbitals. The smaller space $(16\text{e},20\text{o})$, maps onto $40$ qubits and the larger space $(16\text{e},26\text{o})$, maps onto $52$ qubits, both with $2$ frozen orbitals. Table~\ref{tab:subspace_comparison} lists the corresponding determinant-space sizes.

We used the recently released Qiskit-Fermions package~\cite{qiskit-fermions} to generate the qDRIFT circuits. 
To generate the circuits, we fixed the evolution times to $[1,5]$ and the number of circuits per evolution time to $150$, for a total of $300$ circuits per geometry. The number of excitation groups is $2$ for all $40$-qubit simulations and $3$ for the $52$-qubit simulation. We used $2000$ shots per circuit for a total of $600$k samples per geometry.
Circuits were executed on the IBM Quantum processors \texttt{ibm\_boston} (Heron r3) and \texttt{ibm\_pittsburgh} (Heron r3) with the Q-CTRL Fire Opal~\cite{qctrl} Performance Management Qiskit function~\cite{qctrl_perf}. 
Because Fire Opal performs its own layout, routing, and transpilation, the gate counts in Table~\ref{tab:resources} were obtained from an independent transpilation with Qiskit (optimization level 3) and serve as indicative estimates.
The percentage of valid strings improved substantially when using the function compared to enabling error mitigation techniques in Qiskit~\cite{qiskit2024}.
We used Fulqrum~\cite{fulqrum} for the S-CORE workflow with SciPy~\cite{scipy} for diagonalization.

The SqDRIFT results were compared against CISD and CCSD energies. CCSD(T)~\cite{ccsdt} and semi-stochastic heat-bath CI (SHCI) with PT2 extrapolation (using DICE)~\cite{shci1,shci2,shci3} (we use $\epsilon_1=5\cdot10^{-5}~\text{and}~\epsilon_2=10^{-8}$) were also computed to serve as near-exact references.

The molecular-dynamics simulations were carried out in the 40-qubit active space. 
We first used CCSD for simulation to create a baseline benchmark. 
The CCSD benchmark is shown in Fig.~\ref{fig:ccsd_baseline}.
The proton was initialized at $\delta=\SI{-1}{\angstrom}$ with zero initial velocity. The motion was propagated with the velocity-Verlet integrator of Sec.~\ref{sec:velocity}, using a time step of $\SI{4}{\atomicunit}$ for $177$ steps for a total simulation time of $\SI{17.0289}{\fs}$.
For SqDRIFT, the subspace size was fixed at $M=2500$ samples per batch, corresponding to $d=6.25\cdot10^6$ determinants. Ideally, for each subsequent geometry, quantum hardware will be run on the fly to obtain the samples. To avoid queue times and reduce hardware runtime usage, we run SqDRIFT for geometries corresponding to every second step from the CCSD baseline. Then we use the samples from one geometry for the next one too. As the time steps are small, the change in geometry from one step to the next is sufficiently small that this approximation does not yield widely different results. The S-CORE loop is run for each point, which helps in convergence. The effectiveness of the approximation can be verified from the results in Section~\ref{sec:dynamics_result}.

\section{Results}
\label{sec:results}
\begin{table}[b]
\centering
\caption{Computational resources used in this work. For each PES scan and for the AIMD trajectory, the table lists the IBM Quantum Processor (Heron r3) used for SqDRIFT sampling, the total QPU time, the classical hardware and total wall time for the S-CORE post-processing, and the two-qubit (2q) circuit gate depth and gate count of the transpiled qDRIFT circuits, given as minimum, maximum, and median over all circuits. Q-CTRL uses its own hardware mapping and transpilation; thus, the counts reported here were obtained by independent transpilation for the target backend. They indicate the circuit complexity and may differ from the executed circuits. The AIMD row covers the 89 geometries sampled on the QPU.}
\label{tab:resources}
\resizebox{\textwidth}{!}{%
\begin{tabular}{@{}cccccccc@{}}
\toprule
Computation          & System                     & QPU            &  \begin{tabular}[c]{@{}c@{}}Total QPU\\ Time ($\si{\second}$)\end{tabular} & vCPUs, RAM                     & \begin{tabular}[c]{@{}c@{}}Total S-CORE\\ Time ($\si{\second}$)\end{tabular} & \begin{tabular}[c]{@{}c@{}}2q Depth\\ ($\min,\max,\operatorname{median}$)\end{tabular} & \begin{tabular}[c]{@{}c@{}}\#2q Gates\\ ($\min,\max,\operatorname{median}$)\end{tabular} \\ \midrule
\multirow{5}{*}{PES} & 40q (\SI{2.4017}{\angstrom}) & \texttt{ibm\_boston}     &           $1568$                   & 64, 504 GiB   & $19760$                          & $8,1126,116$                                                                           & $10,1190,135$                                                                             \\
                     & 40q (\SI{2.5360}{\angstrom})  & \texttt{ibm\_boston}     &         $1566$                      & 64, 504 GiB   & $19853$                          & $9,971,114$                                                                            & $10,1030,132$                                                                             \\
                     & 40q (\SI{2.6418}{\angstrom}) & \texttt{ibm\_pittsburgh} &           $1757$                    & 64, 504 GiB   & $19716$                          & $3,1162,110$                                                                           & $4,1248,128$                                                                              \\
                     & 40q (\SI{2.7860}{\angstrom})  & \texttt{ibm\_pittsburgh} &           $1585$                    & 64, 504 GiB   & $18987$                          & $8,964,108$                                                                            & $9,1051,127$                                                                              \\
                     & 52q (\SI{2.6418}{\angstrom}) & \texttt{ibm\_boston}     &       $4844$                         & 128, 1008 GiB & $87614$                          &  $9,1436,156$                                                                         &  $13,1556,177$                                                                             \\ \midrule
AIMD                 & 40q  & \texttt{ibm\_boston}     & $15390$                       & 64, 504 GiB   & $231543$                         & $3,1162,110$                                                                           & $4,1255,128$                                                                              \\ \midrule \midrule
Total Time           &                            &                &        $26710$                       &                         & $397473$                         &                                                                                        &                                                                                           \\ \bottomrule
\end{tabular}%
}
\end{table}
We first establish the accuracy of the SqDRIFT method on static potential energy surfaces in the $40$- and $52$- qubit active spaces. Note that all SqDRIFT energies are pure variational subspace eigenvalues, and no perturbative correction or extrapolation is added. The computation resources utilized for the experiments are listed in Table~\ref{tab:resources}.

\subsection{PES in the 40-qubit active space}\label{sec:40pes}

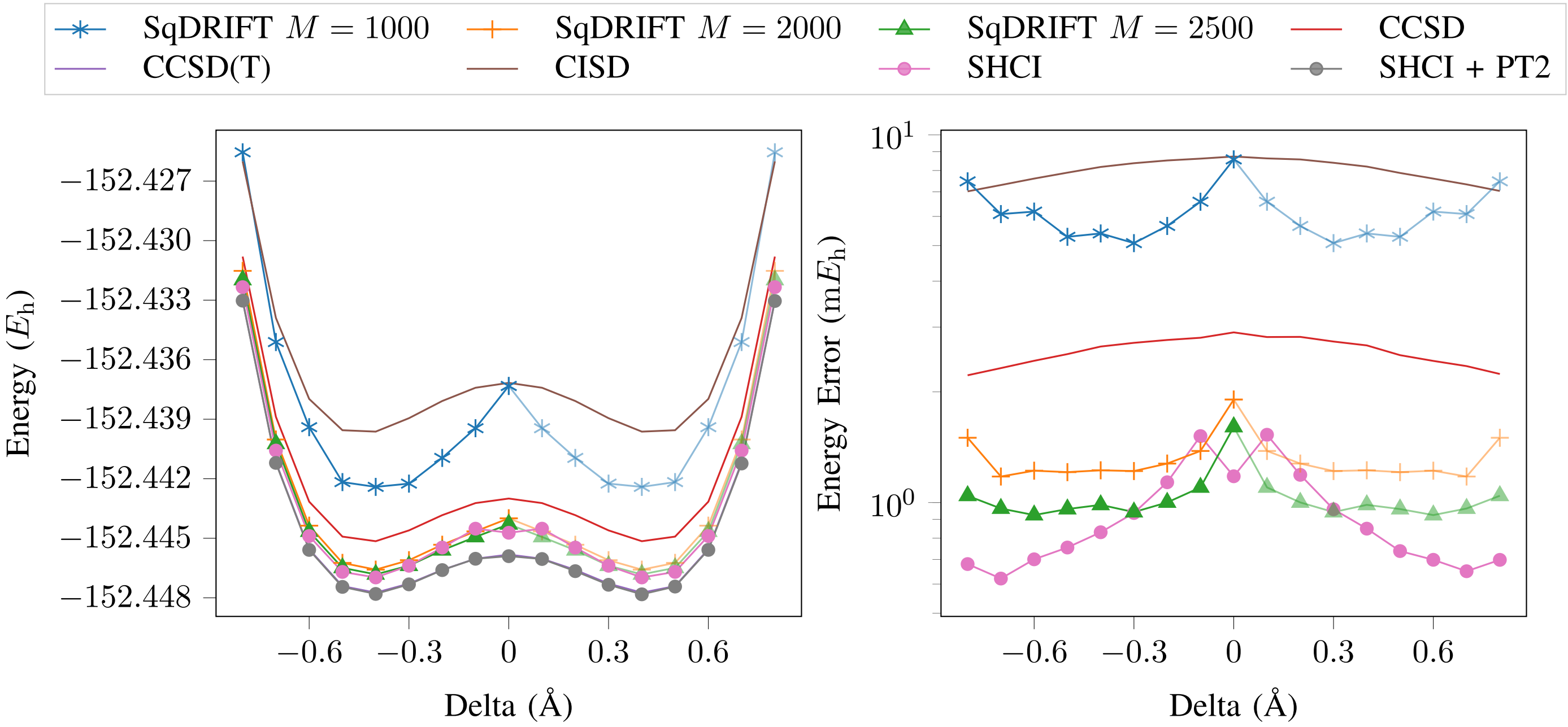
\begin{figure}
    \centering
    \makebox[\textwidth][c]{
    \hspace{20pt}
    \pgfplotslegendfromname{sharedlegend_40q_2.6418}
    }
    \par\vspace{5pt}
    \begin{subfigure}[t]{0.48\textwidth}
        \centering
        \parbox[t][6.5cm][t]{\linewidth}{
        \input{40q_2.6418_energies}
        }
        \par
        \caption{}
        \label{fig:40q_2.6418_energies}
    \end{subfigure}
    \hfill
    \begin{subfigure}[t]{0.48\textwidth}
        \centering
        \input{40q_2.6418_error}
       \caption{}
        \label{fig:40q_2.6418_error}
    \end{subfigure}
    \caption{PES of H$_\text{5}$O$_\text{2}^+$ in the $40$-qubit active space at $r(\text{O}_a\text{O}_b)=\SI{2.6418}{\angstrom}$. (\subref{fig:40q_2.6418_energies}) Energies along $\delta$ from SqDRIFT and the classical benchmarks. The minima moves out to $\delta=\SI{-0.4}{\angstrom}$ with $\Delta E^\ddagger=\SI{1.923}{\milli\hartree}$ for SHCI + PT2 and the double well is clearly developed. (\subref{fig:40q_2.6418_error}) Errors relative to SHCI + PT2.}
    \label{fig:40q_2.6418}
\end{figure}

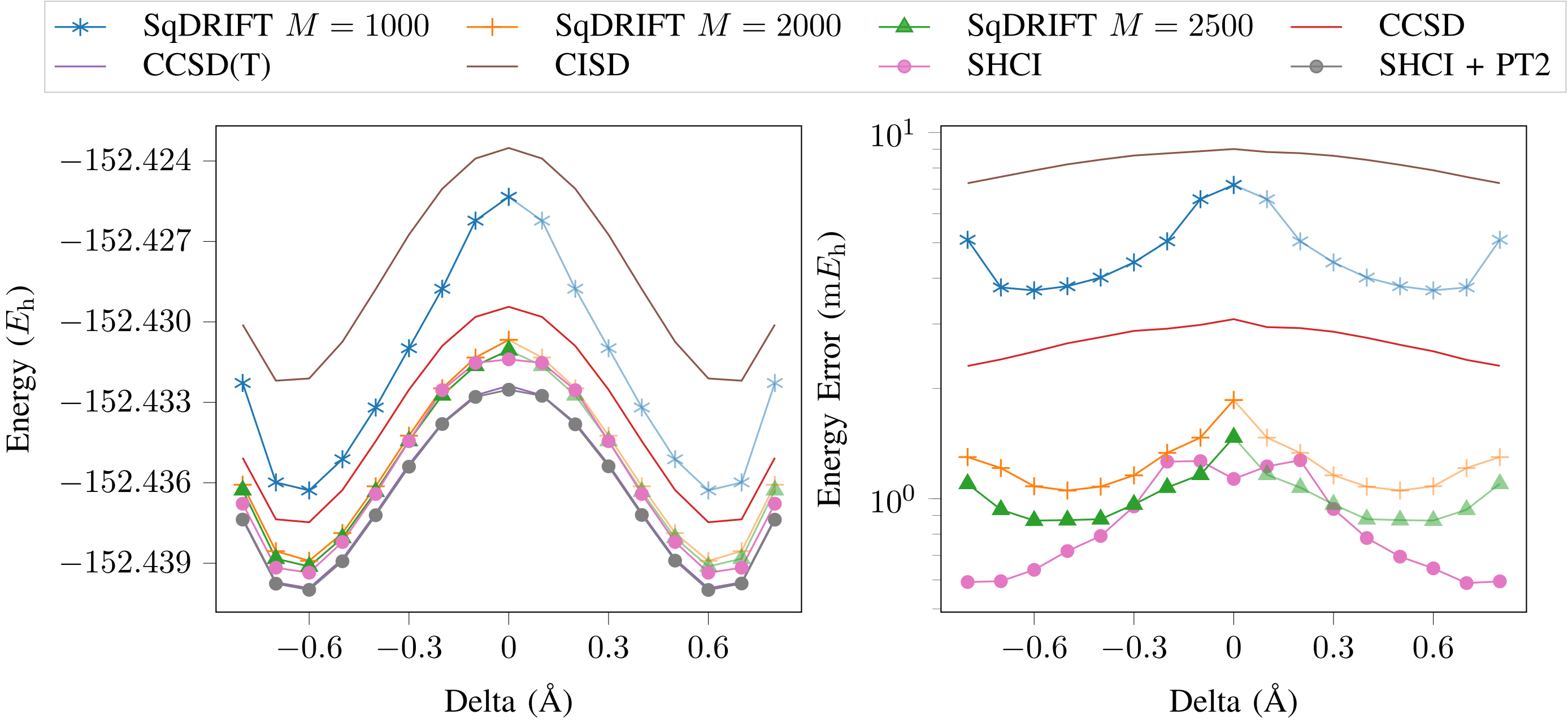
\begin{figure}
    \centering
    \makebox[\textwidth][c]{
    \hspace{20pt}
    \pgfplotslegendfromname{sharedlegend_40q_2.786}
    }
    \par\vspace{5pt}
    \begin{subfigure}[t]{0.48\textwidth}
        \centering
        \parbox[t][6.5cm][t]{\linewidth}{
        \input{40q_2.786_energies}
        }
        \par
        \caption{}
        \label{fig:40q_2.786_energies}
    \end{subfigure}
    \hfill
    \begin{subfigure}[t]{0.48\textwidth}
        \centering
        \input{40q_2.786_error}
       \caption{}
        \label{fig:40q_2.786_error}
    \end{subfigure}
    \caption{PES of H$_\text{5}$O$_\text{2}^+$ in the $40$-qubit active space at $r(\text{O}_a\text{O}_b)=\SI{2.786}{\angstrom}$. (\subref{fig:40q_2.786_energies}) Energies along $\delta$ from SqDRIFT and the classical benchmarks. The minima is now at $\delta=\SI{-0.6}{\angstrom}$ with $\Delta E^\ddagger=\SI{7.468}{\milli\hartree}$ for SHCI + PT2. (\subref{fig:40q_2.786_error}) Errors relative to SHCI + PT2.}
    \label{fig:40q_2.786}
\end{figure}

\begin{figure}
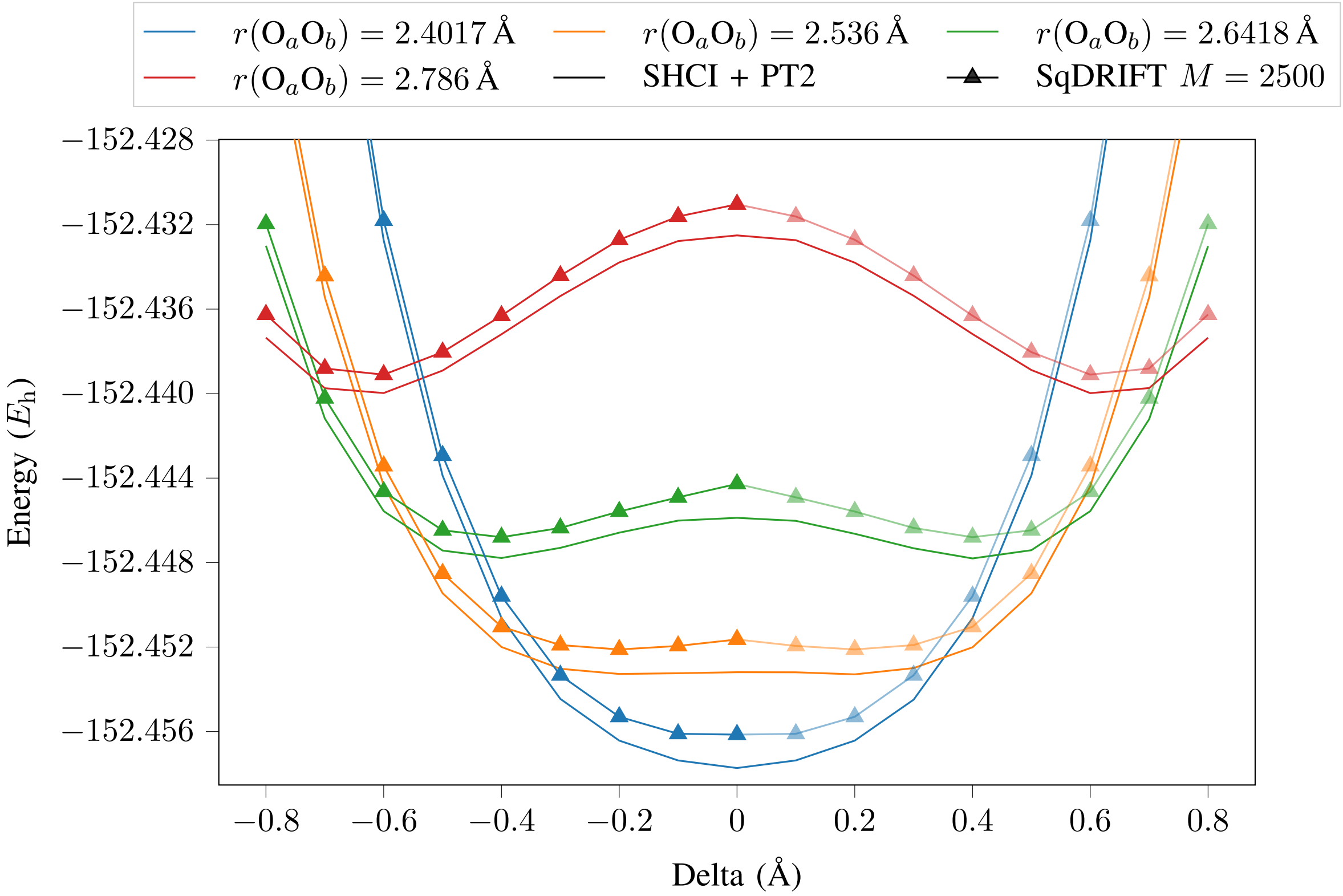

    \centering
    \include{40q_combined_energies}
    \caption{PES of H$_\text{5}$O$_\text{2}^+$ in the $40$-qubit active space for all four $\text{O}_a-\text{O}_b$ separations on a common energy scale computed with SHCI + PT2 and SqDRIFT with $M=2500$. As the separation grows, the barrier goes from a single well to a symmetric double well.}
    \label{fig:40q_combined_energies}
\end{figure}

Figures~\ref{fig:40q_2.4017_energies},~\ref{fig:40q_2.536_energies},~\ref{fig:40q_2.6418_energies} and~\ref{fig:40q_2.786_energies}
show the PES along the proton-transfer coordinate $\delta$ computed with SqDRIFT and the various benchmarks for the four  $r($O$_a$O$_b)$ values. We can see the double-well barrier form as the separation increases. 
Figures~\ref{fig:40q_2.4017_error},~\ref{fig:40q_2.536_error},~\ref{fig:40q_2.6418_error} and~\ref{fig:40q_2.786_error} show the errors in the computed energy values w.r.t. the SHCI + PT2 refernce. We don't include CCSD(T) curves in the error plots for clarity, as CCSD(T) values are within a few $\si{\micro\hartree}$ of the reference. In Fig.~\ref{fig:40q_combined_energies}, we show the PES for all four separation distances on the same scale.

The water molecules are equivalent, so the PES at every fixed $r(\text{O}_a\text{O}_b)$ is symmetric under $
\delta\rightarrow-\delta$. To save both classical and quantum resources, we ran SqDRIFT only for $\delta\in[-0.8,0]\,\si{\angstrom}$ and mirror the energies to complete the PES. This halves the number of sampled geometries. As SqDRIFT sampling is randomized, independent runs at $\pm\delta$ would differ slightly due to statistical fluctuations. Mirroring also makes the computed surface exactly symmetric, matching the physics requirement. We verify the symmetry using CCSD reference, where $|E(\delta)-E(-\delta)|<4\cdot10^{-10}\,\si{\milli\hartree}$ over the whole scan for all separations. In all figures, the mirrored energies are shown with a lighter color shade.

For SqDRIFT, we use $3$ different samples per batch, $M\in\{1000,2000,2500\}$, giving subspace dimensions $d=M^2$ from $10^6$ to $6.25\cdot10^6$ determinants. This corresponds to a fraction $d/D$ of only $6.3-39.3~\cdot10^{-5}$ of the full $D\approx1.587\cdot10^{10}$ dimensional space. At every geometry, the SqDRIFT energies decrease monotonically with $M$, as expected for a variational method with growing subspace dimension.
For $M=2500$, the SqDRIFT energies are lower than those obtained with the CCSD method for all geometries for all  $r(\text{O}_a\text{O}_b)$.

In our experiments, we use the converged eigenvector and subspace from one geometry as the starting point for the next, improving the results. This warm start allows the new geometry to converge to a lower energy in fewer diagonalization steps. We therefore use this warm-start procedure in all our simulations. Specifically, the first subspace for the new geometry is simply the converged subspace from the previous geometry, with no additions or deletions to the component strings. The previous converged eigenvector is used as the initial guess for the eigensolver. After the first batch is diagonalized, the usual S-CORE loop of Fig.~\ref{fig:sqdrift_workflow} is resumed.

\begin{table}[]
\centering
\caption{Characteristics of the potential energy surfaces of the Zundel cation. $E_{\min}$ and $\delta_{\min}$ are the energy and position of the minima. $\Delta E^\ddagger=E(\delta=0)-E_{\min}$ is the barrier height, and NPE is the Non-Parallelity Error measured relative to SHCI + PT2. $E_{\min}$ and $\delta_{\min}$ are only calculated over $\delta\in[-0.8,0]$ for consistency.}
\label{tab:pes_charecteristics}
\resizebox{.9\textwidth}{!}{%
\begin{tabular}{@{}ccccccc@{}}
\toprule
Space                    & $r(\text{O}_a\text{O}_b)\,(\si{\angstrom})$                 & Method           & $E_{\min}\,(\si{\hartree})$ & $\delta_{\min}\,(\si{\angstrom})$ & $\Delta E^\ddagger\,(\si{\milli\hartree})$ & NPE ($\si{\milli\hartree})$ \\ \midrule
\multirow{24}{*}{$40$ q} & \multirow{6}{*}{$2.4017$} & CISD             & $-152.4493$             & $0.0$                           & $0.0$                                & $1.675$                   \\
                         &                                           & CCSD             & $-152.4550$             & $0.0$                           & $0.0$                                & $0.686$                   \\
                         &                                           & CCSD(T)          & $-152.4576$             & $0.0$                           & $0.0$                                & $0.149$                   \\
                         &                                           & SHCI             & $-152.4565$             & $0.0$                           & $0.0$                                & $0.805$                   \\
                         &                                           & SHCI + PT2       & $-152.4578$             & $0.0$                           & $0.0$                                & -                         \\
                         &                                           & SqDRIFT $M=2500$ & $-152.4562$             & $0.0$                           & $0.0$                                & $0.640$                   \\ \cmidrule(l){2-7} 
                         & \multirow{6}{*}{$2.5360$}  & CISD             & $-152.4449$             & $-0.2$                          & $0.251$                              & $1.673$                   \\
                         &                                           & CCSD             & $-152.4506$             & $-0.2$                          & $0.163$                              & $0.638$                   \\
                         &                                           & CCSD(T)          & $-152.4533$             & $-0.2$                          & $0.103$                              & $0.060$                   \\
                         &                                           & SHCI             & $-152.4519$             & $0.0$                           & $0.0$                                & $1.127$                   \\
                         &                                           & SHCI + PT2       & $-152.4533$             & $-0.2$                          & $0.103$                              & -                         \\
                         &                                           & SqDRIFT $M=2500$ & $-152.4521$             & $-0.2$                          & $0.474$                              & $0.616$                   \\ \cmidrule(l){2-7} 
                         & \multirow{6}{*}{$2.6418$} & CISD             & $-152.4396$             & $-0.4$                          & $2.457$                              & $1.712$                   \\
                         &                                           & CCSD             & $-152.4452$             & $-0.4$                          & $2.150$                              & $0.684$                   \\
                         &                                           & CCSD(T)          & $-152.4477$             & $-0.4$                          & $1.919$                              & $0.082$                   \\
                         &                                           & SHCI             & $-152.4470$             & $-0.4$                          & $2.252$                              & $0.908$                   \\
                         &                                           & SHCI + PT2       & $-152.4478$             & $-0.4$                          & $1.923$                              & -                         \\
                         &                                           & SqDRIFT $M=2500$ & $-152.4468$             & $-0.4$                          & $2.524$                              & $0.681$                   \\ \cmidrule(l){2-7} 
                         & \multirow{6}{*}{$2.7860$}  & CISD             & $-152.4322$             & $-0.7$                          & $8.678$                              & $1.744$                   \\
                         &                                           & CCSD             & $-152.4375$             & $-0.6$                          & $8.034$                              & $0.789$                   \\
                         &                                           & CCSD(T)          & $-152.4399$             & $-0.6$                          & $7.548$                              & $0.132$                   \\
                         &                                           & SHCI             & $-152.4394$             & $-0.6$                          & $7.955$                              & $0.685$                   \\
                         &                                           & SHCI + PT2       & $-152.4400$             & $-0.6$                          & $7.468$                              & -                         \\
                         &                                           & SqDRIFT $M=2500$ & $-152.4391$             & $-0.6$                          & $8.061$                              & $0.599$                   \\ \midrule
\multirow{6}{*}{$52$ q}  & \multirow{6}{*}{$2.6418$} & CISD             & $-152.5660$             & $-0.5$                          & $3.788$                              & $2.129$                   \\
                         &                                           & CCSD             & $-152.5863$             & $-0.5$                          & $3.119$                              & $0.713$                   \\
                         &                                           & CCSD(T)          & $-152.5889$             & $-0.5$                          & $2.790$                              & $0.375$                   \\
                         &                                           & SHCI             & $-152.5865$             & $-0.5$                          & $4.103$                              & $3.447$                   \\
                         &                                           & SHCI + PT2       & $-152.5899$             & $-0.5$                          & $2.931$                              & -                         \\
                         &                                           & SqDRIFT $M=5000$ & $-152.5848$             & $-0.4$                          & $5.026$                              & $2.35$                    \\ \bottomrule
\end{tabular}%
}
\end{table}

\subsection{PES in the 52-qubit active space}\label{sec:52pes}
\begin{figure}
    \centering
    \makebox[\textwidth][c]{
    \hspace{20pt}
    \pgfplotslegendfromname{sharedlegend_52q}
    }
    \par\vspace{5pt}
    \begin{subfigure}[t]{0.48\textwidth}
        \centering
        \parbox[t][6.5cm][t]{\linewidth}{
        \input{52q_energies}
        }
        \par
        \caption{}
        \label{fig:52q_pes}
    \end{subfigure}
    \hfill
    \begin{subfigure}[t]{0.48\textwidth}
        \centering
        \input{52q_errors}
       \caption{}
        \label{fig:52q_error}
    \end{subfigure}
    \caption{PES of H$_\text{5}$O$_\text{2}^+$ in the $52$-qubit active space at $r(\text{O}_a\text{O}_b)=\SI{2.6418}{\angstrom}$. (\subref{fig:52q_pes}) Energies along $\delta$ from SqDRIFT and the classical benchmarks. SqDRIFT reproduces the double-well topology and lies \SIrange{1.3}{3.4}{\milli\hartree} above CCSD. (\subref{fig:52q_error}) Errors relative to SHCI + PT2.}
    \label{fig:52q}
\end{figure}
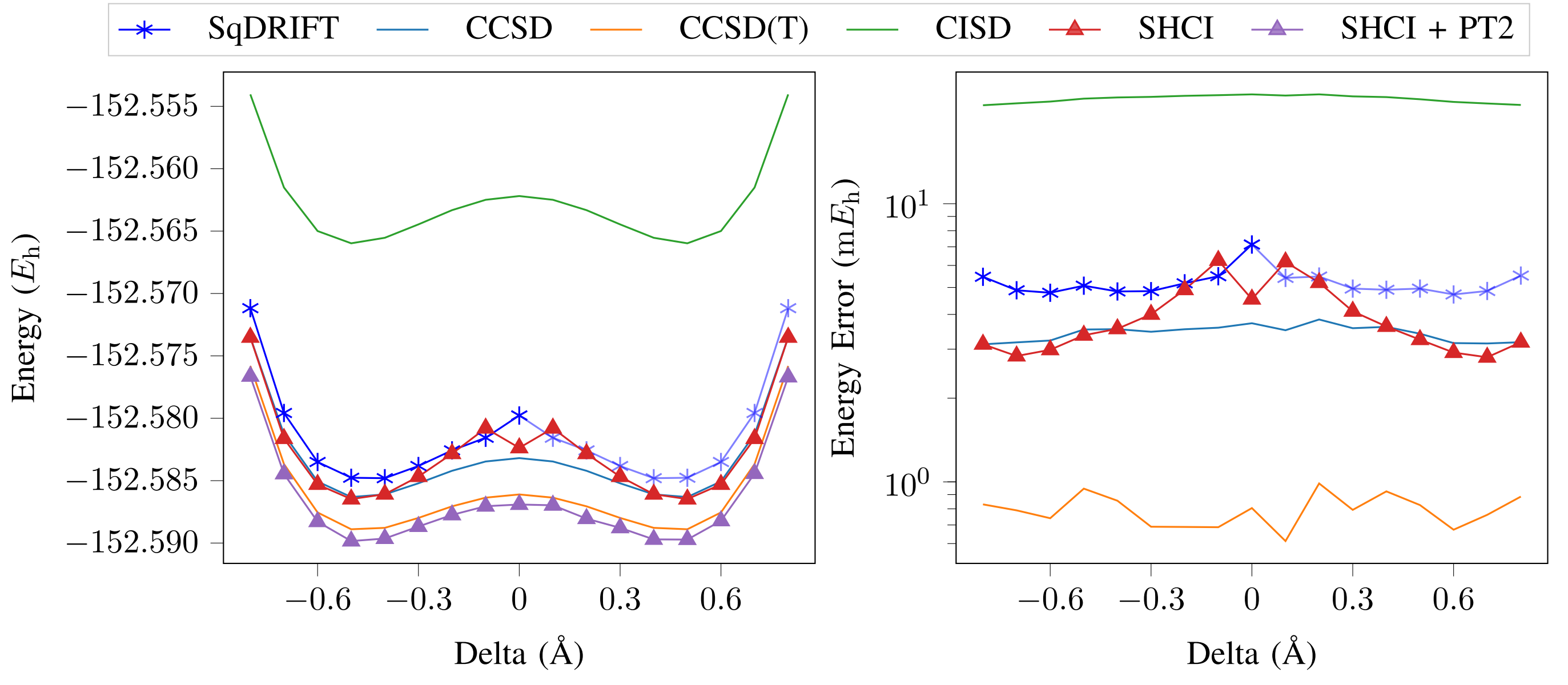

Enlarging the active space to $26$ orbitals ($52$ qubits) increases the full determinant-space dimension to $D\approx2.44\cdot10^{12}$, about $154$ times that of the $40$-qubit space.
Figure~\ref{fig:52q_pes} shows that while SqDRIFT was able to reproduce the double-well structure of the PES, for the $M$ value we tried, it did not reach CCSD values. At the largest subspace considered $(M_{\max}=5000,d/D\approx10^{-5} )$, the SqDRIFT energies lie within the range of $1.3-3.4\,\si{\milli\hartree}$ above CCSD. Figure~\ref{fig:52q_error} shows the energy errors relative to SHCI + PT2.

In Table~\ref{tab:pes_charecteristics} we note some relevant characteristics of the potential energy surfaces.

\subsection{Proton-transfer dynamics simulation in the 40-qubit active space}\label{sec:dynamics_result}

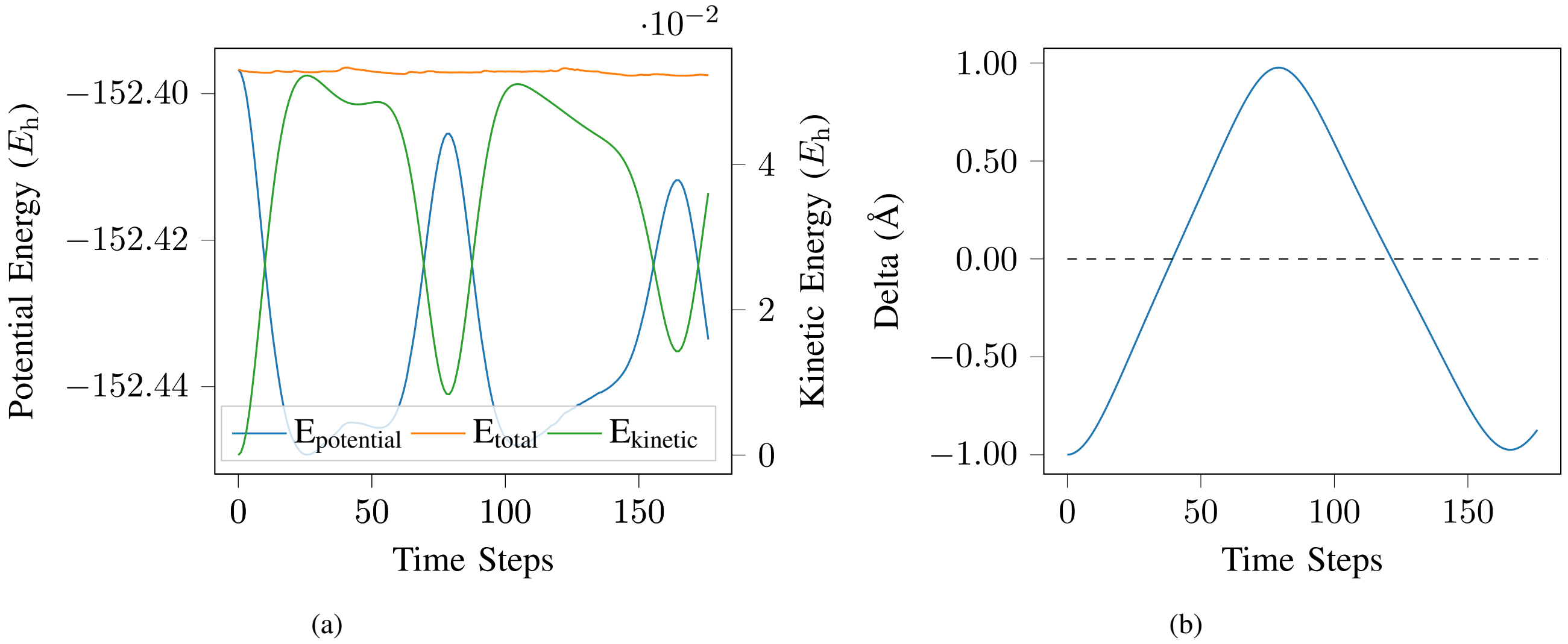
\begin{figure}
    \centering
    \begin{subfigure}[t]{0.4\textwidth}
            \centering
            \parbox[t][4cm][t]{\linewidth}{
            \input{dynamics_energy}
            }
            \par
            \caption{}
            \label{fig:40q_energies}
    \end{subfigure}
    \hspace{19mm}
    \begin{subfigure}[t]{0.4\textwidth}
            \centering  
            \parbox[t][4cm][t]{\linewidth}{
            \input{dynamics_delta}
            }
            \par
            \caption{}
            \label{fig:40q_delta}
    \end{subfigure}
    \caption{SqDRIFT-AIMD of H$_\text{5}$O$_\text{2}^+$ in the $40$-qubit active space with same inital conditions as Fig.~\ref{fig:ccsd_baseline}. (\subref{fig:40q_energies}) Potential, kinetic, and total energies along the simulation. (\subref{fig:40q_delta}) Proton-transfer coordinate $\delta$.}
    \label{fig:sqdrift_dynamics}
\end{figure}

\begin{figure}
    \centering
    \begin{subfigure}[t]{0.4\textwidth}
        \centering
        \input{total_energy_compare3}
        \caption{}
        \label{fig:total_energy_compare3}
    \end{subfigure}
    \hspace{19mm}
    \begin{subfigure}[t]{0.4\textwidth}
        \centering
        \input{potential_compare3}
        \caption{}
        \label{fig:potential_compare3}
    \end{subfigure}

    \begin{subfigure}[t]{0.4\textwidth}
        \centering
        \input{kinetic_compare3}
        \caption{}
        \label{fig:kinetic_compare3}
    \end{subfigure}
    \hspace{19mm}
    \begin{subfigure}[t]{0.4\textwidth}
        \centering
        \input{diff_delta3}
        \caption{}
        \label{fig:diff_delta3}
    \end{subfigure}
    \caption{Comparison between CCSD-, SHCI- and SqDRIFT-AIMD results. (\subref{fig:total_energy_compare3}), (\subref{fig:potential_compare3}) and (\subref{fig:kinetic_compare3}) show the absolute energy errors of CCSD and SqDRIFT relative to SHCI for total, potential and kinetic energy respectively. The black dotted line marks the chemical accuracy value of $\SI{1.6}{\milli\hartree}$. (\subref{fig:diff_delta3}) Shows the absolute delta error of CCSD and SqDRIFT relative to SHCI.}
    \label{fig:dynamice_comparison_all}
\end{figure}
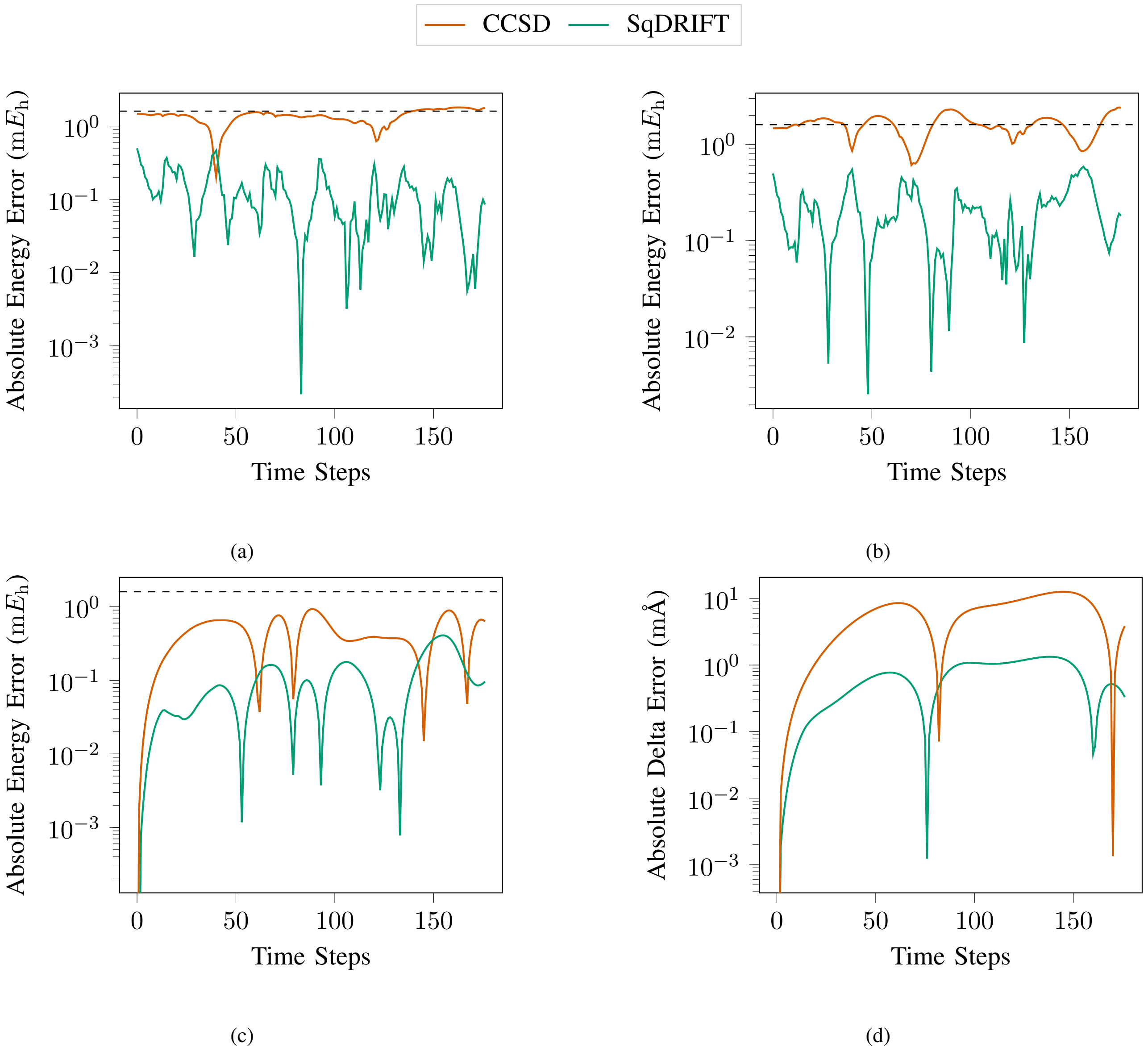

In the $40$-qubit space, the SqDRIFT energies fell below CCSD for a relatively small subspace. We therefore use this active space for the dynamics simulation, with $M=2500$. We choose the initial $r(\text{O}_a\text{O}_b)=\SI{2.6418}{\angstrom}$ and then let all atoms be free to move during the simulation.
As in the PES scans, each MD step was warm-started from the converged subspace and the previous step's eigenvector.

Figure~\ref{fig:40q_delta} shows the evolution of $\delta$ with time. 
The proton transfer starts with $\delta<0$ on O$_a$, a hydronium-like configuration, and accelerates towards the barrier. It crosses the symmetric point $\delta=0$ at $t_1=\SI{3.8702}{\fs}$ and moves into the well on O$_b$, reaching a turning point at $\delta=\SI{0.977}{\angstrom}$ at $t=\SI{7.6437}{\fs}$. It then returns, recrosses the barrier at $t_2=\SI{11.6106}{\fs}$ and relocalizes on O$_a$. This completes a full forward-and-back transfer cycle.

Figure~\ref{fig:40q_energies} shows the potential, kinetic, and total energies over the simulation. Kinetic and potential energies exchange smoothly, and the total energy is conserved with a standard deviation of $\SI{0.245}{\milli\hartree}$. We also run the dynamics simulation with SHCI to have a near FCI benchmark.
In Fig.~\ref{fig:dynamice_comparison_all} we compare the energies and trajectory between CCSD-, SHCI- and SqDRIFT-AIMD. Potential, kinetic and total energies are within chemical accuracy relative to SHCI for the whole duration. $\delta$ for SqDRIFT also stays within $\SI{1.33}{\milli\angstrom}$ of the SHCI values. The consistently small deviations from the SHCI reference demonstrate the high accuracy of the SqDRIFT method throughout the simulated dynamics.

\section{Conclusion}
\label{sec:conclusion}
We have presented, to the best of our knowledge, the first ab initio molecular dynamics simulations driven by SqDRIFT, a randomized sample-based Krylov quantum diagonalization method. In SqDRIFT, the determinant subspaces are generated by qDRIFT-compiled real-time evolution under the molecular Hamiltonian. The sampling circuits therefore need no variational optimization and no amplitudes from classical correlated methods such as CCSD. We combined SqDRIFT with analytical nuclear gradients derived from subspace-reduced density matrices and applied the resulting framework to proton transfer in the Zundel cation, $\text{H}_\text{5}\text{O}_\text{2}^+$. The active spaces we treated have $40$ and $52$ qubits. The $40$-qubit AIMD study is about twice the size of those used in previous sample-based AIMD studies.

On the static potential energy surfaces, SqDRIFT reproduces the double-well profile along the proton-transfer coordinate $\delta$ in both active spaces, and it captures how the barrier rises as the $\text{O}_\text{a}-\text{O}_\text{b}$ separation increases. In the $40$-qubit space, the computed SqDRIFT energies are lower than CCSD energies at every geometry at $M=2500$. Because SqDRIFT energies are rigorous variational upper bounds on the active-space FCI energy, SqDRIFT is more accurate than CCSD throughout the entire scan. In the $52$-qubit space, SqDRIFT reproduces the double-well topology but remains within a few $\si{\milli\hartree}$ of CCSD at $M=5000$. 

In the dynamics, the $40$-qubit SqDRIFT trajectory shows the excess proton moving between the two water molecules and back. We benchmarked the trajectory against a CCSD and SHCI-driven AIMD simulation started from identical initial conditions in the same active space. All three trajectories agree closely while SqDRIFT energies remains within the chemical accuracy threshold with SHCI. The proton transfer coordinate $\delta$ deviates at most $\SI{1.33}{\milli\angstrom}$ from the SHCI reference, and the transfer occurs at the same time.
The total energy is conserved over the whole simulation, with a standard deviation of $\SI{0.245}{\milli\hartree}.$ This shows that SqDRIFT-AIMD is stable and consistent with energy.

Together, these results move quantum-centric AIMD from small active spaces, where FCI is tractable, to larger ones.
Several directions follow naturally. Classical subspace extension, perturbative corrections, and variance extrapolation could close the gap to CCSD in the $52$-qubit space, apart from directly increasing $M$. Fully on-the-fly SqDRIFT-AIMD, in which the quantum processor is sampled at every MD step, is also possible. This will remove the sample-reuse approximation. It can also enable a systematic study of how often fresh samples are needed to maintain a given accuracy, balancing quantum cost against fidelity.
Embedding the Zundel cation in an explicit solvent using a quantum mechanics/molecular mechanics (QM/MM) approach at these sizes would move these simulations toward proton transport in realistic environments, such as hydrogen-bonded networks.
To incorporate nuclear quantum effects, such as quantum tunneling, into the study of proton-transfer dynamics, the Nuclear-Electron Orbital (NEO)~\cite{neo_qc_static, neo_qc_dynamics, dynamics_neo_vqe} approach can be integrated into the SqDRIFT-AIMD workflow. Because the NEO approach adds more terms to the molecular Hamiltonian, constructing qDRIFT circuits on near-term quantum hardware may be challenging and an interesting problem to solve.   

\bibliographystyle{IEEEtran}
\bibliography{IEEEabrv, references.bib}
\end{document}

%% file: ccsd_energy.tex
\begin{tikzpicture}

\definecolor{darkgray176}{RGB}{176,176,176}
\definecolor{darkorange25512714}{RGB}{255,127,14}
\definecolor{forestgreen4416044}{RGB}{44,160,44}
\definecolor{lightgray204}{RGB}{204,204,204}
\definecolor{steelblue31119180}{RGB}{31,119,180}

\begin{axis}[
width=\textwidth,
legend cell align={left},
legend style={
  fill opacity=0.8,
  draw opacity=1,
  text opacity=1,
  legend columns=3,
  at={(0.97,0.03)},
  anchor=south east,
  draw=lightgray204
},
tick align=outside,
tick pos=left,
x grid style={darkgray176},
xlabel={Time Steps},
xmin=-8.8, xmax=184.8,
xtick style={color=black},
y grid style={darkgray176},
ylabel={Potential Energy (\(\displaystyle \si{\hartree}\))},
ymin=-152.45193741084, ymax=-152.39378589016,
ytick style={color=black},
yticklabel style={/pgf/number format/precision=2,
/pgf/number format/fixed zerofill},
]
\addplot [semithick, steelblue31119180]
table {%
0 -152.3957440623
1 -152.3961141117
2 -152.3972073581
3 -152.3989741001
4 -152.4013363441
5 -152.4041955528
6 -152.4074391239
7 -152.4109493929
8 -152.414610939
9 -152.4183162748
10 -152.4219703638
11 -152.4254924949
12 -152.4288195108
13 -152.431902478
14 -152.4347078697
15 -152.4372158027
16 -152.4394178893
17 -152.4413156
18 -152.4429178379
19 -152.4442402084
20 -152.445302097
21 -152.4461258314
22 -152.4467353013
23 -152.4471550847
24 -152.4474095933
25 -152.4475221768
26 -152.4475153452
27 -152.4474100863
28 -152.4472256907
29 -152.4469798227
30 -152.446688364
31 -152.4463656499
32 -152.4460243112
33 -152.4456757516
34 -152.4453295521
35 -152.444994477
36 -152.4446776321
37 -152.4443852915
38 -152.444122271
39 -152.4438928248
40 -152.443699789
41 -152.4435450927
42 -152.443430108
43 -152.4433541003
44 -152.4433161336
45 -152.4433135049
46 -152.4433427529
47 -152.4433984404
48 -152.4434741064
49 -152.4435616893
50 -152.4436515403
51 -152.4437321509
52 -152.4437903906
53 -152.4438112975
54 -152.4437781438
55 -152.4436725262
56 -152.4434744201
57 -152.4431626864
58 -152.4427150929
59 -152.4421091255
60 -152.4413225924
61 -152.4403345819
62 -152.4391266798
63 -152.4376840309
64 -152.4359974768
65 -152.4340649836
66 -152.4318930858
67 -152.4294987749
68 -152.4269108867
69 -152.4241707037
70 -152.4213322224
71 -152.4184614352
72 -152.4156353512
73 -152.4129382658
74 -152.4104595483
75 -152.4082881471
76 -152.4065079108
77 -152.40519191
78 -152.404397441
79 -152.4041613885
80 -152.4044973931
81 -152.4053946374
82 -152.4068187401
83 -152.4087143622
84 -152.4110096667
85 -152.4136215071
86 -152.4164603638
87 -152.4194363993
88 -152.422463699
89 -152.4254637874
90 -152.4283686863
91 -152.4311221208
92 -152.4336802694
93 -152.4360118536
94 -152.4380970615
95 -152.4399267783
96 -152.4415009521
97 -152.4428270631
98 -152.4439183073
99 -152.4447924064
100 -152.4454696247
101 -152.4459717605
102 -152.4463212083
103 -152.4465398892
104 -152.4466482871
105 -152.4466655717
106 -152.4466087804
107 -152.4464929587
108 -152.4463310707
109 -152.4461338919
110 -152.4459102442
111 -152.4456672311
112 -152.4454104722
113 -152.4451440203
114 -152.4448709812
115 -152.4445935294
116 -152.4443131585
117 -152.4440308951
118 -152.4437473429
119 -152.4434630446
120 -152.4431782671
121 -152.4428938411
122 -152.4426098536
123 -152.4423272393
124 -152.4420466394
125 -152.4417689823
126 -152.4414952964
127 -152.4412265567
128 -152.4409640819
129 -152.4407083433
130 -152.440460412
131 -152.4402205533
132 -152.4399890208
133 -152.4397650065
134 -152.4395472973
135 -152.4393338535
136 -152.4391215071
137 -152.4389061484
138 -152.4386825456
139 -152.4384441574
140 -152.4381831142
141 -152.437890202
142 -152.4375548927
143 -152.437165473
144 -152.4367090294
145 -152.4361721187
146 -152.4355407189
147 -152.4348010181
148 -152.4339400212
149 -152.4329464053
150 -152.4318115263
151 -152.4305305087
152 -152.429103498
153 -152.4275370381
154 -152.4258449836
155 -152.424049482
156 -152.4221816479
157 -152.4202806759
158 -152.4183952464
159 -152.4165795099
160 -152.4148930061
161 -152.4133974435
162 -152.4121541015
163 -152.4112211828
164 -152.4106511409
165 -152.4104866824
166 -152.4107559184
167 -152.4114685842
168 -152.4126143624
169 -152.4141638231
170 -152.4160716042
171 -152.4182793905
172 -152.4207195794
173 -152.4233197054
174 -152.4260051268
175 -152.4287036815
176 -152.4313480837
};
\addlegendentry{$\text{E}_{\text{potential}}$}
\addplot [semithick, darkorange25512714]
table {%
0 -152.3957440623
1 -152.3957451212
2 -152.3957483562
3 -152.3957533808
4 -152.3957594249
5 -152.3957664336
6 -152.3957736619
7 -152.3957806082
8 -152.3957869847
9 -152.3957924562
10 -152.3957969021
11 -152.3957997318
12 -152.3958020421
13 -152.3958033691
14 -152.3958037679
15 -152.3958034952
16 -152.3958026319
17 -152.3958013627
18 -152.3957994832
19 -152.3957975054
20 -152.3957954091
21 -152.3957932726
22 -152.3957911214
23 -152.3957890559
24 -152.3957871816
25 -152.3957852955
26 -152.3957835268
27 -152.3957819307
28 -152.3957804742
29 -152.3957791824
30 -152.3957779881
31 -152.3957769302
32 -152.3957759236
33 -152.3957751494
34 -152.3957743669
35 -152.3957738232
36 -152.3957733267
37 -152.3957730016
38 -152.3957726045
39 -152.3957723794
40 -152.3957721793
41 -152.3957719771
42 -152.3957722296
43 -152.395772363
44 -152.3957727306
45 -152.3957731102
46 -152.395773792
47 -152.3957744475
48 -152.3957751714
49 -152.3957759991
50 -152.3957770059
51 -152.3957781088
52 -152.3957793588
53 -152.3957807561
54 -152.3957822977
55 -152.3957839906
56 -152.3957857523
57 -152.3957876668
58 -152.3957896259
59 -152.3957916028
60 -152.3957935359
61 -152.3957953512
62 -152.3957970209
63 -152.3957982424
64 -152.395799001
65 -152.3957993092
66 -152.3957989445
67 -152.3957977514
68 -152.3957957448
69 -152.3957928934
70 -152.395789212
71 -152.3957846694
72 -152.3957797605
73 -152.3957743263
74 -152.395768906
75 -152.3957637233
76 -152.3957591572
77 -152.3957555004
78 -152.3957531421
79 -152.3957522691
80 -152.3957529723
81 -152.3957551723
82 -152.3957586821
83 -152.3957631235
84 -152.395768203
85 -152.3957737599
86 -152.3957792112
87 -152.3957843479
88 -152.3957889825
89 -152.3957928084
90 -152.3957958501
91 -152.3957980693
92 -152.3957994355
93 -152.3958001135
94 -152.3958000722
95 -152.3957994488
96 -152.3957983528
97 -152.3957969308
98 -152.3957951847
99 -152.3957933677
100 -152.3957914657
101 -152.3957894784
102 -152.3957875749
103 -152.3957859013
104 -152.3957842939
105 -152.3957828856
106 -152.3957815992
107 -152.3957804518
108 -152.3957795078
109 -152.3957787184
110 -152.3957780313
111 -152.3957774502
112 -152.395777059
113 -152.3957767168
114 -152.3957764492
115 -152.3957762415
116 -152.3957760817
117 -152.3957759849
118 -152.3957758721
119 -152.3957757898
120 -152.3957755816
121 -152.3957756345
122 -152.3957755093
123 -152.3957754977
124 -152.3957754605
125 -152.3957754185
126 -152.3957753963
127 -152.3957753182
128 -152.3957754704
129 -152.3957753943
130 -152.3957754362
131 -152.3957754491
132 -152.3957757082
133 -152.3957759882
134 -152.3957763489
135 -152.3957768489
136 -152.395777368
137 -152.3957779134
138 -152.3957785485
139 -152.3957792881
140 -152.3957801378
141 -152.3957810973
142 -152.395782162
143 -152.3957833571
144 -152.39578455
145 -152.3957858503
146 -152.395787166
147 -152.3957884605
148 -152.3957896797
149 -152.3957907603
150 -152.3957916335
151 -152.3957921809
152 -152.3957922673
153 -152.3957918936
154 -152.3957909855
155 -152.3957894877
156 -152.3957875363
157 -152.3957845992
158 -152.3957815551
159 -152.3957780335
160 -152.3957743207
161 -152.3957705977
162 -152.3957669941
163 -152.3957636299
164 -152.3957608797
165 -152.3957591337
166 -152.395758545
167 -152.3957592352
168 -152.3957611851
169 -152.3957640731
170 -152.3957677678
171 -152.3957719773
172 -152.3957762996
173 -152.3957808604
174 -152.3957851781
175 -152.3957892173
176 -152.3957926362
};
\addlegendentry{$\text{E}_{\text{total}}$}
\addlegendimage{/pgfplots/refstyle=Ekin_plot}
\addlegendentry{$\text{E}_{\text{kinetic}}$}
\end{axis}

\begin{axis}[
width=\textwidth,
axis y line*=right,
tick align=outside,
x grid style={darkgray176},
xmin=-8.8, xmax=184.8,
xtick pos=left,
xtick style={color=black},
y grid style={darkgray176},
ylabel={Kinetic Energy (\(\displaystyle \si{\hartree}\))},
ymin=-0.002611940928153, ymax=0.056,
ytick pos=right,
ytick style={color=black},
yticklabel style={anchor=west}
]
\addplot [semithick, forestgreen4416044]
table {%
0 0
1 0.0003689905208436
2 0.001459001919805
3 0.003220719271715
4 0.005576919188135
5 0.008429119251681
6 0.01166546203866
7 0.01516878465531
8 0.01882395422867
9 0.022523818632
10 0.02617346167824
11 0.02969276309685
12 0.03301746871287
13 0.03609910887339
14 0.03890410183771
15 0.04141230750648
16 0.0436152573824
17 0.04551423727665
18 0.04711835477815
19 0.04844270298946
20 0.04950668798133
21 0.05033255878309
22 0.05094417994988
23 0.05136602881957
24 0.05162241166078
25 0.05173688123041
26 0.05173181837724
27 0.05162815564407
28 0.05144521647222
29 0.05120064030471
30 0.05091037584792
31 0.05058871967632
32 0.05024838759758
33 0.04990060222181
34 0.04955518519476
35 0.04922065378479
36 0.04890430547751
37 0.04861228986528
38 0.04834966647479
39 0.04812044542704
40 0.04792760972534
41 0.04777311557987
42 0.04765787840513
43 0.0475817372981
44 0.04754340295795
45 0.04754039468576
46 0.04756896085232
47 0.04762399295506
48 0.04769893495648
49 0.04778569013313
50 0.04787453442467
51 0.04795404207579
52 0.048011031816
53 0.04803054136748
54 0.04799584609184
55 0.04788853563903
56 0.04768866787954
57 0.04737501964859
58 0.04692546695373
59 0.0463175226244
60 0.04552905649963
61 0.04453923068227
62 0.04332965894807
63 0.04188578845796
64 0.04019847577831
65 0.03826567443071
66 0.03609414130674
67 0.03370102358937
68 0.03111514194384
69 0.02837781033201
70 0.02554301043496
71 0.02267676580277
72 0.01985559066331
73 0.0171639395284
74 0.01469064228037
75 0.012524423793
76 0.01074875354999
77 0.009436409670656
78 0.008644298920022
79 0.008409119436197
80 0.008744420719628
81 0.009639465089756
82 0.01106005804522
83 0.01295123878745
84 0.01524146368623
85 0.0178477472621
86 0.02068115256501
87 0.02365205136643
88 0.02667471655268
89 0.02967097898766
90 0.03257283619261
91 0.03532405154645
92 0.03788083393607
93 0.04021174005693
94 0.0422969892635
95 0.0441273294998
96 0.04570259932478
97 0.04703013225045
98 0.04812312257723
99 0.04899903865775
100 0.04967815901565
101 0.05018228208065
102 0.05053363337967
103 0.05075398784071
104 0.05086399318934
105 0.05088268605439
106 0.05082718121745
107 0.05071250696733
108 0.05055156288708
109 0.05035517348939
110 0.05013221288541
111 0.04988978099589
112 0.04963341319908
113 0.04936730353739
114 0.04909453201506
115 0.04881728783976
116 0.04853707680321
117 0.04825491016266
118 0.04797147083259
119 0.04768725482904
120 0.04740268546537
121 0.04711820662175
122 0.04683434426869
123 0.04655174155427
124 0.04627117889197
125 0.04599356380641
126 0.04571990008279
127 0.04545123845231
128 0.04518861156004
129 0.0449329490332
130 0.04468497583205
131 0.04444510423397
132 0.04421331258604
133 0.04398901830371
134 0.04377094845986
135 0.04355700467238
136 0.04334413916592
137 0.0431282350349
138 0.04290399706073
139 0.0426648692304
140 0.04240297644905
141 0.04210910471621
142 0.04177273076556
143 0.0413821159063
144 0.04092447944161
145 0.04038626837887
146 0.03975355290751
147 0.03901255758456
148 0.03815034150877
149 0.03715564498214
150 0.03601989281033
151 0.03473832782512
152 0.0333112306661
153 0.03174514453491
154 0.0300539980323
155 0.02825999425086
156 0.02639411154276
157 0.02449607676438
158 0.02261369126745
159 0.02080147644469
160 0.0191186854064
161 0.01762684581831
162 0.01638710740837
163 0.01545755286103
164 0.01489026118543
165 0.01472754869449
166 0.01499737340721
167 0.01570934899946
168 0.01685317732462
169 0.01839975002854
170 0.02030383635524
171 0.02250741321886
172 0.02494327977593
173 0.02753884491735
174 0.03021994869464
175 0.03291446419847
176 0.03555544754519
};
\label{Ekin_plot}
\end{axis}

\end{tikzpicture}

%% file: ccsd_delta.tex
\begin{tikzpicture}

\definecolor{darkgray176}{RGB}{176,176,176}
\definecolor{steelblue31119180}{RGB}{31,119,180}

\begin{axis}[
width=1\linewidth,
tick align=outside,
tick pos=left,
x grid style={darkgray176},
xlabel={Time Steps},
xmin=-8.8, xmax=184.8,
xtick style={color=black},
y grid style={darkgray176},
ylabel={Delta  (\(\displaystyle \si{\angstrom}\))},
ymin=-1.09814854258187, ymax=1.07557355257506,
ytick style={color=black},
yticklabel style={/pgf/number format/precision=2,
/pgf/number format/fixed zerofill},
]
\addplot [semithick, steelblue31119180]
table {%
0 -0.999342992802014
1 -0.998005768499769
2 -0.994009822514713
3 -0.987401615665799
4 -0.978256511945991
5 -0.966675528929606
6 -0.952781432034583
7 -0.9367143302111
8 -0.91862707742833
9 -0.898680658718066
10 -0.877040233763143
11 -0.85387145186329
12 -0.829337389531652
13 -0.803596203148453
14 -0.776799148600091
15 -0.749089117701727
16 -0.720599733462626
17 -0.691454428099415
18 -0.661766300455653
19 -0.631637798785711
20 -0.601160852548939
21 -0.570417133453376
22 -0.539478406687041
23 -0.508406980772824
24 -0.477256278858593
25 -0.446071422404262
26 -0.414889815988018
27 -0.383741735172111
28 -0.352651060673441
29 -0.321635692492814
30 -0.290708349874349
31 -0.259876970312578
32 -0.229145331857912
33 -0.198513510163102
34 -0.167978421632192
35 -0.137534310894787
36 -0.107173183394296
37 -0.0768851128249488
38 -0.0466588814816
39 -0.0164821918621783
40 0.0136578360663266
41 0.0437743222777931
42 0.0738802497591149
43 0.103988002023189
44 0.134109114501491
45 0.164253694445236
46 0.194430210964325
47 0.224644884313653
48 0.254901477372378
49 0.285200553806808
50 0.315539309328627
51 0.345910920317753
52 0.37630408493692
53 0.406702521464734
54 0.437084388939428
55 0.467421806518691
56 0.497680262915945
57 0.527818035969156
58 0.557785722459025
59 0.587525587474014
60 0.616971154132187
61 0.646046793676816
62 0.674667405500004
63 0.702738237138947
64 0.730154985877901
65 0.756803980054152
66 0.782562900398509
67 0.807301603904058
68 0.830883386219915
69 0.853166726302278
70 0.874007171014543
71 0.8932597707029
72 0.910781764367501
73 0.92643561609272
74 0.940092169973863
75 0.951634038672997
76 0.960959006693634
77 0.967983086725768
78 0.972643304774807
79 0.974900044797143
80 0.974738249842061
81 0.972168176329545
82 0.967225002511632
83 0.959967418226169
84 0.950475773242124
85 0.938849251439428
86 0.925202775956209
87 0.909663861851267
88 0.892368979926938
89 0.87346069467232
90 0.85308448017968
91 0.831386114673332
92 0.808509500517737
93 0.784594537890483
94 0.7597756023719
95 0.734180308038563
96 0.707928351525204
97 0.681131007997337
98 0.653890554734485
99 0.626300069284534
100 0.598443468509603
101 0.570395605777752
102 0.542222545015673
103 0.513981869068489
104 0.485723281648182
105 0.45748888977634
106 0.429313955837727
107 0.40122722815698
108 0.373251650393398
109 0.345404754718424
110 0.317699312042693
111 0.290143687182612
112 0.262742423359642
113 0.235496563930513
114 0.208404174043496
115 0.181460607938883
116 0.154658980655255
117 0.127990435923269
118 0.101444424256671
119 0.075009073769428
120 0.0486713660408871
121 0.0224175123582562
122 -0.00376690396029788
123 -0.0298966900035078
124 -0.0559867115853911
125 -0.0820517444222171
126 -0.10810626575879
127 -0.134164171375557
128 -0.160238589551139
129 -0.186341587099516
130 -0.212484011507981
131 -0.238675178998327
132 -0.264922617890897
133 -0.291231790428931
134 -0.317605830765648
135 -0.344045218038014
136 -0.370547436370161
137 -0.397106648168514
138 -0.423713335033469
139 -0.450353920591028
140 -0.477010423825556
141 -0.503659928741267
142 -0.53027428754195
143 -0.556819573961893
144 -0.583255746073726
145 -0.609536139954901
146 -0.635607049090601
147 -0.661407436310355
148 -0.686868452573118
149 -0.711913363063882
150 -0.736457370803625
151 -0.760407657511921
152 -0.783663741154618
153 -0.806117874407825
154 -0.827656031400869
155 -0.848158937969172
156 -0.867503558342052
157 -0.885565019524096
158 -0.90221856909735
159 -0.917341996708991
160 -0.930817945169067
161 -0.942536397618941
162 -0.952396696347212
163 -0.960309624551785
164 -0.966199269577201
165 -0.970005327420583
166 -0.971685767033099
167 -0.971219523191246
168 -0.968607726606006
169 -0.963873664705953
170 -0.957061467587368
171 -0.948234309352719
172 -0.937472368142979
173 -0.924870564095168
174 -0.910535933129736
175 -0.894585137635206
176 -0.87714179974797
};
\addplot[dashed,black,thin,forget plot] coordinates{
(0,0)
(180,0)
};

\end{axis}

\end{tikzpicture}

%% file: 40q_2.4017_energies.tex
\begin{tikzpicture}

\definecolor{crimson2143940}{RGB}{214,39,40}
\definecolor{darkgray176}{RGB}{176,176,176}
\definecolor{darkorange25512714}{RGB}{255,127,14}
\definecolor{forestgreen4416044}{RGB}{44,160,44}
\definecolor{gray127}{RGB}{127,127,127}
\definecolor{lightgray204}{RGB}{204,204,204}
\definecolor{mediumpurple148103189}{RGB}{148,103,189}
\definecolor{orchid227119194}{RGB}{227,119,194}
\definecolor{sienna1408675}{RGB}{140,86,75}
\definecolor{steelblue31119180}{RGB}{31,119,180}

\begin{axis}[
legend cell align={left},
legend to name=sharedlegend_40q_2.4017,
legend columns=4,
legend style={
  fill opacity=0.8,
  draw opacity=1,
  text opacity=1,
  at={(0.5,1.04)},
  anchor=south,
  draw=lightgray204,
  /tikz/column sep=10pt,
},
tick align=outside,
tick pos=left,
x grid style={darkgray176},
xlabel={Delta  (\(\displaystyle \si{\angstrom}\))},
xmin=-0.88, xmax=0.88,
xtick style={color=black},
y grid style={darkgray176},
ylabel={Energy (\(\displaystyle \si{\hartree}\))},
ymin=-152.461536083715, ymax=-152.378297902858,
ytick style={color=black},
yticklabel style={/pgf/number format/precision=2,
/pgf/number format/fixed zerofill},
xtick distance={.3}, 
ytick distance={.01}, 
]
\addplot [semithick, steelblue31119180, mark=asterisk, mark size=3, mark options={solid}]
table {%
-0.8 -152.382978821804
-0.7 -152.410084653035
-0.6 -152.427707319301
-0.5 -152.438693710932
-0.4 -152.444633777537
-0.3 -152.447800163771
-0.2 -152.449529453472
-0.1 -152.44976815105
0 -152.449240378959
};
\addlegendentry{SqDRIFT $M=1000$}
\addplot [semithick, steelblue31119180, opacity=0.5, mark=asterisk, mark size=3, mark options={solid}, forget plot]
table {%
0 -152.449240378959
0.1 -152.44976815105
0.2 -152.449529453472
0.3 -152.447800163771
0.4 -152.444633777537
0.5 -152.438693710932
0.6 -152.427707319301
0.7 -152.410084653035
0.8 -152.382978821804
};
\addplot [semithick, darkorange25512714, mark=+, mark size=3, mark options={solid}]
table {%
-0.8 -152.387463596307
-0.7 -152.414036480684
-0.6 -152.431562460311
-0.5 -152.442670280567
-0.4 -152.449335507778
-0.3 -152.453082769835
-0.2 -152.455066106613
-0.1 -152.455841995246
0 -152.45586486633
};
\addlegendentry{SqDRIFT $M=2000$}
\addplot [semithick, darkorange25512714, opacity=0.5, mark=+, mark size=3, mark options={solid}, forget plot]
table {%
0 -152.45586486633
0.1 -152.455841995246
0.2 -152.455066106613
0.3 -152.453082769835
0.4 -152.449335507778
0.5 -152.442670280567
0.6 -152.431562460311
0.7 -152.414036480684
0.8 -152.387463596307
};
\addplot [semithick, forestgreen4416044, mark=triangle*, mark size=3, mark options={solid}]
table {%
-0.8 -152.387703701408
-0.7 -152.414244623658
-0.6 -152.431832773492
-0.5 -152.442959288851
-0.4 -152.449616431745
-0.3 -152.453361136359
-0.2 -152.455327148446
-0.1 -152.456133024641
0 -152.456170194798
};
\addlegendentry{SqDRIFT $M=2500$}
\addplot [semithick, forestgreen4416044, opacity=0.5, mark=triangle*, mark size=3, mark options={solid}, forget plot]
table {%
0 -152.456170194798
0.1 -152.456133024641
0.2 -152.455327148446
0.3 -152.453361136359
0.4 -152.449616431745
0.5 -152.442959288851
0.6 -152.431832773492
0.7 -152.414244623658
0.8 -152.387703701408
};
\addplot [semithick, crimson2143940]
table {%
-0.8 -152.386783892466
-0.7 -152.413177085479
-0.6 -152.430537701879
-0.5 -152.441568247499
-0.4 -152.448220698605
-0.3 -152.451970561299
-0.2 -152.453906985823
-0.1 -152.454777853181
0 -152.455020214798
0.1 -152.454777853181
0.2 -152.453906985823
0.3 -152.451970561299
0.4 -152.448220698605
0.5 -152.441568247499
0.6 -152.430537701879
0.7 -152.413177085479
0.8 -152.386783892466
};
\addlegendentry{CCSD}
\addplot [semithick, mediumpurple148103189]
table {%
-0.8 -152.38880833575
-0.7 -152.415305756642
-0.6 -152.432769064922
-0.5 -152.443897580373
-0.4 -152.450641066187
-0.3 -152.454471316909
-0.2 -152.456471936682
-0.1 -152.457384633151
0 -152.45764156117
0.1 -152.457384633151
0.2 -152.456471936682
0.3 -152.454471316909
0.4 -152.450641066187
0.5 -152.443897580373
0.6 -152.432769064923
0.7 -152.415305756642
0.8 -152.38880833575
};
\addlegendentry{CCSD(T)}
\addplot [semithick, sienna1408675]
table {%
-0.8 -152.382081456533
-0.7 -152.408373993212
-0.6 -152.425576977191
-0.5 -152.436424420155
-0.4 -152.442898274271
-0.3 -152.446494172873
-0.2 -152.448313560983
-0.1 -152.449111617366
0 -152.449329302631
0.1 -152.449111617366
0.2 -152.448313560984
0.3 -152.446494172873
0.4 -152.442898274271
0.5 -152.436424420154
0.6 -152.425576977192
0.7 -152.408373993211
0.8 -152.382081456533
};
\addlegendentry{CISD}
\addplot [semithick, orchid227119194, mark=*, mark size=2, mark options={solid}]
table {%
-0.8 -152.388056296079
-0.7 -152.414442399236
-0.6 -152.431971880676
-0.5 -152.443084593032
-0.4 -152.449633011921
-0.3 -152.453293410015
-0.2 -152.455108279774
-0.1 -152.455816237448
0 -152.456512608339
0.1 -152.455816237449
0.2 -152.455108279775
0.3 -152.453293410012
0.4 -152.449633011921
0.5 -152.443084593031
0.6 -152.43197188068
0.7 -152.414442399237
0.8 -152.388056296077
};
\addlegendentry{SHCI}
\addplot [semithick, gray127, mark=*, mark size=2, mark options={solid}]
table {%
-0.8 -152.388829837125
-0.7 -152.415268021728
-0.6 -152.432774967218
-0.5 -152.443909899715
-0.4 -152.450641279611
-0.3 -152.454480277051
-0.2 -152.456454433239
-0.1 -152.457395061358
0 -152.45775253004
0.1 -152.457392546413
0.2 -152.456454602333
0.3 -152.454515862461
0.4 -152.450655724337
0.5 -152.44390489437
0.6 -152.432777335475
0.7 -152.415326725
0.8 -152.388840572921
};
\addlegendentry{SHCI + PT2}
\end{axis}

\end{tikzpicture}

%% file: 40q_2.4017_error.tex
\begin{tikzpicture}

\definecolor{crimson2143940}{RGB}{214,39,40}
\definecolor{darkgray176}{RGB}{176,176,176}
\definecolor{darkorange25512714}{RGB}{255,127,14}
\definecolor{forestgreen4416044}{RGB}{44,160,44}
\definecolor{lightgray204}{RGB}{204,204,204}
\definecolor{mediumpurple148103189}{RGB}{148,103,189}
\definecolor{sienna1408675}{RGB}{140,86,75}
\definecolor{steelblue31119180}{RGB}{31,119,180}

\begin{axis}[
legend cell align={left},
legend columns=3,
legend style={
  fill opacity=0.8,
  draw opacity=1,
  text opacity=1,
  at={(0.5,1.04)},
  anchor=south,
  draw=lightgray204
},
tick align=outside,
tick pos=left,
log basis y={10},
x grid style={darkgray176},
xlabel={Delta  (\(\displaystyle \si{\angstrom}\))},
xmin=-0.88, xmax=0.88,
xtick style={color=black},
y grid style={darkgray176},
ylabel={Energy Error (\(\displaystyle \si{\milli\hartree}\))},
ymin=0.49, ymax=10.4254194011427,
ymode=log,
xtick distance={.3}, 
]
\addplot [semithick, crimson2143940]
table {%
-0.8 2.04594465878927
-0.7 2.0909362498287
-0.6 2.23726533963031
-0.5 2.34165221584703
-0.4 2.42058100522513
-0.3 2.50971575127323
-0.2 2.54744741624791
-0.1 2.61720817690048
0 2.73231524221274
0.1 2.61469323129404
0.2 2.54761651032709
0.3 2.54530116148999
0.4 2.43502573204069
0.5 2.3366468711572
0.6 2.23963359658796
0.7 2.14963952123526
0.8 2.05668045467178
};
\addplot [semithick, mediumpurple148103189]
table {%
-0.8 0.0215013742774772
-0.7 -0.0377349132634208
-0.6 0.00590229566910239
-0.5 0.0123193417778111
-0.4 0.000213423192008122
-0.3 0.0089601414003937
-0.2 -0.0175034431038057
-0.1 0.0104282067638906
0 0.110968870131956
0.1 0.00791326124272018
0.2 -0.0173343489962008
0.3 0.0445455519013649
0.4 0.0146581499222975
0.5 0.00731399674691602
0.6 0.00827055296781509
0.7 0.0209683581431364
0.8 0.0322371703589397
};
\addplot [semithick, sienna1408675]
table {%
-0.8 6.74838059129002
-0.7 6.8940285168253
-0.6 7.19799002675359
-0.5 7.48547956004586
-0.4 7.74300533925043
-0.3 7.98610417737677
-0.2 8.1408722557228
-0.1 8.28344399215553
0 8.42322740896861
0.1 8.28092904609434
0.2 8.1410413494325
0.3 8.02168958779248
0.4 7.75745006600914
0.5 7.48047421578235
0.6 7.2003582836544
0.7 6.95273178857292
0.8 6.75911638757043
};
\addplot [semithick, orchid227119194, mark=*, mark size=2, mark options={solid}]
table {%
-0.8 0.773541045958837
-0.7 0.825622492385492
-0.6 0.80308654165151
-0.5 0.825306682429527
-0.4 1.00826768945694
-0.3 1.18686703575577
-0.2 1.34615346502187
-0.1 1.57882391022213
0 1.23992170119891
0.1 1.57630896379146
0.2 1.3463225578505
0.3 1.22245244912733
0.4 1.02271241621565
0.5 0.820301338706031
0.6 0.805454795141713
0.7 0.884325762456228
0.8 0.784276843290854
};
\addplot [semithick, steelblue31119180, mark=asterisk, mark size=3, mark options={solid}]
table {%
-0.8 5.85101532104204
-0.7 5.18336869370728
-0.6 5.06764791680325
-0.5 5.21618878300956
-0.4 6.00750207311762
-0.3 6.68011327994122
-0.2 6.92497976729101
-0.1 7.62691030823248
0 8.51215108133374
};
\addplot [semithick, steelblue31119180, opacity=0.5, mark=asterisk, mark size=3, mark options={solid}, forget plot]
table {%
0 8.51215108133374
0.1 7.62691030823248
0.2 6.92497976729101
0.3 6.68011327994122
0.4 6.00750207311762
0.5 5.21618878300956
0.6 5.06764791680325
0.7 5.18336869370728
0.8 5.85101532104204
};
\addplot [semithick, darkorange25512714, mark=+, mark size=3, mark options={solid}]
table {%
-0.8 1.36624081716263
-0.7 1.23154104400669
-0.6 1.21250690668262
-0.5 1.23961914741244
-0.4 1.30577183270475
-0.3 1.39750721535847
-0.2 1.38832662580057
-0.1 1.55306611208061
0 1.88766371030624
};
\addplot [semithick, darkorange25512714, opacity=0.5, mark=+, mark size=3, mark options={solid}, forget plot]
table {%
0 1.88766371030624
0.1 1.55306611208061
0.2 1.38832662580057
0.3 1.39750721535847
0.4 1.30577183270475
0.5 1.23961914741244
0.6 1.21250690668262
0.7 1.23154104400669
0.8 1.36624081716263
};
\addplot [semithick, forestgreen4416044, mark=triangle*, mark size=3, mark options={solid}]
table {%
-0.8 1.12613571704401
-0.7 1.02339807077101
-0.6 0.94219372604698
-0.5 0.950610863895918
-0.4 1.02484786552282
-0.3 1.11914069148611
-0.2 1.12728479305702
-0.1 1.26203671689495
0 1.5823352422899
};
\addplot [semithick, forestgreen4416044, opacity=0.5, mark=triangle*, mark size=3, mark options={solid}, forget plot]
table {%
0 1.5823352422899
0.1 1.26203671689495
0.2 1.12728479305702
0.3 1.11914069148611
0.4 1.02484786552282
0.5 0.950610863895918
0.6 0.94219372604698
0.7 1.02339807077101
0.8 1.12613571704401
};
\end{axis}

\end{tikzpicture}

%% file: 40q_2.536_energies.tex
\begin{tikzpicture}

\definecolor{crimson2143940}{RGB}{214,39,40}
\definecolor{darkgray176}{RGB}{176,176,176}
\definecolor{darkorange25512714}{RGB}{255,127,14}
\definecolor{forestgreen4416044}{RGB}{44,160,44}
\definecolor{gray127}{RGB}{127,127,127}
\definecolor{lightgray204}{RGB}{204,204,204}
\definecolor{mediumpurple148103189}{RGB}{148,103,189}
\definecolor{orchid227119194}{RGB}{227,119,194}
\definecolor{sienna1408675}{RGB}{140,86,75}
\definecolor{steelblue31119180}{RGB}{31,119,180}

\begin{axis}[
legend cell align={left},
legend to name=sharedlegend_40q_2.536,
legend columns=4,
legend style={
  fill opacity=0.8,
  draw opacity=1,
  text opacity=1,
  at={(0.5,1.04)},
  anchor=south,
  draw=lightgray204,
  /tikz/column sep=10pt,
},
tick align=outside,
tick pos=left,
x grid style={darkgray176},
xlabel={Delta  (\(\displaystyle \si{\angstrom}\))},
xmin=-0.88, xmax=0.88,
xtick style={color=black},
y grid style={darkgray176},
ylabel={Energy (\(\displaystyle \si{\hartree}\))},
ymin=-152.455282315506, ymax=-152.412095051191,
ytick style={color=black},
yticklabel style={/pgf/number format/precision=3,
/pgf/number format/fixed zerofill},
xtick distance={.3}, 
ytick distance={.005}, 
]
\addplot [semithick, steelblue31119180, mark=asterisk, mark size=3, mark options={solid}]
table {%
-0.8 -152.415343395296
-0.7 -152.430026275967
-0.6 -152.439046166152
-0.5 -152.444181021874
-0.4 -152.446600975336
-0.3 -152.446351309603
-0.2 -152.446317998708
-0.1 -152.445974365443
0 -152.446156963292
};
\addlegendentry{SqDRIFT $M=1000$}
\addplot [semithick, steelblue31119180, opacity=0.5, mark=asterisk, mark size=3, mark options={solid}, forget plot]
table {%
0 -152.446156963292
0.1 -152.445974365443
0.2 -152.446317998708
0.3 -152.446351309603
0.4 -152.446600975336
0.5 -152.444181021874
0.6 -152.439046166152
0.7 -152.430026275967
0.8 -152.415343395296
};
\addplot [semithick, darkorange25512714, mark=+, mark size=3, mark options={solid}]
table {%
-0.8 -152.419682380641
-0.7 -152.434258246474
-0.6 -152.443229016564
-0.5 -152.448321581995
-0.4 -152.450855549621
-0.3 -152.451688464516
-0.2 -152.451844216553
-0.1 -152.451692073882
0 -152.451346131279
};
\addlegendentry{SqDRIFT $M=2000$}
\addplot [semithick, darkorange25512714, opacity=0.5, mark=+, mark size=3, mark options={solid}, forget plot]
table {%
0 -152.451346131279
0.1 -152.451692073882
0.2 -152.451844216553
0.3 -152.451688464516
0.4 -152.450855549621
0.5 -152.448321581995
0.6 -152.443229016564
0.7 -152.434258246474
0.8 -152.419682380641
};
\addplot [semithick, forestgreen4416044, mark=triangle*, mark size=3, mark options={solid}]
table {%
-0.8 -152.419874883745
-0.7 -152.434456920068
-0.6 -152.443453553922
-0.5 -152.448542099894
-0.4 -152.451067192166
-0.3 -152.451932461341
-0.2 -152.452140925595
-0.1 -152.451972223016
0 -152.451667265841
};
\addlegendentry{SqDRIFT $M=2500$}
\addplot [semithick, forestgreen4416044, opacity=0.5, mark=triangle*, mark size=3, mark options={solid}, forget plot]
table {%
0 -152.451667265841
0.1 -152.451972223016
0.2 -152.452140925595
0.3 -152.451932461341
0.4 -152.451067192166
0.5 -152.448542099894
0.6 -152.443453553922
0.7 -152.434456920068
0.8 -152.419874883745
};
\addplot [semithick, crimson2143940]
table {%
-0.8 -152.418788943497
-0.7 -152.433212857948
-0.6 -152.442035217539
-0.5 -152.447011606788
-0.4 -152.449468720359
-0.3 -152.450411431659
-0.2 -152.450579491298
-0.1 -152.450481731731
0 -152.450416388313
0.1 -152.45048173173
0.2 -152.450579491297
0.3 -152.450411431659
0.4 -152.449468720359
0.5 -152.447011606788
0.6 -152.442035217538
0.7 -152.433212857948
0.8 -152.418788943497
};
\addlegendentry{CCSD}
\addplot [semithick, mediumpurple148103189]
table {%
-0.8 -152.420926408929
-0.7 -152.435449666572
-0.6 -152.444369647625
-0.5 -152.449441915333
-0.4 -152.451991261964
-0.3 -152.453017727116
-0.2 -152.453253921342
-0.1 -152.453201019628
0 -152.453151366404
0.1 -152.453201019628
0.2 -152.453253921341
0.3 -152.453017727116
0.4 -152.451991261964
0.5 -152.449441915333
0.6 -152.444369647625
0.7 -152.435449666571
0.8 -152.420926408929
};
\addlegendentry{CCSD(T)}
\addplot [semithick, sienna1408675]
table {%
-0.8 -152.414058108661
-0.7 -152.428328276331
-0.6 -152.436966827297
-0.5 -152.441755770316
-0.4 -152.444041204405
-0.3 -152.444840745478
-0.2 -152.444902069281
-0.1 -152.444738665977
0 -152.444651194355
0.1 -152.444738665977
0.2 -152.444902069281
0.3 -152.444840745478
0.4 -152.444041204405
0.5 -152.441755770315
0.6 -152.436966827297
0.7 -152.428328276331
0.8 -152.41405810866
};
\addlegendentry{CISD}
\addplot [semithick, orchid227119194, mark=*, mark size=2, mark options={solid}]
table {%
-0.8 -152.420264104005
-0.7 -152.434782532611
-0.6 -152.443684203393
-0.5 -152.448680150143
-0.4 -152.45106601484
-0.3 -152.451908619607
-0.2 -152.451900161172
-0.1 -152.451454752043
0 -152.451933528692
0.1 -152.451454752041
0.2 -152.451900161172
0.3 -152.451908619608
0.4 -152.451066014839
0.5 -152.448680150142
0.6 -152.443684203395
0.7 -152.43478253261
0.8 -152.420264104005
};
\addlegendentry{SHCI}
\addplot [semithick, gray127, mark=*, mark size=2, mark options={solid}]
table {%
-0.8 -152.420950760505
-0.7 -152.435474261979
-0.6 -152.444386971327
-0.5 -152.449479466491
-0.4 -152.452023515482
-0.3 -152.453054614158
-0.2 -152.453299742533
-0.1 -152.453264647146
0 -152.453216676905
0.1 -152.453221441579
0.2 -152.453319258037
0.3 -152.453022804827
0.4 -152.452036691306
0.5 -152.449482902423
0.6 -152.444413329993
0.7 -152.43546569728
0.8 -152.420960182956
};
\addlegendentry{SHCI + PT2}
\end{axis}

\end{tikzpicture}

%% file: 40q_2.536_error.tex
\begin{tikzpicture}

``\definecolor{crimson2143940}{RGB}{214,39,40}
\definecolor{darkgray176}{RGB}{176,176,176}
\definecolor{darkorange25512714}{RGB}{255,127,14}
\definecolor{forestgreen4416044}{RGB}{44,160,44}
\definecolor{lightgray204}{RGB}{204,204,204}
\definecolor{mediumpurple148103189}{RGB}{148,103,189}
\definecolor{sienna1408675}{RGB}{140,86,75}
\definecolor{steelblue31119180}{RGB}{31,119,180}

\begin{axis}[
legend cell align={left},
legend columns=3,
legend style={
  fill opacity=0.8,
  draw opacity=1,
  text opacity=1,
  at={(0.5,1.04)},
  anchor=south,
  draw=lightgray204,
  /tikz/column sep=10pt,
},
log basis y={10},
tick align=outside,
tick pos=left,
x grid style={darkgray176},
xlabel={Delta  (\(\displaystyle \si{\angstrom}\))},
xmin=-0.88, xmax=0.88,
xtick style={color=black},
y grid style={darkgray176},
ylabel={Energy Error (\(\displaystyle \si{\milli\hartree}\))},
ymin=0.4923368250483, ymax=10.24653181266275,
ymode=log,
xtick distance={.3}, 
]
\addplot [semithick, crimson2143940]
table {%
-0.8 2.16181700787388
-0.7 2.26140403100317
-0.6 2.35175378804797
-0.5 2.46785970324481
-0.4 2.55479512256329
-0.3 2.6431824998383
-0.2 2.72025123535968
-0.1 2.78291541525277
0 2.80028859270942
0.1 2.73970984892458
0.2 2.73976674009191
0.3 2.61137316880422
0.4 2.56797094638728
0.5 2.47129563581439
0.6 2.3781124543234
0.7 2.25283933241371
0.8 2.17123945921571
};
\addplot [semithick, mediumpurple148103189]
table {%
-0.8 0.0243515758029389
-0.7 0.0245954075523969
-0.6 0.0173237013143535
-0.5 0.0375511574759457
-0.4 0.0322535182988304
-0.3 0.0368870427394086
-0.2 0.04582119152019
-0.1 0.0636275176759682
0 0.0653105009860155
0.1 0.0204219510067105
0.2 0.0653366960818857
0.3 0.00507771167690407
0.4 0.0454293421796592
0.5 0.0409870899034104
0.6 0.0436823677034681
0.7 0.0160307089913658
0.8 0.0337740273437248
};
\addplot [semithick, sienna1408675]
table {%
-0.8 6.89265184436749
-0.7 7.14598564792368
-0.6 7.42014402962354
-0.5 7.72369617516233
-0.4 7.98231107671654
-0.3 8.21386868014429
-0.2 8.39767325160778
-0.1 8.52598116892977
0 8.56548255040934
0.1 8.48277560254473
0.2 8.41718875611264
0.3 8.18205934913863
0.4 7.99548690059737
0.5 7.72713210804454
0.6 7.44650269589897
0.7 7.13742094950476
0.8 6.90207429582301
};
\addplot [semithick, orchid227119194, mark=*, mark size=2, mark options={solid}]
table {%
-0.8 0.686656499766514
-0.7 0.691729368583083
-0.6 0.702767933603354
-0.5 0.799316347553258
-0.4 0.957500641959541
-0.3 1.14599455173447
-0.2 1.39958136094265
-0.1 1.80989510280938
0 1.28314821364484
0.1 1.7666895380728
0.2 1.41909686499275
0.3 1.11418521893825
0.4 0.970676466351961
0.5 0.802752281799712
0.6 0.729126597775576
0.7 0.683164670846281
0.8 0.696078950767287
};
\addplot [semithick, steelblue31119180, mark=asterisk, mark size=3, mark options={solid}]
table {%
-0.8 5.60736520921523
-0.7 5.44798601211482
-0.6 5.34080517473967
-0.5 5.29844461701146
-0.4 5.42254014584387
-0.3 6.70330455577073
-0.2 6.98174382523575
-0.1 7.29028170258061
0 7.05971361361435
};
\addplot [semithick, steelblue31119180, opacity=0.5, mark=asterisk, mark size=3, mark options={solid}, forget plot]
table {%
0 7.05971361361435
0.1 7.29028170258061
0.2 6.98174382523575
0.3 6.70330455577073
0.4 5.42254014584387
0.5 5.29844461701146
0.6 5.34080517473967
0.7 5.44798601211482
0.8 5.60736520921523
};
\addplot [semithick, darkorange25512714, mark=+, mark size=3, mark options={solid}]
table {%
-0.8 1.26837986405803
-0.7 1.21601550523565
-0.6 1.15795476256153
-0.5 1.15788449565457
-0.4 1.16796586118539
-0.3 1.36614964233672
-0.2 1.45552598047516
-0.1 1.57257326409876
0 1.87054562655931
};
\addplot [semithick, darkorange25512714, opacity=0.5, mark=+, mark size=3, mark options={solid}, forget plot]
table {%
0 1.87054562655931
0.1 1.57257326409876
0.2 1.45552598047516
0.3 1.36614964233672
0.4 1.16796586118539
0.5 1.15788449565457
0.6 1.15795476256153
0.7 1.21601550523565
0.8 1.26837986405803
};
\addplot [semithick, forestgreen4416044, mark=triangle*, mark size=3, mark options={solid}]
table {%
-0.8 1.07587675947229
-0.7 1.0173419107673
-0.6 0.933417405178716
-0.5 0.93736659687238
-0.4 0.956323315676855
-0.3 1.12215281731665
-0.2 1.15881693776032
-0.1 1.29242413046882
0 1.54941106407591
};
\addplot [semithick, forestgreen4416044, opacity=0.5, mark=triangle*, mark size=3, mark options={solid}, forget plot]
table {%
0 1.54941106407591
0.1 1.29242413046882
0.2 1.15881693776032
0.3 1.12215281731665
0.4 0.956323315676855
0.5 0.93736659687238
0.6 0.933417405178716
0.7 1.0173419107673
0.8 1.07587675947229
};
\end{axis}

\end{tikzpicture}

%% file: 40q_2.6418_energies.tex
\begin{tikzpicture}

\definecolor{crimson2143940}{RGB}{214,39,40}
\definecolor{darkgray176}{RGB}{176,176,176}
\definecolor{darkorange25512714}{RGB}{255,127,14}
\definecolor{forestgreen4416044}{RGB}{44,160,44}
\definecolor{gray127}{RGB}{127,127,127}
\definecolor{lightgray204}{RGB}{204,204,204}
\definecolor{mediumpurple148103189}{RGB}{148,103,189}
\definecolor{orchid227119194}{RGB}{227,119,194}
\definecolor{sienna1408675}{RGB}{140,86,75}
\definecolor{steelblue31119180}{RGB}{31,119,180}

\begin{axis}[
legend cell align={left},
legend to name=sharedlegend_40q_2.6418,
legend columns=4,
legend style={
  fill opacity=0.8,
  draw opacity=1,
  text opacity=1,
  at={(0.5,1.04)},
  anchor=south,
  draw=lightgray204,
  /tikz/column sep=10pt,
},
tick align=outside,
tick pos=left,
x grid style={darkgray176},
xlabel={Delta  (\(\displaystyle \si{\angstrom}\))},
xmin=-0.88, xmax=0.88,
xtick style={color=black},
y grid style={darkgray176},
ylabel={Energy (\(\displaystyle \si{\hartree}\))},
ymin=-152.448945117494, ymax=-152.424443538123,
ytick style={color=black},
yticklabel style={/pgf/number format/precision=3,
/pgf/number format/fixed zerofill},
xtick distance={.3}, 
ytick distance={.003}, 
]
\addplot [semithick, steelblue31119180, mark=asterisk, mark size=3, mark options={solid}]
table {%
-0.8 -152.425557246276
-0.7 -152.435119706827
-0.6 -152.439408132581
-0.5 -152.442176573944
-0.4 -152.442418504641
-0.3 -152.44225362354
-0.2 -152.440957901171
-0.1 -152.439461206636
0 -152.437331970678
};
\addlegendentry{SqDRIFT $M=1000$}
\addplot [semithick, steelblue31119180, opacity=0.5, mark=asterisk, mark size=3, mark options={solid}, forget plot]
table {%
0 -152.437331970678
0.1 -152.439461206636
0.2 -152.440957901171
0.3 -152.44225362354
0.4 -152.442418504641
0.5 -152.442176573944
0.6 -152.439408132581
0.7 -152.435119706827
0.8 -152.425557246276
};
\addplot [semithick, darkorange25512714, mark=+, mark size=3, mark options={solid}]
table {%
-0.8 -152.431532743978
-0.7 -152.440032133019
-0.6 -152.444372662393
-0.5 -152.446247856066
-0.4 -152.446588694969
-0.3 -152.446111702086
-0.2 -152.445334480104
-0.1 -152.444658931381
0 -152.444002406625
};
\addlegendentry{SqDRIFT $M=2000$}
\addplot [semithick, darkorange25512714, opacity=0.5, mark=+, mark size=3, mark options={solid}, forget plot]
table {%
0 -152.444002406625
0.1 -152.444658931381
0.2 -152.445334480104
0.3 -152.446111702086
0.4 -152.446588694969
0.5 -152.446247856066
0.6 -152.444372662393
0.7 -152.440032133019
0.8 -152.431532743978
};
\addplot [semithick, forestgreen4416044, mark=triangle*, mark size=3, mark options={solid}]
table {%
-0.8 -152.431989797654
-0.7 -152.4402458411
-0.6 -152.444668901965
-0.5 -152.446497662352
-0.4 -152.446826847248
-0.3 -152.446387703819
-0.2 -152.445610251682
-0.1 -152.444940755519
0 -152.444302706461
};
\addlegendentry{SqDRIFT $M=2500$}
\addplot [semithick, forestgreen4416044, opacity=0.5, mark=triangle*, mark size=3, mark options={solid}, forget plot]
table {%
0 -152.444302706461
0.1 -152.444940755519
0.2 -152.445610251682
0.3 -152.446387703819
0.4 -152.446826847248
0.5 -152.446497662352
0.6 -152.444668901965
0.7 -152.4402458411
0.8 -152.431989797654
};
\addplot [semithick, crimson2143940]
table {%
-0.8 -152.430814225434
-0.7 -152.438888276405
-0.6 -152.443163592101
-0.5 -152.444924141587
-0.4 -152.445156020094
-0.3 -152.444610727345
-0.2 -152.443842798379
-0.1 -152.443233502976
0 -152.443005934165
0.1 -152.443233502976
0.2 -152.443842798379
0.3 -152.444610727345
0.4 -152.445156020094
0.5 -152.444924141587
0.6 -152.443163592101
0.7 -152.438888276405
0.8 -152.430814225434
};
\addlegendentry{CCSD}
\addplot [semithick, mediumpurple148103189]
table {%
-0.8 -152.433012823755
-0.7 -152.44118402004
-0.6 -152.44555767299
-0.5 -152.447418067888
-0.4 -152.447748376513
-0.3 -152.447293694694
-0.2 -152.446599906149
-0.1 -152.446039501549
0 -152.445829036091
0.1 -152.446039501549
0.2 -152.44659990615
0.3 -152.447293694694
0.4 -152.447748376512
0.5 -152.447418067889
0.6 -152.44555767299
0.7 -152.441184020041
0.8 -152.433012823755
};
\addlegendentry{CCSD(T)}
\addplot [semithick, sienna1408675]
table {%
-0.8 -152.426012350073
-0.7 -152.433904604076
-0.6 -152.437986815638
-0.5 -152.439563012095
-0.4 -152.439632684978
-0.3 -152.438955583145
-0.2 -152.438091028118
-0.1 -152.437422912767
0 -152.437175613545
0.1 -152.437422912767
0.2 -152.438091028119
0.3 -152.438955583145
0.4 -152.439632684978
0.5 -152.439563012096
0.6 -152.437986815637
0.7 -152.433904604076
0.8 -152.426012350073
};
\addlegendentry{CISD}
\addplot [semithick, orchid227119194, mark=*, mark size=2, mark options={solid}]
table {%
-0.8 -152.43235283584
-0.7 -152.440587861963
-0.6 -152.444892553738
-0.5 -152.446703006153
-0.4 -152.446982281596
-0.3 -152.44639300939
-0.2 -152.445475416507
-0.1 -152.444523986467
0 -152.444729798504
0.1 -152.444523986466
0.2 -152.445475416506
0.3 -152.446393009388
0.4 -152.446982281594
0.5 -152.446703006157
0.6 -152.444892553738
0.7 -152.440587861961
0.8 -152.432352835841
};
\addlegendentry{SHCI}
\addplot [semithick, gray127, mark=*, mark size=2, mark options={solid}]
table {%
-0.8 -152.433033073441
-0.7 -152.441209062216
-0.6 -152.445593346917
-0.5 -152.447457552897
-0.4 -152.447812432312
-0.3 -152.447329647859
-0.2 -152.446610895902
-0.1 -152.446040198517
0 -152.445908583438
0.1 -152.446052887275
0.2 -152.44666475498
0.3 -152.447351671494
0.4 -152.447831409341
0.5 -152.44744096979
0.6 -152.445591198875
0.7 -152.441238243664
0.8 -152.433051207076
};
\addlegendentry{SHCI + PT2}
\end{axis}

\end{tikzpicture}

%% file: 40q_2.6418_error.tex
\begin{tikzpicture}

\definecolor{crimson2143940}{RGB}{214,39,40}
\definecolor{darkgray176}{RGB}{176,176,176}
\definecolor{darkorange25512714}{RGB}{255,127,14}
\definecolor{forestgreen4416044}{RGB}{44,160,44}
\definecolor{lightgray204}{RGB}{204,204,204}
\definecolor{mediumpurple148103189}{RGB}{148,103,189}
\definecolor{orchid227119194}{RGB}{227,119,194}
\definecolor{sienna1408675}{RGB}{140,86,75}
\definecolor{steelblue31119180}{RGB}{31,119,180}

\begin{axis}[
legend cell align={left},
legend columns=3,
legend style={
  fill opacity=0.8,
  draw opacity=1,
  text opacity=1,
  at={(0.5,1.04)},
  anchor=south,
  draw=lightgray204
},
log basis y={10},
tick align=outside,
tick pos=left,
x grid style={darkgray176},
xlabel={Delta  (\(\displaystyle \si{\angstrom}\))},
xmin=-0.88, xmax=0.88,
xtick style={color=black},
y grid style={darkgray176},
ylabel={Energy Error (\(\displaystyle \si{\milli\hartree}\))},
ymin=0.4900434825981362582, ymax=10.2977880811164,
ymode=log,
xtick distance={.3}, 
]
\addplot [semithick, crimson2143940]
table {%
-0.8 2.21884800717476
-0.7 2.32078581140627
-0.6 2.42975481600638
-0.5 2.53341131059415
-0.4 2.65641221756141
-0.3 2.71892051407008
-0.2 2.7680975234432
-0.1 2.80669554135216
0 2.90264927298267
0.1 2.81938429932893
0.2 2.82195660136608
0.3 2.74094414865544
0.4 2.67538924683208
0.5 2.51682820311316
0.6 2.42760677375031
0.7 2.34996725916403
0.8 2.2369816421417
};
\addplot [semithick, mediumpurple148103189]
table {%
-0.8 0.0202496861732016
-0.7 0.0250421755652042
-0.6 0.0356739274423035
-0.5 0.0394850093812238
-0.4 0.0640557988447199
-0.3 0.0359531652236456
-0.2 0.0109897527806879
-0.1 0.000696968157853917
0 0.0795473472976482
0.1 0.0133857260209425
0.2 0.0648488308172546
0.3 0.0579767999511205
0.4 0.0830328286269832
0.5 0.0229019014454934
0.6 0.0335258851009712
0.7 0.0542236231240167
0.8 0.0383833210833018
};
\addplot [semithick, sienna1408675]
table {%
-0.8 7.02072336804349
-0.7 7.30445814002678
-0.6 7.60653127935029
-0.5 7.89454080202745
-0.4 8.17974733348592
-0.3 8.37406471396207
-0.2 8.5198677837468
-0.1 8.61728575060283
0 8.73296989351502
0.1 8.62997450815328
0.2 8.57372686189706
0.3 8.39608834860428
0.4 8.19872436304081
0.5 7.87795769448962
0.6 7.60438323717949
0.7 7.33363958795508
0.8 7.03885700318096
};
\addplot [semithick, orchid227119194, mark=*, mark size=2, mark options={solid}]
table {%
-0.8 0.680237601358158
-0.7 0.621200253249299
-0.6 0.700793178708636
-0.5 0.754546743991114
-0.4 0.830150715700029
-0.3 0.936638469227091
-0.2 1.13547939491809
-0.1 1.51621204980756
0 1.17878493395551
0.1 1.52890080960333
0.2 1.18933847465996
0.3 0.958662105773556
0.4 0.849127747471812
0.5 0.737963633355321
0.6 0.698645136651521
0.7 0.650381702598679
0.8 0.698371235728246
};
\addplot [semithick, steelblue31119180, mark=asterisk, mark size=3, mark options={solid}]
table {%
-0.8 7.47582716519446
-0.7 6.08935538915034
-0.6 6.18521433599994
-0.5 5.28097895355017
-0.4 5.39392767146296
-0.3 5.07602431937926
-0.2 5.65299473106506
-0.1 6.57899188161082
0 8.57661275972532
};
\addplot [semithick, steelblue31119180, opacity=0.5, mark=asterisk, mark size=3, mark options={solid}, forget plot]
table {%
0 8.57661275972532
0.1 6.57899188161082
0.2 5.65299473106506
0.3 5.07602431937926
0.4 5.39392767146296
0.5 5.28097895355017
0.6 6.18521433599994
0.7 6.08935538915034
0.8 7.47582716519446
};
\addplot [semithick, darkorange25512714, mark=+, mark size=3, mark options={solid}]
table {%
-0.8 1.5003294635676
-0.7 1.17692919690171
-0.6 1.2206845236733
-0.5 1.20969683155181
-0.4 1.22373734328107
-0.3 1.21794577333389
-0.2 1.27641579780402
-0.1 1.38126713656561
0 1.90617681306549
};
\addplot [semithick, darkorange25512714, opacity=0.5, mark=+, mark size=3, mark options={solid}, forget plot]
table {%
0 1.90617681306549
0.1 1.38126713656561
0.2 1.27641579780402
0.3 1.21794577333389
0.4 1.22373734328107
0.5 1.20969683155181
0.6 1.2206845236733
0.7 1.17692919690171
0.8 1.5003294635676
};
\addplot [semithick, forestgreen4416044, mark=triangle*, mark size=3, mark options={solid}]
table {%
-0.8 1.04327578793573
-0.7 0.963221115966917
-0.6 0.924444952403292
-0.5 0.959890545175313
-0.4 0.985585063602912
-0.3 0.941944040022236
-0.2 1.00064422008472
-0.1 1.09944299842368
0 1.60587697689607
};
\addplot [semithick, forestgreen4416044, opacity=0.5, mark=triangle*, mark size=3, mark options={solid}, forget plot]
table {%
0 1.60587697689607
0.1 1.09944299842368
0.2 1.00064422008472
0.3 0.941944040022236
0.4 0.985585063602912
0.5 0.959890545175313
0.6 0.924444952403292
0.7 0.963221115966917
0.8 1.04327578793573
};
\end{axis}

\end{tikzpicture}

%% file: 40q_2.786_energies.tex
\begin{tikzpicture}

\definecolor{crimson2143940}{RGB}{214,39,40}
\definecolor{darkgray176}{RGB}{176,176,176}
\definecolor{darkorange25512714}{RGB}{255,127,14}
\definecolor{forestgreen4416044}{RGB}{44,160,44}
\definecolor{gray127}{RGB}{127,127,127}
\definecolor{lightgray204}{RGB}{204,204,204}
\definecolor{mediumpurple148103189}{RGB}{148,103,189}
\definecolor{orchid227119194}{RGB}{227,119,194}
\definecolor{sienna1408675}{RGB}{140,86,75}
\definecolor{steelblue31119180}{RGB}{31,119,180}

\begin{axis}[
legend cell align={left},
legend to name=sharedlegend_40q_2.786,
legend columns=4,
legend style={
  fill opacity=0.8,
  draw opacity=1,
  text opacity=1,
  at={(0.5,1.04)},
  anchor=south,
  draw=lightgray204,
  /tikz/column sep=10pt,
},
tick align=outside,
tick pos=left,
x grid style={darkgray176},
xlabel={Delta  (\(\displaystyle \si{\angstrom}\))},
xmin=-0.88, xmax=0.88,
xtick style={color=black},
y grid style={darkgray176},
ylabel={Energy (\(\displaystyle \si{\hartree}\))},
ymin=-152.440831795524, ymax=-152.422701776777,
ytick style={color=black},
yticklabel style={/pgf/number format/precision=3,
/pgf/number format/fixed zerofill},
xtick distance={.3}, 
ytick distance={.003}, 
]
\addplot [semithick, steelblue31119180, mark=asterisk, mark size=3, mark options={solid}]
table {%
-0.8 -152.432288906655
-0.7 -152.435997669709
-0.6 -152.436296602821
-0.5 -152.435133209208
-0.4 -152.433203738056
-0.3 -152.430992443675
-0.2 -152.428775464184
-0.1 -152.426233818141
0 -152.425348872022
};
\addlegendentry{SqDRIFT $M=1000$}
\addplot [semithick, steelblue31119180, opacity=0.5, mark=asterisk, mark size=3, mark options={solid}, forget plot]
table {%
0 -152.425348872022
0.1 -152.426233818141
0.2 -152.428775464184
0.3 -152.430992443675
0.4 -152.433203738056
0.5 -152.435133209208
0.6 -152.436296602821
0.7 -152.435997669709
0.8 -152.432288906655
};
\addplot [semithick, darkorange25512714, mark=+, mark size=3, mark options={solid}]
table {%
-0.8 -152.43608341816
-0.7 -152.438560785462
-0.6 -152.438920564178
-0.5 -152.43788539183
-0.4 -152.436141398236
-0.3 -152.434253718167
-0.2 -152.432485381065
-0.1 -152.431339673996
0 -152.430680467612
};
\addlegendentry{SqDRIFT $M=2000$}
\addplot [semithick, darkorange25512714, opacity=0.5, mark=+, mark size=3, mark options={solid}, forget plot]
table {%
0 -152.430680467612
0.1 -152.431339673996
0.2 -152.432485381065
0.3 -152.434253718167
0.4 -152.436141398236
0.5 -152.43788539183
0.6 -152.438920564178
0.7 -152.438560785462
0.8 -152.43608341816
};
\addplot [semithick, forestgreen4416044, mark=triangle*, mark size=3, mark options={solid}]
table {%
-0.8 -152.436284055527
-0.7 -152.438840249184
-0.6 -152.439130464166
-0.5 -152.438064062178
-0.4 -152.436341053939
-0.3 -152.43444757171
-0.2 -152.432746822833
-0.1 -152.431645009455
0 -152.431069058879
};
\addlegendentry{SqDRIFT $M=2500$}
\addplot [semithick, forestgreen4416044, opacity=0.5, mark=triangle*, mark size=3, mark options={solid}, forget plot]
table {%
0 -152.431069058879
0.1 -152.431645009455
0.2 -152.432746822833
0.3 -152.43444757171
0.4 -152.436341053939
0.5 -152.438064062178
0.6 -152.439130464166
0.7 -152.438840249184
0.8 -152.436284055527
};
\addplot [semithick, crimson2143940]
table {%
-0.8 -152.435078989938
-0.7 -152.437372807493
-0.6 -152.437481325148
-0.5 -152.436281522476
-0.4 -152.43445954427
-0.3 -152.432539649208
-0.2 -152.430907004148
-0.1 -152.429825115868
0 -152.429447650119
0.1 -152.429825115868
0.2 -152.430907004148
0.3 -152.432539649208
0.4 -152.43445954427
0.5 -152.436281522476
0.6 -152.437481325148
0.7 -152.437372807493
0.8 -152.435078989939
};
\addlegendentry{CCSD}
\addplot [semithick, mediumpurple148103189]
table {%
-0.8 -152.437332737163
-0.7 -152.439725060625
-0.6 -152.439939260957
-0.5 -152.438850989372
-0.4 -152.437140777195
-0.3 -152.435324056297
-0.2 -152.43377559554
-0.1 -152.432749018675
0 -152.432390858567
0.1 -152.432749018675
0.2 -152.43377559554
0.3 -152.435324056297
0.4 -152.437140777195
0.5 -152.438850989372
0.6 -152.439939260957
0.7 -152.439725060625
0.8 -152.437332737163
};
\addlegendentry{CCSD(T)}
\addplot [semithick, sienna1408675]
table {%
-0.8 -152.430112218075
-0.7 -152.43220360259
-0.6 -152.432121066467
-0.5 -152.430752449926
-0.4 -152.428789580652
-0.3 -152.426759379482
-0.2 -152.42504798649
-0.1 -152.423918990481
0 -152.423525868538
0.1 -152.42391899048
0.2 -152.425047986491
0.3 -152.426759379482
0.4 -152.428789580652
0.5 -152.430752449926
0.6 -152.432121066467
0.7 -152.432203602591
0.8 -152.430112218075
};
\addlegendentry{CISD}
\addplot [semithick, orchid227119194, mark=*, mark size=2, mark options={solid}]
table {%
-0.8 -152.436790265988
-0.7 -152.439178072369
-0.6 -152.439362534196
-0.5 -152.43821850665
-0.4 -152.436429203137
-0.3 -152.43445973427
-0.2 -152.432556842643
-0.1 -152.431541131252
0 -152.431407487013
0.1 -152.431541131252
0.2 -152.432556842647
0.3 -152.434459734268
0.4 -152.436429203137
0.5 -152.438218506649
0.6 -152.439362534196
0.7 -152.439178072368
0.8 -152.436790265988
};
\addlegendentry{SHCI}
\addplot [semithick, gray127, mark=*, mark size=2, mark options={solid}]
table {%
-0.8 -152.437382324268
-0.7 -152.439773223615
-0.6 -152.440001722588
-0.5 -152.438937816468
-0.4 -152.437220033605
-0.3 -152.435411752955
-0.2 -152.433819309137
-0.1 -152.432807748765
0 -152.4325397863
0.1 -152.432766185091
0.2 -152.433830106049
0.3 -152.435396867253
0.4 -152.437209965808
0.5 -152.438913075876
0.6 -152.440007703763
0.7 -152.439766171955
0.8 -152.437384843308
};
\addlegendentry{SHCI + PT2}
\end{axis}

\end{tikzpicture}

%% file: 40q_2.786_error.tex
\begin{tikzpicture}

\definecolor{crimson2143940}{RGB}{214,39,40}
\definecolor{darkgray176}{RGB}{176,176,176}
\definecolor{darkorange25512714}{RGB}{255,127,14}
\definecolor{forestgreen4416044}{RGB}{44,160,44}
\definecolor{lightgray204}{RGB}{204,204,204}
\definecolor{mediumpurple148103189}{RGB}{148,103,189}
\definecolor{orchid227119194}{RGB}{227,119,194}
\definecolor{sienna1408675}{RGB}{140,86,75}
\definecolor{steelblue31119180}{RGB}{31,119,180}

\begin{axis}[
legend cell align={left},
legend columns=3,
legend style={
  fill opacity=0.8,
  draw opacity=1,
  text opacity=1,
  at={(0.5,1.04)},
  anchor=south,
  draw=lightgray204
},
tick align=outside,
tick pos=left,
log basis y={10},
x grid style={darkgray176},
xlabel={Delta  (\(\displaystyle \si{\angstrom}\))},
xmin=-0.88, xmax=0.88,
xtick style={color=black},
y grid style={darkgray176},
ylabel={Energy Error (\(\displaystyle \si{\milli\hartree}\))},
ymin=0.49, ymax=10.4254194011427,
ymode=log,
xtick distance={.3}, 
]
\addplot [semithick, crimson2143940]
table {%
-0.8 2.3033343292127
-0.7 2.40041612173059
-0.6 2.52039743983801
-0.5 2.65629399166301
-0.4 2.76048933471884
-0.3 2.87210374719393
-0.2 2.91230498851291
-0.1 2.98263289678857
0 3.09213618052695
0.1 2.94106922260085
0.2 2.92310190130252
0.3 2.85721804507943
0.4 2.75042153859317
0.5 2.6315534000787
0.6 2.52637861510152
0.7 2.39336446151128
0.8 2.30585336888112
};
\addplot [semithick, mediumpurple148103189]
table {%
-0.8 0.0495871044847718
-0.7 0.0481629897990388
-0.6 0.0624616305628933
-0.5 0.086827095941544
-0.4 0.0792564096059323
-0.3 0.0876966578289284
-0.2 0.043713596426187
-0.1 0.0587300892789244
0 0.148927732823267
0.1 0.0171664151764617
0.2 0.0545105094431619
0.3 0.0728109558281176
0.4 0.0691886133665776
0.5 0.0620865046698782
0.6 0.0684428057695641
0.7 0.0411113294944698
0.8 0.0521061446363547
};
\addplot [semithick, sienna1408675]
table {%
-0.8 7.2701061929763
-0.7 7.56962102491343
-0.6 7.88065612096034
-0.5 8.1853665417384
-0.4 8.43045295303568
-0.3 8.65237347275638
-0.2 8.77132264611191
-0.1 8.88875828405844
0 9.01391776139349
0.1 8.84719461015493
0.2 8.78211955887309
0.3 8.63748777061346
0.4 8.42038515679633
0.5 8.16062595040989
0.6 7.88663729608174
0.7 7.56256936446675
0.8 7.2726252327584
};
\addplot [semithick, orchid227119194, mark=*, mark size=2, mark options={solid}]
table {%
-0.8 0.59205827986375
-0.7 0.595151246272962
-0.6 0.639188391062362
-0.5 0.719309817384328
-0.4 0.790830467479964
-0.3 0.952018685495659
-0.2 1.26246649313089
-0.1 1.26661751261281
0 1.13229928626879
0.1 1.22505383893667
0.2 1.27326340276568
0.3 0.937132984631717
0.4 0.780762671354296
0.5 0.694569227704278
0.6 0.645169566524828
0.7 0.588099586792623
0.8 0.594577319930067
};
\addplot [semithick, steelblue31119180, mark=asterisk, mark size=3, mark options={solid}]
table {%
-0.8 5.09341761207338
-0.7 3.77555390610951
-0.6 3.7051197661242
-0.5 3.80460726003662
-0.4 4.01629554900751
-0.3 4.41930928050738
-0.2 5.04384495263821
-0.1 6.57393062354572
0 7.19091427714602
};
\addplot [semithick, steelblue31119180, opacity=0.5, mark=asterisk, mark size=3, mark options={solid}, forget plot]
table {%
0 7.19091427714602
0.1 6.57393062354572
0.2 5.04384495263821
0.3 4.41930928050738
0.4 4.01629554900751
0.5 3.80460726003662
0.6 3.7051197661242
0.7 3.77555390610951
0.8 5.09341761207338
};
\addplot [semithick, darkorange25512714, mark=+, mark size=3, mark options={solid}]
table {%
-0.8 1.29890610756433
-0.7 1.21243815340222
-0.6 1.08115840941991
-0.5 1.05242463729383
-0.4 1.07863536857167
-0.3 1.15803478797716
-0.2 1.33392807163091
-0.1 1.46807476892263
0 1.85931868770695
};
\addplot [semithick, darkorange25512714, opacity=0.5, mark=+, mark size=3, mark options={solid}, forget plot]
table {%
0 1.85931868770695
0.1 1.46807476892263
0.2 1.33392807163091
0.3 1.15803478797716
0.4 1.07863536857167
0.5 1.05242463729383
0.6 1.08115840941991
0.7 1.21243815340222
0.8 1.29890610756433
};
\addplot [semithick, forestgreen4416044, mark=triangle*, mark size=3, mark options={solid}]
table {%
-0.8 1.09826874029295
-0.7 0.932974431037792
-0.6 0.871258421142329
-0.5 0.873754290040551
-0.4 0.878979665543511
-0.3 0.964181245336704
-0.2 1.07248630379786
-0.1 1.16273931004685
0 1.47072742015553
};
\addplot [semithick, forestgreen4416044, opacity=0.5, mark=triangle*, mark size=3, mark options={solid}, forget plot]
table {%
0 1.47072742015553
0.1 1.16273931004685
0.2 1.07248630379786
0.3 0.964181245336704
0.4 0.878979665543511
0.5 0.873754290040551
0.6 0.871258421142329
0.7 0.932974431037792
0.8 1.09826874029295
};
\end{axis}

\end{tikzpicture}

%% file: 40q_combined_energies.tex
\begin{tikzpicture}

\definecolor{crimson2143940}{RGB}{214,39,40}
\definecolor{darkgray176}{RGB}{176,176,176}
\definecolor{darkorange25512714}{RGB}{255,127,14}
\definecolor{forestgreen4416044}{RGB}{44,160,44}
\definecolor{lightgray204}{RGB}{204,204,204}
\definecolor{steelblue31119180}{RGB}{31,119,180}

\begin{axis}[
width=.75\textwidth,
height=.5\textwidth,
legend cell align={left},
legend columns=3,
legend style={
  fill opacity=0.8,
  draw opacity=1,
  text opacity=1,
  at={(0.5,1.05)},
  anchor=south,
  draw=lightgray204,
  /tikz/column sep=10pt,
},
tick align=outside,
tick pos=left,
x grid style={darkgray176},
xlabel={Delta  (\(\displaystyle \si{\angstrom}\))},
xmin=-0.88, xmax=0.88,
xtick style={color=black},
y grid style={darkgray176},
ylabel={Energy (\(\displaystyle \si{\hartree}\))},
ymin=-152.458554971472, ymax=-152.428,
ytick style={color=black},
xtick distance={.2}, 
ytick distance={.004}, 
yticklabel style={/pgf/number format/precision=3,
/pgf/number format/fixed zerofill},
]
\addplot [semithick, steelblue31119180]
table {%
-0.8 -152.388829837125
-0.7 -152.415268021728
-0.6 -152.432774967218
-0.5 -152.443909899715
-0.4 -152.450641279611
-0.3 -152.454480277051
-0.2 -152.456454433239
-0.1 -152.457395061358
0 -152.45775253004
0.1 -152.457392546413
0.2 -152.456454602333
0.3 -152.454515862461
0.4 -152.450655724337
0.5 -152.44390489437
0.6 -152.432777335475
0.7 -152.415326725
0.8 -152.388840572921
};
\addlegendentry{$r(\text{O}_a\text{O}_b)=2.4017\,\si{\angstrom}$}
\addplot [semithick, darkorange25512714]
table {%
-0.8 -152.420950760505
-0.7 -152.435474261979
-0.6 -152.444386971327
-0.5 -152.449479466491
-0.4 -152.452023515482
-0.3 -152.453054614158
-0.2 -152.453299742533
-0.1 -152.453264647146
0 -152.453216676905
0.1 -152.453221441579
0.2 -152.453319258037
0.3 -152.453022804827
0.4 -152.452036691306
0.5 -152.449482902423
0.6 -152.444413329993
0.7 -152.43546569728
0.8 -152.420960182956
};
\addlegendentry{$r(\text{O}_a\text{O}_b)=2.536\,\si{\angstrom}$}
\addplot [semithick, forestgreen4416044]
table {%
-0.8 -152.433033073441
-0.7 -152.441209062216
-0.6 -152.445593346917
-0.5 -152.447457552897
-0.4 -152.447812432312
-0.3 -152.447329647859
-0.2 -152.446610895902
-0.1 -152.446040198517
0 -152.445908583438
0.1 -152.446052887275
0.2 -152.44666475498
0.3 -152.447351671494
0.4 -152.447831409341
0.5 -152.44744096979
0.6 -152.445591198875
0.7 -152.441238243664
0.8 -152.433051207076
};
\addlegendentry{$r(\text{O}_a\text{O}_b)=2.6418\,\si{\angstrom}$}
\addplot [semithick, crimson2143940]
table {%
-0.8 -152.437382324268
-0.7 -152.439773223615
-0.6 -152.440001722588
-0.5 -152.438937816468
-0.4 -152.437220033605
-0.3 -152.435411752955
-0.2 -152.433819309137
-0.1 -152.432807748765
0 -152.4325397863
0.1 -152.432766185091
0.2 -152.433830106049
0.3 -152.435396867253
0.4 -152.437209965808
0.5 -152.438913075876
0.6 -152.440007703763
0.7 -152.439766171955
0.8 -152.437384843308
};
\addlegendentry{$r(\text{O}_a\text{O}_b)=2.786\,\si{\angstrom}$}
\addplot [semithick, steelblue31119180, mark=triangle*, mark size=3, mark options={solid}, forget plot]
table {%
-0.8 -152.387703701408
-0.7 -152.414244623658
-0.6 -152.431832773492
-0.5 -152.442959288851
-0.4 -152.449616431745
-0.3 -152.453361136359
-0.2 -152.455327148446
-0.1 -152.456133024641
0 -152.456170194798
};
\addplot [semithick, steelblue31119180, opacity=0.5, mark=triangle*, mark size=3, mark options={solid}, forget plot]
table {%
0 -152.456170194798
0.1 -152.456133024641
0.2 -152.455327148446
0.3 -152.453361136359
0.4 -152.449616431745
0.5 -152.442959288851
0.6 -152.431832773492
0.7 -152.414244623658
0.8 -152.387703701408
};
\addplot [semithick, darkorange25512714, mark=triangle*, mark size=3, mark options={solid}, forget plot]
table {%
-0.8 -152.419874883745
-0.7 -152.434456920068
-0.6 -152.443453553922
-0.5 -152.448542099894
-0.4 -152.451067192166
-0.3 -152.451932461341
-0.2 -152.452140925595
-0.1 -152.451972223016
0 -152.451667265841
};
\addplot [semithick, darkorange25512714, opacity=0.5, mark=triangle*, mark size=3, mark options={solid}, forget plot]
table {%
0 -152.451667265841
0.1 -152.451972223016
0.2 -152.452140925595
0.3 -152.451932461341
0.4 -152.451067192166
0.5 -152.448542099894
0.6 -152.443453553922
0.7 -152.434456920068
0.8 -152.419874883745
};
\addplot [semithick, forestgreen4416044, mark=triangle*, mark size=3, mark options={solid}, forget plot]
table {%
-0.8 -152.431989797654
-0.7 -152.4402458411
-0.6 -152.444668901965
-0.5 -152.446497662352
-0.4 -152.446826847248
-0.3 -152.446387703819
-0.2 -152.445610251682
-0.1 -152.444940755519
0 -152.444302706461
};
\addplot [semithick, forestgreen4416044, opacity=0.5, mark=triangle*, mark size=3, mark options={solid}, forget plot]
table {%
0 -152.444302706461
0.1 -152.444940755519
0.2 -152.445610251682
0.3 -152.446387703819
0.4 -152.446826847248
0.5 -152.446497662352
0.6 -152.444668901965
0.7 -152.4402458411
0.8 -152.431989797654
};
\addplot [semithick, crimson2143940, mark=triangle*, mark size=3, mark options={solid}, forget plot]
table {%
-0.8 -152.436284055527
-0.7 -152.438840249184
-0.6 -152.439130464166
-0.5 -152.438064062178
-0.4 -152.436341053939
-0.3 -152.43444757171
-0.2 -152.432746822833
-0.1 -152.431645009455
0 -152.431069058879
};
\addplot [semithick, crimson2143940, opacity=0.5, mark=triangle*, mark size=3, mark options={solid}, forget plot]
table {%
0 -152.431069058879
0.1 -152.431645009455
0.2 -152.432746822833
0.3 -152.43444757171
0.4 -152.436341053939
0.5 -152.438064062178
0.6 -152.439130464166
0.7 -152.438840249184
0.8 -152.436284055527
};

\addlegendimage{black,semithick}
\addlegendentry{SHCI + PT2}

\addlegendimage{black,semithick,mark=triangle*, mark size=3, mark options={solid}}
\addlegendentry{SqDRIFT $M=2500$}
\end{axis}

\end{tikzpicture}

%% file: 52q_energies.tex
\begin{tikzpicture}

\definecolor{crimson2143940}{RGB}{214,39,40}
\definecolor{darkgray176}{RGB}{176,176,176}
\definecolor{darkorange25512714}{RGB}{255,127,14}
\definecolor{forestgreen4416044}{RGB}{44,160,44}
\definecolor{lightgray204}{RGB}{204,204,204}
\definecolor{mediumpurple148103189}{RGB}{148,103,189}
\definecolor{steelblue31119180}{RGB}{31,119,180}

\begin{axis}[
legend cell align={left},
legend to name=sharedlegend_52q,
legend columns=6,
legend style={
  fill opacity=0.8,
  draw opacity=1,
  text opacity=1,
  at={(0.5,1.04)},
  anchor=south,
  draw=lightgray204,
  /tikz/column sep=10pt,
},
tick align=outside,
tick pos=left,
x grid style={darkgray176},
xlabel={Delta  (\(\displaystyle \si{\angstrom}\))},
xmin=-0.88, xmax=0.88,
xtick style={color=black},
y grid style={darkgray176},
ylabel={Energy (\(\displaystyle \si{\hartree}\))},
ymin=-152.591657909502, ymax=-152.552275781611,
ytick style={color=black},
yticklabel style={/pgf/number format/precision=3,
/pgf/number format/fixed zerofill},
xtick distance={.3}, 
ytick distance={.005}, 
]
\addplot [semithick, blue, mark=asterisk, mark size=3, mark options={solid}]
table {%
-0.8 -152.571201420589
-0.7 -152.579601707194
-0.6 -152.583529487197
-0.5 -152.5847936762
-0.4 -152.584828016589
-0.3 -152.583852954776
-0.2 -152.582596810898
-0.1 -152.581587682832
0 -152.579802175871
};
\addlegendentry{SqDRIFT}
\addplot [semithick, blue, opacity=0.5, mark=asterisk, mark size=3, mark options={solid}, forget plot]
table {%
0 -152.579802175871
0.1 -152.581587682832
0.2 -152.582596810898
0.3 -152.583852954776
0.4 -152.584828016589
0.5 -152.5847936762
0.6 -152.583529487197
0.7 -152.579601707194
0.8 -152.571201420589
};
\addplot [semithick, steelblue31119180]
table {%
-0.8 -152.573540207854
-0.7 -152.581312934425
-0.6 -152.585092470656
-0.5 -152.586338053368
-0.4 -152.58612472737
-0.3 -152.58523749474
-0.2 -152.584234846951
-0.1 -152.583490406908
0 -152.583218903929
0.1 -152.583490406908
0.2 -152.584234846951
0.3 -152.58523749474
0.4 -152.586124727369
0.5 -152.586338053369
0.6 -152.585092470656
0.7 -152.581312934425
0.8 -152.573540207854
};
\addlegendentry{CCSD}
\addplot [semithick, darkorange25512714]
table {%
-0.8 -152.575833354299
-0.7 -152.583696756354
-0.6 -152.587574583105
-0.5 -152.588922280894
-0.4 -152.588809196908
-0.3 -152.588012906651
-0.2 -152.587083551233
-0.1 -152.586386897795
0 -152.586132013374
0.1 -152.586386897795
0.2 -152.587083551234
0.3 -152.588012906651
0.4 -152.588809196908
0.5 -152.588922280894
0.6 -152.587574583105
0.7 -152.583696756354
0.8 -152.575833354299
};
\addlegendentry{CCSD(T)}
\addplot [semithick, forestgreen4416044]
table {%
-0.8 -152.554065878334
-0.7 -152.561529470783
-0.6 -152.565019444392
-0.5 -152.566002188665
-0.4 -152.565560272003
-0.3 -152.564486051759
-0.2 -152.563344579591
-0.1 -152.56251464369
0 -152.562214266864
0.1 -152.562514643691
0.2 -152.563344579591
0.3 -152.564486051759
0.4 -152.565560272003
0.5 -152.566002188665
0.6 -152.565019444392
0.7 -152.561529470784
0.8 -152.554065878333
};
\addlegendentry{CISD}
\addplot [semithick, crimson2143940, mark=triangle*, mark size=3, mark options={solid}]
table {%
-0.8 -152.573539854028
-0.7 -152.581652734941
-0.6 -152.585330695147
-0.5 -152.586501890697
-0.4 -152.586113557938
-0.3 -152.584701426647
-0.2 -152.58286557401
-0.1 -152.580821696224
0 -152.582399307484
0.1 -152.580821696222
0.2 -152.582865574011
0.3 -152.58470142665
0.4 -152.586113557937
0.5 -152.586501890696
0.6 -152.585330695146
0.7 -152.581652734941
0.8 -152.573539854029
};
\addlegendentry{SHCI}
\addplot [semithick, mediumpurple148103189, mark=triangle*, mark size=3, mark options={solid}]
table {%
-0.8 -152.576663429661
-0.7 -152.584486703255
-0.6 -152.588314690793
-0.5 -152.58986781278
-0.4 -152.589664785096
-0.3 -152.588702762355
-0.2 -152.587772178219
-0.1 -152.587074079593
0 -152.586936906752
0.1 -152.586999510019
0.2 -152.588070679115
0.3 -152.588806135594
0.4 -152.589733643124
0.5 -152.589747572435
0.6 -152.588247624007
0.7 -152.584458009618
0.8 -152.576718668018
};
\addlegendentry{SHCI + PT2}
\end{axis}

\end{tikzpicture}

%% file: 52q_errors.tex
\begin{tikzpicture}

\definecolor{crimson2143940}{RGB}{214,39,40}
\definecolor{darkgray176}{RGB}{176,176,176}
\definecolor{darkorange25512714}{RGB}{255,127,14}
\definecolor{forestgreen4416044}{RGB}{44,160,44}
\definecolor{lightgray204}{RGB}{204,204,204}
\definecolor{steelblue31119180}{RGB}{31,119,180}

\begin{axis}[
legend cell align={left},
legend columns=3,
legend style={
  fill opacity=0.8,
  draw opacity=1,
  text opacity=1,
  at={(0.5,1.04)},
  anchor=south,
  draw=lightgray204,
  /tikz/column sep=10pt,
},
log basis y={10},
tick align=outside,
tick pos=left,
x grid style={darkgray176},
xlabel={Delta  (\(\displaystyle \si{\angstrom}\))},
xmin=-0.88, xmax=0.88,
xtick style={color=black},
y grid style={darkgray176},
ylabel={Energy Error (\(\displaystyle \si{\milli\hartree}\))},
ymin=0.509198552896693, ymax=29.7477491554564,
ymode=log,
ytick style={color=black},
xtick distance={.3}, 
]
\addplot [semithick, blue, mark=asterisk, mark size=3, mark options={solid}]
table {%
-0.8 5.46200907197658
-0.7 4.88499606137793
-0.6 4.78520359558843
-0.5 5.07413658039013
-0.4 4.83676850723214
-0.3 4.8498075790917
-0.2 5.17536732141366
-0.1 5.48639676125617
0 7.13473088114824
};
\addplot [semithick, blue, opacity=0.5, mark=asterisk, mark size=3, mark options={solid}, forget plot]
table {%
0 7.13473088114824
0.1 5.41182718728805
0.2 5.47386821702389
0.3 4.95318081817686
0.4 4.90562653564552
0.5 4.95389623574738
0.6 4.71813680982791
0.7 4.85630242474144
0.8 5.51724742942383
};
\addplot [semithick, steelblue31119180]
table {%
-0.8 3.12322180661795
-0.7 3.17376882998133
-0.6 3.22222013659257
-0.5 3.52975941186173
-0.4 3.54005772643973
-0.3 3.46526761475729
-0.2 3.53733126783595
-0.1 3.58367268520965
0 3.71800282340473
0.1 3.50910311109942
0.2 3.83583216330408
0.3 3.56864085372877
0.4 3.6089157549668
0.5 3.40951906667897
0.6 3.15515335086047
0.7 3.14507519334484
0.8 3.17846016423573
};
\addplot [semithick, darkorange25512714]
table {%
-0.8 0.830075362358684
-0.7 0.789946900908944
-0.6 0.74010768830135
-0.5 0.945531886259232
-0.4 0.85558818753384
-0.3 0.689855703541298
-0.2 0.688626985692053
-0.1 0.687181798468828
0 0.804893378443694
0.1 0.612612224074383
0.2 0.987127880762273
0.3 0.793228942598034
0.4 0.924446216203023
0.5 0.825291541246997
0.6 0.673040902398725
0.7 0.761253263931394
0.8 0.885313719862779
};
\addplot [semithick, forestgreen4416044]
table {%
-0.8 22.5975513274079
-0.7 22.9572324715548
-0.6 23.2952464008349
-0.5 23.8656241147623
-0.4 24.1045130929081
-0.3 24.2167105956526
-0.2 24.4275986283355
-0.1 24.559435902745
0 24.7226398884948
0.1 24.4848663286064
0.2 24.72609952369
0.3 24.3200838347377
0.4 24.1733711215204
0.5 23.7453837700343
0.6 23.2281796149607
0.7 22.9285388347478
0.8 22.6527896849973
};
\addplot [semithick, crimson2143940, mark=triangle*, mark size=3, mark options={solid}]
table {%
-0.8 3.1235756329977
-0.7 2.83396831400751
-0.6 2.98399564599094
-0.5 3.3659220829918
-0.4 3.55122715799894
-0.3 4.00133570801131
-0.2 4.906604209026
-0.1 6.25238336948541
0 4.53759926853081
0.1 6.17781379685312
0.2 5.20510510395411
0.3 4.10470894456694
0.4 3.62008518709445
0.5 3.24568173905959
0.6 2.91692886131045
0.7 2.80527467751313
0.8 3.17881398964914
};
\end{axis}

\end{tikzpicture}

%% file: dynamics_energy.tex
\begin{tikzpicture}

\definecolor{darkgray176}{RGB}{176,176,176}
\definecolor{darkorange25512714}{RGB}{255,127,14}
\definecolor{forestgreen4416044}{RGB}{44,160,44}
\definecolor{lightgray204}{RGB}{204,204,204}
\definecolor{steelblue31119180}{RGB}{31,119,180}

\begin{axis}[
width=1\textwidth,
legend cell align={left},
legend style={
  fill opacity=0.8,
  draw opacity=1,
  text opacity=1,
  at={(0.97,0.03)},
  anchor=south east,
  legend columns=3,
  draw=lightgray204
},
tick align=outside,
tick pos=left,
x grid style={darkgray176},
xlabel={Time Steps},
xmin=-8.8, xmax=184.8,
xtick style={color=black},
y grid style={darkgray176},
ylabel={Potential Energy (\(\displaystyle \si{\hartree}\))},
ymin=-152.45193741084, ymax=-152.39378589016,
ytick style={color=black},
yticklabel style={/pgf/number format/precision=2,
/pgf/number format/fixed zerofill},
]
\addplot [semithick, steelblue31119180]
table {%
0 -152.3967150405
1 -152.3971834461
2 -152.398380763
3 -152.4001711404
4 -152.4026113677
5 -152.4054909322
6 -152.4087766073
7 -152.4123016537
8 -152.4160320279
9 -152.4197735188
10 -152.423463715
11 -152.4269984182
12 -152.4303659792
13 -152.4333533961
14 -152.4360344744
15 -152.4385516516
16 -152.4408714132
17 -152.4428122523
18 -152.4444711626
19 -152.4458044687
20 -152.4468925575
21 -152.4476053825
22 -152.4482903233
23 -152.4487729863
24 -152.4491115478
25 -152.4492700978
26 -152.4492941599
27 -152.4492272077
28 -152.4490614815
29 -152.4488324827
30 -152.448534127
31 -152.4481764763
32 -152.4478255017
33 -152.447520588
34 -152.4471935763
35 -152.4468725113
36 -152.4465609569
37 -152.4461948774
38 -152.4458409355
39 -152.4453676542
40 -152.4451014509
41 -152.4449414091
42 -152.444919449
43 -152.4449145894
44 -152.4449622254
45 -152.4449630073
46 -152.4450456721
47 -152.4451069315
48 -152.4452548944
49 -152.4453545633
50 -152.4454919546
51 -152.4455749454
52 -152.4456274758
53 -152.4456174482
54 -152.4456011992
55 -152.4454827435
56 -152.4452558639
57 -152.444868142
58 -152.4443988464
59 -152.4437155132
60 -152.4428449732
61 -152.4417640379
62 -152.4404644144
63 -152.4388631839
64 -152.4368466466
65 -152.4347952936
66 -152.4325641962
67 -152.4300748703
68 -152.4274792973
69 -152.424635395
70 -152.4216734392
71 -152.4186754478
72 -152.4158875131
73 -152.4132375419
74 -152.4109380046
75 -152.4088800959
76 -152.4072538699
77 -152.4060959922
78 -152.4054860493
79 -152.4054538102
80 -152.4059795347
81 -152.4070467314
82 -152.4086404671
83 -152.4106815703
84 -152.4131028008
85 -152.41581217
86 -152.4187422336
87 -152.4217494945
88 -152.4247832809
89 -152.4277445836
90 -152.4306290064
91 -152.4333214116
92 -152.4356023502
93 -152.4378680487
94 -152.4399756497
95 -152.4417289162
96 -152.4432728537
97 -152.4444790488
98 -152.4455100466
99 -152.4463251919
100 -152.446975829
101 -152.4474073408
102 -152.4477402398
103 -152.4479324818
104 -152.4480270402
105 -152.4480245007
106 -152.4479995004
107 -152.4478622714
108 -152.447712565
109 -152.4474879552
110 -152.4472876076
111 -152.4470157446
112 -152.4468069516
113 -152.4465606561
114 -152.4463255148
115 -152.4460446829
116 -152.4457298582
117 -152.4453696863
118 -152.445203776
119 -152.4449553315
120 -152.4445999273
121 -152.4440842043
122 -152.4437178884
123 -152.4434668099
124 -152.4433024647
125 -152.4430329153
126 -152.4429077507
127 -152.4425138148
128 -152.4423929392
129 -152.4421571229
130 -152.4419708386
131 -152.4417502725
132 -152.4415508808
133 -152.4413241255
134 -152.4410894173
135 -152.4408630942
136 -152.4407600671
137 -152.4405484043
138 -152.4403415758
139 -152.440077094
140 -152.4398028081
141 -152.4394613202
142 -152.4391187043
143 -152.4386869293
144 -152.4382147453
145 -152.4376488154
146 -152.4369380445
147 -152.4361112011
148 -152.435134327
149 -152.4340050313
150 -152.4327022213
151 -152.4312662954
152 -152.4297889323
153 -152.4281090224
154 -152.4263460997
155 -152.4244067094
156 -152.4224684997
157 -152.4205464311
158 -152.4187146952
159 -152.4169402279
160 -152.4153791957
161 -152.4139885295
162 -152.4129301551
163 -152.4121824467
164 -152.4118001024
165 -152.4118222768
166 -152.4122753118
167 -152.4131613018
168 -152.4144694807
169 -152.4161507901
170 -152.4181649708
171 -152.4204168019
172 -152.422891799
173 -152.4254959657
174 -152.4282166141
175 -152.4309284363
176 -152.4335612788
};
\addlegendentry{$\text{E}_{\text{potential}}$}
\addplot [semithick, darkorange25512714]
table {%
0 -152.3967150405
1 -152.3968126353
2 -152.3969145558
3 -152.3969343606
4 -152.3970062306
5 -152.3970182731
6 -152.3970498616
7 -152.3970519098
8 -152.3971053829
9 -152.3971235381
10 -152.3971396753
11 -152.3971300447
12 -152.3971478686
13 -152.3970300883
14 -152.3968853667
15 -152.3968750836
16 -152.3969713798
17 -152.39699116
18 -152.3970234886
19 -152.3970097971
20 -152.3970115921
21 -152.3968772407
22 -152.3969297246
23 -152.3969690726
24 -152.3970287315
25 -152.397049587
26 -152.3970553413
27 -152.3970687316
28 -152.3970629529
29 -152.3970559772
30 -152.3970262123
31 -152.3969700413
32 -152.3969405838
33 -152.3969654596
34 -152.3969666689
35 -152.396964285
36 -152.3969548514
37 -152.3968685469
38 -152.3967672725
39 -152.3965157897
40 -152.3964375406
41 -152.3964291411
42 -152.396521295
43 -152.3965932156
44 -152.3966821
45 -152.396691492
46 -152.3967543405
47 -152.3967726957
48 -152.3968615074
49 -152.3968943371
50 -152.3969673463
51 -152.3970002396
52 -152.3970307073
53 -152.397042472
54 -152.3971088964
55 -152.3971528621
56 -152.3971887771
57 -152.3971855599
58 -152.3972450338
59 -152.3972571692
60 -152.3972701773
61 -152.3972809863
62 -152.3972985794
63 -152.3972525077
64 -152.397034083
65 -152.3970234181
66 -152.3970678701
67 -152.3970680308
68 -152.3971424727
69 -152.3971035297
70 -152.3970226544
71 -152.3969116047
72 -152.3969367691
73 -152.396939519
74 -152.3970429844
75 -152.3970501534
76 -152.3970695979
77 -152.397069276
78 -152.3970775325
79 -152.3970947455
80 -152.3970957351
81 -152.3970827003
82 -152.3970827207
83 -152.3970784634
84 -152.3970794287
85 -152.3970800041
86 -152.3971031879
87 -152.3970952479
88 -152.3970895718
89 -152.3970624355
90 -152.3970746431
91 -152.3970635263
92 -152.3968491796
93 -152.3968554694
94 -152.3969550275
95 -152.3969571112
96 -152.397003174
97 -152.3969549656
98 -152.3969604714
99 -152.3969600438
100 -152.3969839308
101 -152.3969551781
102 -152.3969722797
103 -152.3969718212
104 -152.396976293
105 -152.3969680632
106 -152.397004921
107 -152.3969828545
108 -152.3969899688
109 -152.3969539409
110 -152.3969659244
111 -152.3969235891
112 -152.3969564479
113 -152.3969599777
114 -152.3969798734
115 -152.3969576147
116 -152.3969046323
117 -152.3968091667
118 -152.3969095387
119 -152.3969284678
120 -152.3968424592
121 -152.3965988486
122 -152.3965066682
123 -152.3965301291
124 -152.3966392576
125 -152.3966413761
126 -152.396785419
127 -152.3966578995
128 -152.3967997716
129 -152.3968214778
130 -152.3968870945
131 -152.396912909
132 -152.3969550622
133 -152.3969661293
134 -152.3969670894
135 -152.3969765321
136 -152.3971130631
137 -152.397149566
138 -152.3972049423
139 -152.3972235868
140 -152.3972621538
141 -152.3972730227
142 -152.3973326482
143 -152.3973646266
144 -152.3974305847
145 -152.3974913661
146 -152.3975092773
147 -152.3975254906
148 -152.3975177873
149 -152.3974931155
150 -152.3974367272
151 -152.3973909201
152 -152.3974448778
153 -152.3974275397
154 -152.3974397226
155 -152.3973608446
156 -152.3973329285
157 -152.3973277719
158 -152.3973687957
159 -152.3973665641
160 -152.3974173145
161 -152.3974171367
162 -152.3974695745
163 -152.3975001403
164 -152.3975173269
165 -152.397524595
166 -152.397527978
167 -152.3975274209
168 -152.3975289455
169 -152.3975180613
170 -152.3975044506
171 -152.3974541865
172 -152.3974223287
173 -152.3973885896
174 -152.3974137485
175 -152.3974430743
176 -152.3974700333
};
\addlegendentry{$\text{E}_{\text{total}}$}
\addlegendimage{/pgfplots/refstyle=Ekin_plot}
\addlegendentry{$\text{E}_{\text{kinetic}}$}
\end{axis}

\begin{axis}[
width=1\textwidth,
axis y line*=right,
tick align=outside,
x grid style={darkgray176},
xmin=-8.8, xmax=184.8,
xtick pos=left,
xtick style={color=black},
y grid style={darkgray176},
ylabel={Kinetic Energy (\(\displaystyle \si{\hartree}\))},
ymin=-0.002611940928153, ymax=0.056,
ytick pos=right,
ytick style={color=black},
yticklabel style={anchor=west}
]
\addplot [semithick, forestgreen4416044]
table {%
0 0
1 0.0003708107955039
2 0.001466207185823
3 0.003236779746829
4 0.005605137088299
5 0.008472659067019
6 0.01172674566569
7 0.01524974389066
8 0.01892664500179
9 0.02264998070728
10 0.02632403966471
11 0.02986837358896
12 0.0332181105619
13 0.03632330779184
14 0.03914910766536
15 0.04167656795774
16 0.04390003338558
17 0.04582109238299
18 0.0474476739436
19 0.04879467161041
20 0.04988096540323
21 0.05072814181945
22 0.05136059867985
23 0.05180391365148
24 0.05208281628379
25 0.05222051075949
26 0.05223881856306
27 0.05215847605259
28 0.0519985285673
29 0.05177650549759
30 0.05150791479073
31 0.05120643505311
32 0.05088491790903
33 0.05055512836318
34 0.05022690739659
35 0.04990822638801
36 0.04960610552035
37 0.04932633048878
38 0.04907366296793
39 0.04885186451696
40 0.04866391032808
41 0.04851226796801
42 0.04839815403472
43 0.04832137380807
44 0.04828012549469
45 0.04827151528301
46 0.04829133158087
47 0.0483342358384
48 0.04839338701259
49 0.04846022617267
50 0.04852460829183
51 0.04857470577358
52 0.04859676852293
53 0.04857497624709
54 0.0484923027415
55 0.04832988137187
56 0.04806708681485
57 0.04768258204084
58 0.04715381257188
59 0.04645834397242
60 0.0455747958993
61 0.04448305158236
62 0.04316583499677
63 0.04161067613004
64 0.0398125635403
65 0.03777187555206
66 0.03549632607004
67 0.03300683950491
68 0.0303368246613
69 0.02753186527515
70 0.02465078477915
71 0.02176384304263
72 0.01895074396816
73 0.01629802288076
74 0.01389502022382
75 0.01182994247177
76 0.01018427196754
77 0.009026716194513
78 0.008408516877874
79 0.008359064686813
80 0.008883799664623
81 0.009964031107878
82 0.01155774645418
83 0.01360310690328
84 0.01602337213004
85 0.01873216592333
86 0.02163904569952
87 0.02465424658996
88 0.02769370904894
89 0.03068214811586
90 0.03355436335376
91 0.03625788530373
92 0.03875317059436
93 0.04101257928808
94 0.04302062214889
95 0.04477180503516
96 0.04626967969123
97 0.04752408318468
98 0.04854957514478
99 0.04936514812953
100 0.0499918982502
101 0.05045216275607
102 0.05076796005161
103 0.05096066066017
104 0.05105074719803
105 0.05105643751307
106 0.05099457936789
107 0.05087941682696
108 0.05072259622841
109 0.05053401424981
110 0.05032168323303
111 0.0500921555165
112 0.0498505036821
113 0.04960067836592
114 0.04934564144267
115 0.04908706826588
116 0.04882522584106
117 0.04856051955381
118 0.04829423724728
119 0.04802686372518
120 0.04775746808144
121 0.0474853556483
122 0.04721122022299
123 0.04693668075654
124 0.0466632071421
125 0.04639153927456
126 0.04612233173613
127 0.04585591523295
128 0.04559316761748
129 0.04533564510323
130 0.04508374408788
131 0.04483736347464
132 0.04459581862727
133 0.04435799623719
134 0.04412232790697
135 0.043886562103
136 0.04364700398879
137 0.04339883837247
138 0.04313663349101
139 0.04285350721008
140 0.04254065428594
141 0.04218829758001
142 0.04178605613017
143 0.04132230265594
144 0.04078416056344
145 0.040157449344
146 0.03942876717964
147 0.03858571049277
148 0.03761653967181
149 0.03651191586558
150 0.03526549418172
151 0.03387537527873
152 0.03234405448931
153 0.03068148269669
154 0.02890637713405
155 0.02704586482162
156 0.02513557114565
157 0.02321865925568
158 0.0213458994176
159 0.01957366381654
160 0.01796188119209
161 0.01657139279942
162 0.01546058064749
163 0.01468230636628
164 0.01428277548966
165 0.0142976818495
166 0.01474733384476
167 0.0156338808523
168 0.01694053514554
169 0.01863272882224
170 0.02066052017455
171 0.02296261537726
172 0.02546947031568
173 0.02810737606229
174 0.03080286562014
175 0.03348536195637
176 0.03609124545364
};
\label{Ekin_plot}
\end{axis}

\end{tikzpicture}

%% file: dynamics_delta.tex
\begin{tikzpicture}

\definecolor{darkgray176}{RGB}{176,176,176}
\definecolor{steelblue31119180}{RGB}{31,119,180}

\begin{axis}[
width=1\textwidth,
tick align=outside,
tick pos=left,
x grid style={darkgray176},
xlabel={Time Steps},
xmin=-8.8, xmax=184.8,
xtick style={color=black},
y grid style={darkgray176},
ylabel={Delta  (\(\displaystyle \si{\angstrom}\))},
ymin=-1.09814854258187, ymax=1.07557355257506,
ytick style={color=black},
yticklabel style={/pgf/number format/precision=2,
/pgf/number format/fixed zerofill},
]
\addplot [semithick, steelblue31119180]
table {%
0 -0.999342992802014
1 -0.998002111798883
2 -0.99399582098898
3 -0.987370285639644
4 -0.978201123942425
5 -0.966589014711646
6 -0.952656897916636
7 -0.936545203960371
8 -0.918406931527599
9 -0.898402790849261
10 -0.876698126473182
11 -0.853458702468514
12 -0.82884776289201
13 -0.803023802325932
14 -0.776138913174844
15 -0.748336833202501
16 -0.719750958418302
17 -0.690503725237568
18 -0.660707866413916
19 -0.630465559044052
20 -0.599868552450784
21 -0.568998514930538
22 -0.537927294915852
23 -0.506716830435694
24 -0.475420163340809
25 -0.444081924889286
26 -0.412739299183008
27 -0.381422335856927
28 -0.350154797327297
29 -0.318954537560936
30 -0.287834229186691
31 -0.256802016310701
32 -0.225861966165043
33 -0.19501436481194
34 -0.164256437568832
35 -0.133582682452853
36 -0.102985595633895
37 -0.0724558484347042
38 -0.0419828982112256
39 -0.0115551017640645
40 0.0188396215919104
41 0.0492136246454469
42 0.0795790324125858
43 0.109947392538629
44 0.140329249683284
45 0.170733648838363
46 0.201167934035127
47 0.231637009952636
48 0.262143394989114
49 0.292686245098354
50 0.3232613222934
51 0.353860189062882
52 0.384469970811684
53 0.415072470451578
54 0.445643815553444
55 0.476154071875009
56 0.506566390727862
57 0.536836536215531
58 0.566912543056071
59 0.596733597771234
60 0.626230408471887
61 0.65532426461578
62 0.683927011970151
63 0.711940689815479
64 0.739258186938999
65 0.765763941196094
66 0.791332942725161
67 0.815832271136836
68 0.839123369666037
69 0.86106335074881
70 0.881507024954516
71 0.900309688733787
72 0.917329398262828
73 0.932430764400361
74 0.945487638655775
75 0.95638666828278
76 0.965030899603022
77 0.971342349325161
78 0.975265338301729
79 0.976768002795204
80 0.975843681621438
81 0.972511101880761
82 0.96681362148478
83 0.958817904863711
84 0.948611245815159
85 0.936298936751758
86 0.922001032027309
87 0.905848783938393
88 0.887981868478521
89 0.868544472887182
90 0.847682923850215
91 0.82554325392615
92 0.80226850223833
93 0.7779978115075
94 0.752863605093225
95 0.726990585030983
96 0.700495620239902
97 0.673486238205171
98 0.646061541967424
99 0.618311063041396
100 0.590315317642127
101 0.562145715669417
102 0.533865003303672
103 0.505527878278991
104 0.477181328293787
105 0.448864777804591
106 0.420611336437391
107 0.392447272597696
108 0.364393778455775
109 0.336466803925681
110 0.308677909574999
111 0.281034242325702
112 0.253539414225145
113 0.226193764631281
114 0.198994754011259
115 0.171937158560607
116 0.145013797579808
117 0.118215885446749
118 0.0915327734897149
119 0.0649525875511114
120 0.0384626024458024
121 0.0120494317784412
122 -0.0143008825297166
123 -0.0406027765321157
124 -0.0668705577759348
125 -0.0931184320403697
126 -0.119360148539206
127 -0.145608638788599
128 -0.17187606837302
129 -0.198173442909838
130 -0.224510668422292
131 -0.250895958285467
132 -0.27733557400151
133 -0.303833496931291
134 -0.330391277041673
135 -0.357007766655768
136 -0.383678775203616
137 -0.41039609501683
138 -0.437148078737295
139 -0.463918954478481
140 -0.490688414506123
141 -0.517430777839682
142 -0.544115131643521
143 -0.570704761330347
144 -0.597156738558137
145 -0.623421242131753
146 -0.649440977618949
147 -0.675151611647479
148 -0.700481010721629
149 -0.725349208245517
150 -0.749668371934274
151 -0.773342801155641
152 -0.796269566968459
153 -0.818338066709892
154 -0.839432591684982
155 -0.859432665065297
156 -0.878215773788744
157 -0.895658238340732
158 -0.911637768164012
159 -0.926035973856763
160 -0.938740599675438
161 -0.949648104153937
162 -0.958665973360511
163 -0.965713965299131
164 -0.970725170565248
165 -0.97364950632104
166 -0.974454442584183
167 -0.973127515828204
168 -0.969677387490208
169 -0.964133882304166
170 -0.956547022168502
171 -0.9469855424145
172 -0.935534136050724
173 -0.92229190185463
174 -0.90736895502067
175 -0.890883454293788
176 -0.872959894182556
};
\addplot[dashed,black,thin,forget plot] coordinates{
(0,0)
(180,0)
};

\end{axis}

\end{tikzpicture}

%% file: total_energy_compare3.tex
\begin{tikzpicture}

\definecolor{chocolate213940}{RGB}{213,94,0}
\definecolor{darkcyan0158115}{RGB}{0,158,115}
\definecolor{darkgray176}{RGB}{176,176,176}
\definecolor{lightgray204}{RGB}{204,204,204}

\begin{axis}[
width=\textwidth,
legend cell align={left},
legend columns=3,
legend style={
  fill opacity=0.8,
  draw opacity=1,
  text opacity=1,
  at={(1.2,1.15)},
  anchor=south,
  draw=lightgray204,
    /tikz/column sep=5pt,
},
log basis y={10},
tick align=outside,
tick pos=left,
x grid style={darkgray176},
xlabel={Time Steps},
xmin=-8.8, xmax=184.8,
xtick style={color=black},
y grid style={darkgray176},
ylabel={Absolute Energy Error (\(\displaystyle \si{\milli\hartree}\))},
ymin=0.000139548416815224, ymax=2.82194128377656,
ymode=log,
ytick style={color=black}
]
\addplot [thick, chocolate213940]
table {%
0 1.46831930001667
1 1.46698890000607
2 1.463409500019
3 1.45761800001765
4 1.44943589998547
5 1.43596710000793
6 1.41769320001117
7 1.4034181999989
8 1.41820599998255
9 1.43935359997727
10 1.45464400000606
11 1.45810149999193
12 1.44092470000601
13 1.36948239997992
14 1.41490870001348
15 1.43983699999239
16 1.45431449999478
17 1.46329209999863
18 1.45747260000917
19 1.44913229999588
20 1.40364820001082
21 1.37947069998745
22 1.41898699999388
23 1.42410419999806
24 1.41856839999832
25 1.40649150000627
26 1.38565130001211
27 1.35371120001082
28 1.31297609999592
29 1.26048820001756
30 1.19753359999208
31 1.13835080000513
32 1.10282009998741
33 1.08732670000222
34 1.07056949997286
35 1.03512930002125
36 0.965544899997894
37 0.83558260001837
38 0.597500800012085
39 0.32148010001265
40 0.202111899994861
41 0.346218799990083
42 0.560041299991099
43 0.705908400021826
44 0.795011100024112
45 0.869885599996678
46 0.95678130000465
47 1.05054369998925
48 1.14117260000057
49 1.22347909999121
50 1.29359870001622
51 1.34729830000424
52 1.39246649999336
53 1.42897399999242
54 1.45660219999399
55 1.47958670001458
56 1.49915970001757
57 1.51769409998792
58 1.53260630000318
59 1.54308090000654
60 1.54916740001454
61 1.54854369998247
62 1.53589799998599
63 1.49829250000266
64 1.43855420000705
65 1.52177700002198
66 1.52767439999479
67 1.51286559997743
68 1.48623249998536
69 1.44537560001368
70 1.3427993999926
71 1.40112750000299
72 1.394076299988
73 1.4043522999998
74 1.41309209999463
75 1.4171814000008
76 1.41703720001374
77 1.4123552999763
78 1.40445819999968
79 1.39358659998834
80 1.37577419999957
81 1.35396440001045
82 1.32950040000424
83 1.31555890001778
84 1.32562030000827
85 1.33863459998906
86 1.35233820000735
87 1.3586885000052
88 1.35510760000557
89 1.36014280002428
90 1.38555769999016
91 1.40339599997219
92 1.40676039998766
93 1.40881289999584
94 1.39938809999762
95 1.3770265999824
96 1.3439685000094
97 1.3069195999833
98 1.2774587000024
99 1.26185860000305
100 1.25044000000685
101 1.24195880002276
102 1.23968950001085
103 1.23930609998979
104 1.23791949999941
105 1.23331509999502
106 1.2200982999957
107 1.19517340002062
108 1.16134670000179
109 1.12118199999145
110 1.09447689999342
111 1.11008429999515
112 1.1491293999768
113 1.17744600001402
114 1.1833075999732
115 1.15413080001758
116 1.07562209998946
117 1.05902079999964
118 1.03347750001603
119 0.949192599989601
120 0.769289000004392
121 0.621665499977553
122 0.653638500011766
123 0.807420300020567
124 0.931830399991895
125 0.983297400011907
126 0.893840800017642
127 0.921759799979327
128 1.09116409998933
129 1.14786590000904
130 1.17704429999321
131 1.22970819998613
132 1.3036745000079
133 1.37485319999087
134 1.43287340000597
135 1.47805509999444
136 1.51476930000172
137 1.5453821000051
138 1.57120110000619
139 1.5949556999999
140 1.61502030002225
141 1.63384180001458
142 1.64965740000866
143 1.66554070000302
144 1.68003839999642
145 1.69122980000225
146 1.70049890002133
147 1.70548680000593
148 1.70237260002182
149 1.68788670001163
150 1.67387139998709
151 1.69953479999663
152 1.72101580000117
153 1.7225460999839
154 1.70899969998572
155 1.69110799998862
156 1.70418100000802
157 1.73635830000762
158 1.76169229999346
159 1.77809460001299
160 1.7891543999724
161 1.79542230000607
162 1.7981617000089
163 1.79782890000979
164 1.79521900000168
165 1.79039489998445
166 1.78363690000083
167 1.77383530001407
168 1.76063349999822
169 1.74296410000352
170 1.71872859999667
171 1.68819059999237
172 1.66203670002574
173 1.64552190000222
174 1.70837460001394
175 1.75339230000304
176 1.76250239999831
};
\addlegendentry{CCSD}
\addplot [thick, darkcyan0158115]
table {%
0 0.497341100015092
1 0.399474800019561
2 0.297209899997597
3 0.276638200006118
4 0.202630199993337
5 0.184127599993644
6 0.141493499995704
7 0.132116599985466
8 0.0998077999838642
9 0.108271699986062
10 0.11187079999786
11 0.127788599996848
12 0.095098200006305
13 0.142763199988849
14 0.333309900014456
15 0.36824859998319
16 0.285566600012999
17 0.273494800012486
18 0.2334671999904
19 0.236840599995958
20 0.187465200014003
21 0.295502599982456
22 0.280383800003392
23 0.244087499993384
24 0.177018500011172
25 0.142199999999093
26 0.113836800011313
27 0.0669103000063842
28 0.0304973999902813
29 0.0163066000027356
30 0.0506906000055096
31 0.054760300002954
32 0.0618400999883306
33 0.102983499999709
34 0.121732500019789
35 0.155332499986116
36 0.215979800003652
37 0.259962699999505
38 0.397167199992055
39 0.421930199991039
40 0.46324940001341
41 0.310945199998969
42 0.189024100023971
43 0.114944199992806
44 0.114358299981632
45 0.0484962000086853
46 0.0237671999911981
47 0.0522954999837566
48 0.0548366000145961
49 0.105141099993489
50 0.103258300015341
51 0.125167499987811
52 0.141118000016149
53 0.167258099992296
54 0.13000350000425
55 0.110715200008826
56 0.0961348999908296
57 0.119800999982544
58 0.0771983999925396
59 0.0775144999920485
60 0.0725260000251637
61 0.0629086000003554
62 0.0343394999902102
63 0.0440272000048481
64 0.203472200013266
65 0.297668100017745
66 0.258748800007425
67 0.242586199988182
68 0.139504599985685
69 0.134739300023057
70 0.109357000013688
71 0.274192200009793
72 0.237067699998761
73 0.239159599999539
74 0.139013699993029
75 0.130751300019938
76 0.106596499989564
77 0.0985796999941613
78 0.0800678000132393
79 0.0511101999904895
80 0.0330113999780224
81 0.0264364000202022
82 0.00546179998650587
83 0.000219000014567428
84 0.0143945999866446
85 0.032390400008353
86 0.0283615000000736
87 0.0477884999838807
88 0.0545183000042471
89 0.0905157000090639
90 0.106764699978612
91 0.137938999984044
92 0.357016300000623
93 0.353456999988566
94 0.244432799973993
95 0.219364199978145
96 0.139147300018294
97 0.148884799983762
98 0.112172000001465
99 0.0951825000186091
100 0.0579749000166885
101 0.0762591000125212
102 0.0549847000002046
103 0.0533861999940655
104 0.0459203999980673
105 0.0481374999878881
106 0.00322349998782556
107 0.00722929999597
108 0.0491142999976546
109 0.0540405000037936
110 0.0934162000021388
111 0.0360546000024442
112 0.0302595000221118
113 0.00581489999262885
114 0.0201166000124431
115 0.0272423999945204
116 0.0529284999970514
117 0.0258390000169584
118 0.100189099981662
119 0.203485399993042
120 0.297588600005838
121 0.201548600017531
122 0.0775203999978658
123 0.0527889000068171
124 0.0680333000104838
125 0.117339799999172
126 0.116181899983303
127 0.0391784999749234
128 0.0668628999846987
129 0.101782400008688
130 0.0653860000170425
131 0.0922482999783369
132 0.124320500020758
133 0.184712099979834
134 0.242132900012848
135 0.278371899980812
136 0.179074200019613
137 0.173729500005493
138 0.144807300017646
139 0.150656999977627
140 0.133004300010953
141 0.141916399996944
142 0.0991711999915879
143 0.0842711999951007
144 0.0340036999944004
145 0.0142860000096334
146 0.0216123999905449
147 0.0315432999968834
148 0.0257349999799317
149 0.0144684999838773
150 0.0287777000096412
151 0.100795599990988
152 0.0684052999986307
153 0.0868999999852349
154 0.0602625999874817
155 0.119751099987297
156 0.158788800007414
157 0.193185600011248
158 0.174451699990641
159 0.189564000010023
160 0.146160599996392
161 0.148883299999625
162 0.0955813000018679
163 0.0613185000020167
164 0.0387718000069981
165 0.0249336000024414
166 0.0142039000081695
167 0.00564960001270265
168 0.00712689998749738
169 0.0110240999902089
170 0.0179541999898447
171 0.00598139999397063
172 0.0160076000099707
173 0.0377926999988176
174 0.0798041999985344
175 0.0995352999950683
176 0.0851053000019419
};
\addlegendentry{SqDRIFT}
\addplot [line width=0.48pt, black, dashed, forget plot]
table {%
-8.79999999999999 1.6
184.8 1.6
};
\end{axis}

\end{tikzpicture}

%% file: potential_compare3.tex
\begin{tikzpicture}

\definecolor{chocolate213940}{RGB}{213,94,0}
\definecolor{darkcyan0158115}{RGB}{0,158,115}
\definecolor{darkgray176}{RGB}{176,176,176}

\begin{axis}[
width=\textwidth,
legend cell align={left},
legend style={
  fill opacity=0.8,
  draw opacity=1,
  text opacity=1,
  at={(0.97,0.03)},
  anchor=south east,
  draw=none
},
log basis y={10},
tick align=outside,
tick pos=left,
x grid style={darkgray176},
xlabel={Time Steps},
xmin=-8.8, xmax=184.8,
xtick style={color=black},
y grid style={darkgray176},
ylabel={Absolute Energy Error (\(\displaystyle \si{\milli\hartree}\))},
ymin=0.00180391148196563, ymax=3.40281821481422,
ymode=log,
ytick style={color=black}
]
\addplot [thick, chocolate213940]
table {%
0 1.46831930001667
1 1.46860629999424
2 1.46980300002042
3 1.47173060000227
4 1.47389640000029
5 1.4730077000138
6 1.46914230001016
7 1.47080460001803
8 1.50292800000784
9 1.54271220000624
10 1.57774959998846
11 1.601793399999
12 1.60560559999112
13 1.55473249998295
14 1.62055219999502
15 1.66650549999758
16 1.70298560001925
17 1.7350914000076
18 1.75296789998924
19 1.76816889998577
20 1.74503760001699
21 1.74238900001455
22 1.8040962999919
23 1.83208519999312
24 1.84944430000655
25 1.85998930001574
26 1.86129829998549
27 1.85080890000222
28 1.83052199997746
29 1.79716700000654
30 1.75179519999347
31 1.7086006999989
32 1.68768929998464
33 1.68574730000159
34 1.68152669999699
35 1.65751270000669
36 1.59810010001138
37 1.47678370001358
38 1.2453459000028
39 0.973320199989303
40 0.855424400015181
41 1.00000060001548
42 1.21451749998869
43 1.36104619997468
44 1.44998700000087
45 1.52327730000934
46 1.6068487999803
47 1.69529130002388
48 1.77824639999358
49 1.85008540000808
50 1.90631399999575
51 1.9419861000074
52 1.96424570000886
53 1.97222890000148
54 1.96510890000923
55 1.94658079999499
56 1.91737130001002
57 1.87947559999202
58 1.83001400000649
59 1.76805899999977
60 1.69381210000097
61 1.60545500000353
62 1.49877869998249
63 1.36262680001664
64 1.20220990001485
65 1.18378420000909
66 1.08943309999177
67 0.980453500005751
68 0.870050499997888
69 0.760144799983209
70 0.607736900008149
71 0.638634500006674
72 0.629185900010043
73 0.66458310001849
74 0.726855999999998
75 0.812661300017226
76 0.920625899993865
77 1.04699459998869
78 1.18818670000564
79 1.33831060000489
80 1.48648819998698
81 1.62851139998565
82 1.75887680001097
83 1.88470060001578
84 2.01450899999145
85 2.12399680000885
86 2.20950579998203
87 2.26317519999952
88 2.2838671000045
89 2.29226980002295
90 2.30280990001575
91 2.29083069999092
92 2.25295719999963
93 2.20590189999825
94 2.14279200000078
95 2.06492579999917
96 1.97698089999676
97 1.88764410000886
98 1.80988899998624
99 1.75092619997486
100 1.70175450000443
101 1.6615660000241
102 1.63375440001801
103 1.61372169998231
104 1.59807699998282
105 1.58401939998498
106 1.56539299999281
107 1.53824830002236
108 1.50454229998331
109 1.46598680001375
110 1.44185389999052
111 1.46097550000945
112 1.5047081000148
113 1.53886409998449
114 1.55125370000064
115 1.52860039997904
116 1.45558350001807
117 1.44327029997271
118 1.42137770001227
119 1.33977999999502
120 1.16052679999257
121 1.01073350001002
122 1.03882080000517
123 1.18915370001105
124 1.31143920000909
125 1.36161410000568
126 1.2703507999845
127 1.29594739999561
128 1.46418309998353
129 1.52086779999649
130 1.55017730000395
131 1.6026963999991
132 1.67575570000622
133 1.74460979999935
134 1.79815420000295
135 1.83603710001989
136 1.86207519999471
137 1.87811870000587
138 1.88495190002413
139 1.88475179999159
140 1.87533720000488
141 1.85866580000038
142 1.83253199998035
143 1.79960890000075
144 1.75815840000837
145 1.70616549999636
146 1.64513629999874
147 1.57306020000192
148 1.48691280000435
149 1.38470680002456
150 1.27991069999212
151 1.21360090000167
152 1.14437790000466
153 1.05983479997462
154 0.968704400008846
155 0.885625800009393
156 0.849559199991745
157 0.851838000016869
158 0.86944040000958
159 0.902773199982221
160 0.956871900001488
161 1.03243079999515
162 1.12955450001095
163 1.24628070000199
164 1.37968530000876
165 1.52464559999999
166 1.67482939997399
167 1.82174910000299
168 1.95780769999487
169 2.07555719998709
170 2.1677734999912
171 2.23096069998974
172 2.2734435000018
173 2.29952800000888
174 2.3786868000002
175 2.41516479999859
176 2.39439110001172
};
\addplot [thick, darkcyan0158115]
table {%
0 0.497341100015092
1 0.399271899993892
2 0.296398100005035
3 0.274690299988833
4 0.19887280001285
5 0.177628300008337
6 0.131658899988452
7 0.118543799999316
8 0.0818391000052543
9 0.0854681999840068
10 0.0843983999914144
11 0.0958701000115525
12 0.0591371999973944
13 0.103814400006286
14 0.29394749998346
15 0.330656599999202
16 0.249461700008169
17 0.238439100002097
18 0.199643200005539
19 0.203908599985425
20 0.154577100005326
21 0.262837900010027
22 0.249074300000984
23 0.214183599979378
24 0.147489800014
25 0.112068300012425
26 0.0824835999821971
27 0.0336875000073178
28 0.00526880000961683
29 0.0554929999907472
30 0.0939677999838295
31 0.10222570000451
32 0.113501200019073
33 0.159089100009169
34 0.182497500020418
35 0.220521599999302
36 0.285224699979381
37 0.332802199977777
38 0.473318600000994
39 0.501509200006467
40 0.546237500003599
41 0.396315800003322
42 0.27482350000696
43 0.19944290002627
44 0.196104799982777
45 0.126225099990052
46 0.0960704000192436
47 0.013199799980157
48 0.00254159999713011
49 0.0572114000192414
50 0.0658997000186901
51 0.0991915999861703
52 0.127160499999945
53 0.166078199981712
54 0.142053499985195
55 0.136363499990466
56 0.135927499997024
57 0.174020000002884
58 0.146260500002882
59 0.161671299991895
60 0.171431300003633
61 0.175999000020965
62 0.161044099996843
63 0.18347380000705
64 0.353040100009139
65 0.453474200014625
66 0.418322699999862
67 0.404358100013269
68 0.301639899987549
69 0.295453500001486
70 0.266520100012713
71 0.424621899981048
72 0.37702400001649
73 0.365307000009807
74 0.248399699984247
75 0.220712500009768
76 0.174666800006662
77 0.142912400008299
78 0.0995783999826472
79 0.0458889000185536
80 0.00434659997949893
81 0.0235825999936878
82 0.0628502000097342
83 0.0825074999966091
84 0.0786251000022276
85 0.0666660999968371
86 0.0723640000046544
87 0.0499200000092515
88 0.0357147999920926
89 0.0114736000114135
90 0.0424897999948826
91 0.0915399000120942
92 0.330876399999624
93 0.349706799994465
94 0.264203800014684
95 0.262787899998784
96 0.205079300002353
97 0.235658400015382
98 0.218149700003778
99 0.21814069998527
100 0.195550200004391
101 0.225985700012643
102 0.214722900011566
103 0.221129099998052
104 0.219323899983692
105 0.225090399993633
106 0.174673000003622
107 0.168935600015629
108 0.123048000006065
109 0.111923500014655
110 0.0644904999944629
111 0.112462000004143
112 0.108228700014479
113 0.122228300000415
114 0.0967201000037221
115 0.0774468999793498
116 0.038883800016265
117 0.104479099974242
118 0.0350553999908243
119 0.15250690000812
120 0.261133400016433
121 0.17962969999985
122 0.0692139999785013
123 0.0495831000080216
124 0.0556139000025269
125 0.097681100015734
126 0.142103499996438
127 0.00868929998887324
128 0.0353257999847756
129 0.0720881999995981
130 0.0397506999831876
131 0.0729772000056528
132 0.113895699996647
133 0.185490800021171
134 0.256034200020849
135 0.306796399996756
136 0.223515199991198
137 0.235862800025188
138 0.225921700007348
139 0.251815199987959
140 0.255643299993835
141 0.28754760000993
142 0.268720399986933
143 0.278152600003523
144 0.252442499999006
145 0.229468800000632
146 0.247810699988804
147 0.262877199986633
148 0.292607000005773
149 0.326080800022055
150 0.389215700010936
151 0.477814199996374
152 0.458943599994655
153 0.487850499979459
154 0.467588300011812
155 0.528398400007291
156 0.562707400007412
157 0.586082799998167
158 0.549991599996247
159 0.54205519998618
160 0.470682299976488
161 0.441344800009347
162 0.353500899990422
163 0.285016800006588
164 0.230723800001442
165 0.189051200010226
166 0.155436000000009
167 0.129031500023302
168 0.102689399994915
169 0.0885901999936323
170 0.0744068999836145
171 0.0935493000042698
172 0.101223899974912
173 0.123267699990492
174 0.167199499998105
175 0.190409999987651
176 0.181195999999773
};
\addplot [line width=0.48pt, black, dashed, forget plot]
table {%
-8.79999999999999 1.6
184.8 1.6
};
\end{axis}

\end{tikzpicture}

%% file: kinetic_compare3.tex
\begin{tikzpicture}

\definecolor{chocolate213940}{RGB}{213,94,0}
\definecolor{darkcyan0158115}{RGB}{0,158,115}
\definecolor{darkgray176}{RGB}{176,176,176}

\begin{axis}[
width=\textwidth,
legend cell align={left},
legend style={
  fill opacity=0.8,
  draw opacity=1,
  text opacity=1,
  at={(0.97,0.03)},
  anchor=south east,
  draw=none
},
log basis y={10},
tick align=outside,
tick pos=left,
x grid style={darkgray176},
xlabel={Time Steps},
xmin=-8.8, xmax=184.8,
xtick style={color=black},
y grid style={darkgray176},
ylabel={Absolute Energy Error (\(\displaystyle \si{\milli\hartree}\))},
ymin=0.000129548535068621, ymax=2.50589138710133,
ymode=log,
ytick style={color=black}
]
\addplot [thick, chocolate213940]
table {%
0 0.00000001
1 0.00161737806139998
2 0.00639349194999999
3 0.0141126737600002
4 0.0244604225670001
5 0.0370404561940013
6 0.0514490580200005
7 0.0673864026499993
8 0.0847220511599987
9 0.103358568590003
10 0.123105692809999
11 0.143691962819999
12 0.164680931199999
13 0.185250155620001
14 0.205643492089999
15 0.226668441640002
16 0.248671197259999
17 0.271799396529999
18 0.295495246140004
19 0.319036579370005
20 0.341389300290006
21 0.362918384399995
22 0.385109241619996
23 0.407980929489994
24 0.43087592287
25 0.453497829149996
26 0.475647029009998
27 0.49709759963
28 0.517545942039996
29 0.536678737920002
30 0.554261585260001
31 0.570249921490004
32 0.584869139840002
33 0.598420662549996
34 0.610957188009999
35 0.622383462669997
36 0.632555159980001
37 0.641201134759996
38 0.64784512853
39 0.651840133060001
40 0.653312469450003
41 0.6537817812
42 0.65447619635
43 0.65513779773
44 0.654975942239996
45 0.653391720610005
46 0.650067596530003
47 0.644747468250002
48 0.637073866000003
49 0.626606345580004
50 0.61271532591
51 0.594687858330001
52 0.57177921413
53 0.54325498215
54 0.50850669145
55 0.466994061310004
56 0.41821150396
57 0.361781484800004
58 0.297407789140003
59 0.224978129480004
60 0.144644657929995
61 0.0569112835600016
62 0.0371193076399989
63 0.135665701320002
64 0.236344256599998
65 0.337992787080001
66 0.438241319099995
67 0.532412205350004
68 0.616182073630001
69 0.68523086076
70 0.735062556219999
71 0.76249308259
72 0.764890417820002
73 0.739769272599999
74 0.68623606042
75 0.604520083730001
76 0.496411216759999
77 0.365360830692999
78 0.21627148618
79 0.0552760081250006
80 0.110714076520999
81 0.274547002757001
82 0.42937629124
83 0.569141634460001
84 0.68888868605
85 0.785362143160002
86 0.857167649210001
87 0.904486810129997
88 0.928759482209999
89 0.93212701572
90 0.917252299740004
91 0.887434583440004
92 0.84619676416
93 0.797089085739999
94 0.743403922549998
95 0.687899174429997
96 0.633012363770002
97 0.580724587449996
98 0.532430304680004
99 0.489067572259999
100 0.451314436790001
101 0.419607177639995
102 0.394064862640001
103 0.374415612439997
104 0.360157513309999
105 0.350704344659995
106 0.34529467187
107 0.343074838819998
108 0.34319557576
109 0.344804802160002
110 0.34737698976
111 0.350891175430003
112 0.35557869563
113 0.36141806672
114 0.367946059900003
115 0.374469720760003
116 0.37996147878
117 0.384249476419997
118 0.387900188990001
119 0.390587430809997
120 0.391237884479999
121 0.389067920469999
122 0.385182330259998
123 0.381733411030004
124 0.379608833830002
125 0.378316719329995
126 0.376510034809999
127 0.374187605630003
128 0.373018930390004
129 0.373001862000004
130 0.373132990789997
131 0.3729881813
132 0.372081265620001
133 0.369756626710002
134 0.365280786610002
135 0.35798193988
136 0.347305783290001
137 0.332736601700002
138 0.313750925120003
139 0.28979617061
140 0.260316804449996
141 0.224824017479995
142 0.182874540200002
143 0.134068179910005
144 0.0781199696400017
145 0.0149357336399966
146 0.0553625404100047
147 0.132426572200001
148 0.215459791940001
149 0.303179911870004
150 0.393960715580004
151 0.485933956299998
152 0.576637867209996
153 0.662711337000001
154 0.740295310140002
155 0.805482121649999
156 0.854621731070003
157 0.884520359409998
158 0.892251851310002
159 0.875321434289998
160 0.83228242955
161 0.762991470650003
162 0.668607163189998
163 0.551548134250001
164 0.41553375211
165 0.26574937543
166 0.108807498019999
167 0.0479137773200002
168 0.197174156769998
169 0.332593019979999
170 0.449044982449999
171 0.54277003586
172 0.611406804220001
173 0.654006164069999
174 0.670312240959998
175 0.661772432269996
176 0.63188871981
};
\addplot [thick, darkcyan0158115]
table {%
0 0.00000001 
1 0.000202896598900033
2 0.000811774067999886
3 0.00194780135400006
4 0.00375747759699934
5 0.00649935914399935
6 0.00983456900999907
7 0.0135728327000009
8 0.0179687219600023
9 0.0228035066899973
10 0.0274722936599997
11 0.0319185292899996
12 0.035960917830001
13 0.0389487628300045
14 0.0393623355599976
15 0.0375920096200022
16 0.0361048059200064
17 0.0350557098100016
18 0.0338239193099985
19 0.0329320415799958
20 0.0328881216099947
21 0.0326646519600016
22 0.0313094883500065
23 0.0299039024200046
24 0.0295287001400024
25 0.0301316999300055
26 0.0313531568100003
27 0.0332228088899961
28 0.0357661530400039
29 0.0391864549599968
30 0.0432773575499978
31 0.0474654552999956
32 0.0516611716099985
33 0.0561054788199983
34 0.060765013819998
35 0.0651891405500024
36 0.0692448828599965
37 0.072839488739998
38 0.0761513646099984
39 0.0795789568599978
40 0.0829881332899962
41 0.0853706069400065
42 0.0857994332400039
43 0.0844987122399984
44 0.0817465944999984
45 0.0777288766399958
46 0.0723031320199949
47 0.0654954150899958
48 0.0573781901100009
49 0.0479296939599957
50 0.0373585412500044
51 0.025975839459999
52 0.0139574928000041
53 0.00117989745999914
54 0.0120500417900016
55 0.0256483284700043
56 0.0397925686499989
57 0.0542190925500036
58 0.0690621709900016
59 0.0841567814600028
60 0.09890525826
61 0.113090383469995
62 0.126704643659999
63 0.139446626600002
64 0.149567981409998
65 0.155806091570003
66 0.159573917599999
67 0.161771879109998
68 0.162135208910001
69 0.1607141961
70 0.157163099590001
71 0.150429677549999
72 0.139956277329999
73 0.126147375040003
74 0.10938599613
75 0.0899612375000001
76 0.0680703656900008
77 0.0443326454500001
78 0.0195105559679994
79 0.00522125874100055
80 0.0286648684740005
81 0.0500190153649993
82 0.0683121177199996
83 0.0827264813699994
84 0.0930197577600003
85 0.0990565180700002
86 0.100725485300001
87 0.0977084134000002
88 0.0902330140499993
89 0.0790421124800024
90 0.064274861409995
91 0.0463991738400005
92 0.0261398941300006
93 0.00375014541000029
94 0.0197710371599991
95 0.0434236390699991
96 0.0659319973200026
97 0.0867736532199972
98 0.105977737130003
99 0.122958100479999
100 0.137575202240003
101 0.149726502219998
102 0.159738190699997
103 0.16774279298
104 0.173403504619994
105 0.176952885979999
106 0.17789652143
107 0.176164979189999
108 0.17216223443
109 0.165964041739998
110 0.15790664214
111 0.148516654819998
112 0.138488212609998
113 0.128043238190004
114 0.11683663229
115 0.104689294639998
116 0.0918124409299975
117 0.0786400852699973
118 0.0651337743000011
119 0.0509785346699959
120 0.0364552684100003
121 0.0219188939200032
122 0.00830637595999884
123 0.00320579123999631
124 0.0124194162999941
125 0.0196587488200048
126 0.0259216185300032
127 0.0304891750099942
128 0.0315371270500006
129 0.0296942080299997
130 0.0256352650399999
131 0.0192710593699963
132 0.0104247756100018
133 0.000778693230001071
134 0.0139013395000012
135 0.0284245092600013
136 0.0444409604200002
137 0.0621332641300015
138 0.0811144948400025
139 0.101158190929998
140 0.122638967559999
141 0.14563115368
142 0.169549175589999
143 0.19388143027
144 0.218438847809997
145 0.243754768509997
146 0.269423187459997
147 0.29442051959
148 0.318342045020001
149 0.340549204689997
150 0.360437913029997
151 0.377018590090004
152 0.390538309580003
153 0.40095050122
154 0.407325588109998
155 0.40864730759
156 0.403918666039997
157 0.39289714929
158 0.375539998539998
159 0.352491193860001
160 0.324521784760001
161 0.292461548239999
162 0.257919597690001
163 0.223698360499999
164 0.19195194366
165 0.16411746956
166 0.141232064430001
167 0.123381924479998
168 0.109816335849999
169 0.0996142262799996
170 0.0923611631399972
171 0.0875678774600007
172 0.0852162644700005
173 0.0854750191300012
174 0.0873953154599989
175 0.0908746743699951
176 0.0960908113600001
};
\addplot [line width=0.48pt, black, dashed, forget plot]
table {%
-8.79999999999999 1.6
184.8 1.6
};
\end{axis}

\end{tikzpicture}

%% file: diff_delta3.tex
\begin{tikzpicture}

\definecolor{chocolate213940}{RGB}{213,94,0}
\definecolor{darkcyan0158115}{RGB}{0,158,115}
\definecolor{darkgray176}{RGB}{176,176,176}

\begin{axis}[
width=\textwidth,
legend cell align={left},
legend style={
  fill opacity=0.8,
  draw opacity=1,
  text opacity=1,
  at={(0.03,0.97)},
  anchor=north west,
  draw=lightgray204
},
log basis y={10},
tick align=outside,
tick pos=left,
x grid style={darkgray176},
xlabel={Time Steps},
xmin=-8.8, xmax=184.8,
xtick style={color=black},
y grid style={darkgray176},
ylabel={Absolute Delta Error ($\si{\milli\angstrom}$)},
ymin=0.000378696757845086, ymax=20.7683021453191,
ymode=log,
ytick style={color=black}
]
\addplot [thick, darkcyan0158115]
table {%
0 0.00000001 
1 0.000621878655460151
2 0.00189302248598811
3 0.00424840928003878
4 0.00753282611853923
5 0.0122124589890804
6 0.0184122383579677
7 0.0258874813382493
8 0.034572420735457
9 0.0446949291350585
10 0.0560575957159193
11 0.0684298808257822
12 0.0816707517556914
13 0.0954878992798225
14 0.109312739688283
15 0.122154642017369
16 0.13394914219933
17 0.145235316699077
18 0.155907475161143
19 0.166096790964465
20 0.176033133366005
21 0.186049061363214
22 0.195800859272177
23 0.205267416726329
24 0.214662114009112
25 0.224342671561129
26 0.234461951190923
27 0.245223429503616
28 0.256666941425454
29 0.269019525871217
30 0.282448310408689
31 0.297007498082591
32 0.312629101347772
33 0.329414981201204
34 0.347368162731065
35 0.366597784623668
36 0.386954019221397
37 0.408312058778115
38 0.430359462603391
39 0.452783220957986
40 0.475325984087505
41 0.498051133855038
42 0.521043655855147
43 0.544232557528002
44 0.567499907161428
45 0.590697444281307
46 0.613693331289911
47 0.636095994957397
48 0.657805177733506
49 0.67837699037776
50 0.69759249578194
51 0.715141643620631
52 0.730866310396872
53 0.744455534956723
54 0.755632755430335
55 0.764210039630786
56 0.769895077583849
57 0.772310051682323
58 0.771278497787353
59 0.766217755806098
60 0.756958494662952
61 0.743244806779564
62 0.724938444894185
63 0.701685892877002
64 0.673422346847197
65 0.640790476611475
66 0.604271409237134
67 0.563449177892861
68 0.51829877179399
69 0.468719676838369
70 0.414537441394969
71 0.355717954738877
72 0.292378422331319
73 0.224910315053894
74 0.153611791628316
75 0.0788265911370045
76 0.0012318412389023
77 0.0787452106303377
78 0.160169184441727
79 0.242178739112586
80 0.323830220962606
81 0.404226627900184
82 0.482631135816924
83 0.558131040456322
84 0.630110696676955
85 0.697894640809493
86 0.760843689248669
87 0.81860797591593
88 0.8705731197719
89 0.916745556499676
90 0.957102118023467
91 0.991550923886364
92 1.02018307223639
93 1.04264172678725
94 1.05915419596614
95 1.07052007537234
96 1.07728741216473
97 1.08062735785985
98 1.08102343880279
99 1.07930925505917
100 1.07604348888946
101 1.07184429584639
102 1.06721724996794
103 1.06236158649553
104 1.0574400099026
105 1.05284087147117
106 1.04857318000806
107 1.04529500910311
108 1.04309609315045
109 1.04221051020414
110 1.04262616364958
111 1.04455742430365
112 1.04798797958061
113 1.05275748958333
114 1.05875650171217
115 1.06603929864857
116 1.07458704729768
117 1.08420227443773
118 1.09484819691308
119 1.10646227141387
120 1.11883944729385
121 1.13174805324046
122 1.1450640616173
123 1.15894711590148
124 1.17339595124832
125 1.18846725771271
126 1.20406320424449
127 1.21981552788819
128 1.23544528928132
129 1.2506746350498
130 1.26535757120272
131 1.27920277816806
132 1.29196673354715
133 1.30334800051468
134 1.31317356214633
135 1.32138377327196
136 1.3279574041305
137 1.33221641146641
138 1.33380652701387
139 1.33243803832617
140 1.32784857473855
141 1.31938054341041
142 1.30653878820874
143 1.28896533187595
144 1.26631321084614
145 1.23811933007745
146 1.2036180427526
147 1.16251050596805
148 1.11450365421872
149 1.05944265786462
150 0.997301955829699
151 0.928120837171642
152 0.852364478335699
153 0.769580685229809
154 0.680124050787434
155 0.584015640584079
156 0.482340557602656
157 0.376148755459216
158 0.266864322806204
159 0.156315506472215
160 0.0463578190293967
161 0.0607271343305849
162 0.162012721591664
163 0.254438622148911
164 0.335605365077818
165 0.40260199953801
166 0.453938671914123
167 0.489591766647313
168 0.510609065218826
169 0.518790584653761
170 0.515793867877701
171 0.503016160620873
172 0.481844740468862
173 0.452993541334878
174 0.417379399785256
175 0.376186174368187
176 0.329994306896175
};
\addplot [thick, chocolate213940]
table {%
0 0.00000001 
1 0.00303482223074081
2 0.0121085032470658
3 0.0270816168750354
4 0.0478551774473202
5 0.0743017589708295
6 0.106121879589804
7 0.143238769390597
8 0.185573479995482
9 0.233172939669957
10 0.286049694245949
11 0.344319513950198
12 0.407955887886335
13 0.476912923241479
14 0.550922685558564
15 0.630129857208339
16 0.714825902125127
17 0.805467545147875
18 0.902526566575412
19 1.00614295069423
20 1.11626696478906
21 1.23256946147476
22 1.3553109119171
23 1.48488292040416
24 1.62145340377551
25 1.76515484341544
26 1.91605485381863
27 2.07417588567993
28 2.23959640471927
29 2.41213540600715
30 2.59167237724856
31 2.77794650379404
32 2.97073659152192
33 3.16973036996115
34 3.37461590062871
35 3.5850306573102
36 3.80063374117978
37 4.02095233146649
38 4.24562380777105
39 4.47430687715578
40 4.70645954149629
41 4.94125123379874
42 5.17773899761576
43 5.41515795791203
44 5.65263527463178
45 5.88925694884535
46 6.12402973951265
47 6.35602964402571
48 6.58411243900292
49 6.80731430116777
50 7.02442046899154
51 7.23412710150795
52 7.4350195643671
53 7.6254934518869
54 7.80379385858532
55 7.96805531668676
56 8.11623273433337
57 8.24619019469242
58 8.35554209925871
59 8.44179254141308
60 8.50229584503692
61 8.53422613218468
62 8.53466802525271
63 8.50076678365541
64 8.42977871425077
65 8.31917066533094
66 8.16577091741466
67 7.96721805488476
68 7.72168467432821
69 7.42790476969368
70 7.08531649857846
71 6.69420007614741
72 6.25525547299621
73 5.77023799258725
74 5.24185689028411
75 4.67380301864628
76 4.07066106814979
77 3.43800781002412
78 2.78220271136331
79 2.11013673717309
80 1.42926200033933
81 0.74715217911614
82 0.0712501089644713
83 0.591382322002509
84 1.23441673028835
85 1.85242004686026
86 2.44090023965127
87 2.99646993695868
88 3.51653832864485
89 3.9994762286385
90 4.44445421144124
91 4.85130982329562
92 5.2208152071711
93 5.554084656195
94 5.85284308270806
95 6.11920293220736
96 6.35544387313802
97 6.5641424343057
98 6.74798932825826
99 6.90969698807886
100 7.0521073785863
101 7.1780458124886
102 7.29032446203282
103 7.39162920300185
104 7.48451334449252
105 7.57127110027822
106 7.6540462203285
107 7.73466055018113
108 7.81477584447288
109 7.89574028253903
110 7.97877630404464
111 8.06488743260636
112 8.15502115491551
113 8.25004180964828
114 8.35066353052483
115 8.45741007962819
116 8.57059602814925
117 8.69034820208237
118 8.81680257004347
119 8.9500239469027
120 9.08992414779086
121 9.23633252657452
122 9.38891450780144
123 9.54713941270646
124 9.71045023929529
125 9.87822036043995
126 10.0498195761711
127 10.2246518851532
128 10.4020335325998
129 10.5811811752721
130 10.7612993431083
131 10.9415765089713
132 11.1209893770661
133 11.2983585018449
134 11.4722727138783
135 11.6411648444819
136 11.8033814293241
137 11.9572304368498
138 12.1009371768117
139 12.2325958491274
140 12.3501421058292
141 12.4514685550048
142 12.5343053133617
143 12.5962220365781
144 12.6346792735645
145 12.6469828467741
146 12.6303104855963
147 12.5816648311563
148 12.4980544942926
149 12.376402523771
150 12.2136991748194
151 12.0070228065478
152 11.7534613355046
153 11.4506116168378
154 11.0964362333256
155 10.6897114555411
156 10.2298748890894
157 9.71707006117672
158 9.15233474385524
159 8.53766164129965
160 7.87629668734224
161 7.17243366932696
162 6.43128973489038
163 5.65877936949466
164 4.86150635312466
165 4.04678089999533
166 3.22261422299819
167 2.39758440360571
168 1.58026994942018
169 0.779008182866581
170 0.00134844901122921
171 0.745750777597753
172 1.45638735178599
173 2.12566869920272
174 2.74959870928071
175 3.32549716704977
176 3.85191125851769
};
\end{axis}

\end{tikzpicture}